\documentclass[12pt]{report}

\usepackage[utf8]{inputenc}
\usepackage{amssymb} 
\usepackage{calrsfs} 
\usepackage{cancel}
\usepackage{tabularx}
\usepackage[hyphens]{url}
\usepackage{booktabs}
\usepackage{adjustbox}
\usepackage[justification=centering]{caption}
\usepackage{graphicx}
\usepackage{svg}
\usepackage{amsmath}
\usepackage{comment}
\usepackage{colortbl} 
\usepackage{gensymb}
\usepackage{xcolor}
\usepackage{xltabular, booktabs}
\usepackage{rotating}
\usepackage{pbox}
\usepackage[titletoc,title]{appendix}
\DeclareMathAlphabet{\pazocal}{OMS}{zplm}{m}{n} 
\usepackage{epsfig, float,array,tabu,longtable,}
\usepackage{hyperref}
\usepackage{wrapfig}
\usepackage{enumerate}
\usepackage{graphicx,psfrag}
\usepackage{sectsty}
\usepackage{epstopdf}
\usepackage{subfig}
\usepackage{amsmath,esint, setspace, fancyhdr, amsfonts, bookmark, blindtext}
\usepackage[normalem]{ulem}
\usepackage{tikz}
\usepackage{rotating}
\usepackage[americanvoltages,fulldiodes,siunitx]{circuitikz}
\usepackage{stackengine}
\usetikzlibrary{matrix}
\usepackage{multirow}
\usepackage{multicol}
\usepackage{pdfpages}
\usepackage{apalike}

\usepackage[square, sort]{natbib}
\usepackage{attachfile2}
\usepackage{acronym} 

\usetikzlibrary{shapes,backgrounds,patterns}
\usetikzlibrary{mindmap,trees,decorations.markings}
\usetikzlibrary{quotes,angles}
\usepackage{verbatim}

\usepackage{graphicx}
\usepackage{listings}
\newcommand{\blue}[1]{\textcolor{black}{#1}}

\usepackage{color} 

\newcommand\MyBox[2]{
  \fbox{\lower0.75cm
    \vbox to 1.7cm{\vfil
      \hbox to 1.7cm{\hfil\parbox{1.4cm}{#1\\#2}\hfil}
      \vfil}%
  }%
}

\definecolor{mygreen}{RGB}{28,172,0} 
\definecolor{mylilas}{RGB}{170,55,241}

\usepackage[utf8]{inputenc}
\usepackage[italian, main = english]{babel} 			
\usepackage[T1]{fontenc}							

\usepackage[ 
		layoutoffset=0pt,						
		bottom=4cm, 
		top=2.5cm,	
		left=4cm, 
		right=2.5cm,			
		bindingoffset=0pt,						
		headheight=26pt				
		]{geometry}	
					
\usepackage{microtype, fancyhdr}	 
\usepackage{setspace}					 
\usepackage{pdflscape} 					 

\usepackage{amsfonts,amsmath,amsthm,amssymb, mathrsfs, amscd}	 
\numberwithin{equation}{chapter}	
\usepackage{faktor}
\allowdisplaybreaks

\usepackage{url, enumitem} 			
\usepackage{epigraph}						
\usepackage[nottoc]{tocbibind}			
\usepackage{blindtext}						
\usepackage{imakeidx}					
\makeindex

\usepackage{graphicx} 				
\graphicspath{{Pictures/}}				
\usepackage{xcolor}
\usepackage{longtable} 				
\usepackage{lscape} 					
\usepackage{booktabs} 				
\usepackage{multicol} 					
\usepackage{multirow}		
\usepackage{wrapfig} 					

\let\headruleORIG\headrule
\renewcommand{\headrule}{\color{black} \headruleORIG}

\usepackage{colortbl}
\arrayrulecolor{black}

\fancypagestyle{plain}{
  \fancyhead{}
  \fancyfoot{}
  
}

\makeatletter
\def\cleardoublepage{\clearpage\if@twoside \ifodd\c@page\else%
    \hbox{}%
    \thispagestyle{empty}
    \newpage%
    \if@twocolumn\hbox{}\newpage\fi\fi\fi}
\makeatother

\usepackage{etoolbox}
\patchcmd{\chapter}{plain}{empty}{}{}
\makeatletter
  \let\ps@plain\ps@empty
\makeatother

\usepackage{breqn} 
\usepackage{pdfpages}

\begin{document}

\begin{titlepage}

\newcommand{\HRule}{\rule{\linewidth}{0.5mm}} 

\center 
 

\textsc{\LARGE Queen Mary University of London}\\[1.5cm] 
\textsc{\Large Centre For Digital Music}\\[0.5cm] 
\textsc{\large PhD Thesis}\\[0.5cm] 


\HRule \\[0.6cm]
{ \huge \bfseries Towards Context-Aware Neural Performance-Score Synchronisation}\\[0.4cm] 
\HRule \\[1.0cm]
 

\begin{center} \large
\medskip
Author: \quad {\textsc{\textbf{Ruchit Agrawal} }} \\
Supervisors: \quad  {\textsc{\textbf{Professor Simon Dixon}}} 	\quad
{\textsc{\textbf{Dr Daniel Wolff} }} \\
Independent Assessor: \quad 
{\textsc{\textbf{Dr Emmanouil Benetos} }}   
\end{center}

~


\begin{center}
{\large 30 April 2022}
\end{center}

\newcommand*{\plogo}{\includegraphics[width=0.25\textwidth]{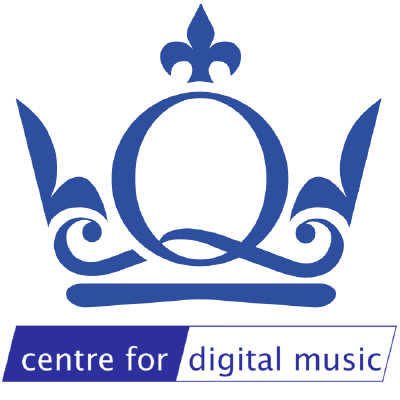}}

\plogo\\[1cm] 
 

\vfill 
\end{titlepage}

\newpage
\newpage

\chapter*{Statement of originality}
\label{statement}

I, Ruchit Rajeshkumar Agrawal, confirm that the research included
within this thesis is my own work or that where it has been carried out in
collaboration with, or supported by others, that this is duly acknowledged
below and my contribution indicated. Previously published material is also
acknowledged below.

\bigskip
\noindent
I attest that I have exercised reasonable care to ensure that the work is
original, and does not to the best of my knowledge break any UK law, infringe
any third party’s copyright or other Intellectual Property Right, or contain any
confidential material.

\bigskip

\noindent
I accept that the College has the right to use plagiarism detection software to
check the electronic version of the thesis.

\bigskip

\noindent
I confirm that this thesis has not been previously submitted for the award of a
degree by this or any other university.

\bigskip

\noindent
The copyright of this thesis rests with the author and no quotation from it or
information derived from it may be published without the prior written consent
of the author.

\bigskip

\noindent
Signature: Ruchit Agrawal

\noindent
Date: 30 April 2022

\noindent
Details on collaborations and publications: Please see Section 1.3

\chapter*{Abstract}
\label{abstract}
\vspace{-1cm}
Music can be represented in multiple forms, such as in the audio form as a recording of a performance, in the symbolic form as a computer readable score, or in the image form as a scan of the sheet music. Music synchronisation provides a way to navigate among  multiple representations of music in a unified manner by generating an accurate mapping between them, lending itself applicable to a myriad of domains like music education, performance analysis, automatic accompaniment and music editing. Traditional synchronisation methods compute alignment using knowledge-driven and stochastic approaches, typically employing handcrafted features. These methods are often unable to generalise well to different instruments, acoustic environments and recording conditions, and normally assume complete structural agreement between the performances and the scores. This PhD furthers the development of performance-score synchronisation research by proposing data-driven, context-aware alignment approaches, on three fronts: Firstly, I replace the handcrafted features by employing a metric learning based approach that is adaptable to different acoustic settings and performs well in data-scarce conditions. Secondly, I address the handling of structural differences between the performances and scores, which is a common limitation of standard alignment methods. Finally, I eschew the reliance on both feature engineering and dynamic programming, and propose a completely data-driven synchronisation method that computes alignments using a neural framework, whilst also being robust to structural differences between the performances and scores.
\cleardoublepage
\thispagestyle{empty}
\epigraph{The real voyage of discovery lies not in seeking new landscapes,
but in seeing with new eyes.}
{\textit{À la recherche du temps perdu}\\
\textsc{Marcel Proust}}

\cleardoublepage
\thispagestyle{empty}
\begin{center}
\vspace*{9cm}
\textit{Dedicated to my parents, Dr Rajesh Agrawal and Krishna Agrawal}
\end{center}

 
\chapter*{Acknowledgements}
\label{acknowledgements}

\medskip
A number of people have played a pivotal role in the development of the work presented in this thesis. Above all, I would like to express my deepest gratitude to Professor Simon Dixon, the primary supervisor of my PhD.  His feedback at different stages of the PhD was very relevant and encouraging, at the same time being pragmatic, which helped me maintain fortitude throughout the PhD and keep making progress, howsoever small, eventually leading to the completion of this PhD. 
\par This dissertation would not have been possible without
my secondary supervisor, Dr Daniel Wolff, who has also provided great guidance on various aspects of this project. I am very grateful for his guidance and support throughout this PhD. The regular meetings with my supervisors not only kept my research on track but helped broadening my horizons by shedding light on domains I was not familiar with. 
I also wish to extend thanks to my two examiners, Dr Huy Phan and Dr Cynthia Liem, for the attention and feedback they provided to my research, which helped to improve the quality of the dissertation.

\par In addition to my supervisors, I would also like to thank my independent assessor, Dr Emmanouil Benetos, who was always available for help and guidance throughout my PhD. His attentiveness and prompt guidance helped me to develop my acumen and research judgment.

\par I would also like to thank my colleagues and friends at Queen Mary, including (but not limited to) Saumitra Mishra, Adrien Ycart, Changhong Wang, Daniel Stoller,  Arjun Pankajakshan, Delia Fano Yela, as well as the MIP-Frontiers fellows Vinod Subramanian, Alejandro Delgado, Emir Demirel and Carlos Lordelo for the thought-provoking discussions we shared (in addition to the jovial ones), which not only fostered research-related activities, but also ensured a pleasant experience that lightened up the PhD journey.  

\par This research was carried out as part of the MIP-Frontiers programme, funded by the European Union's Horizon 2020 research and innovation programme under the Marie Skłodowska-Curie grant agreement No. 765068. The extraordinary avenues for collaboration and the  trainings offered by this programme as part of various meetups and workshops made it a particularly unique experience, despite the dampening effect of Covid-19. 

\par I am grateful to the MIP-Frontiers programme for bringing various eminent researchers together and providing the intellectually stimulating environment that enabled the development of this research. I would also like to thank the amicable Mr Alvaro Bort, the programme manager for MIP-Frontiers, who ensured a smooth ride throughout the project. 

\par Last but not the least, this research would never have existed without the unconditional support and love of my parents. I am profoundly indebted to their immense contributions in my personal and professional development.

\chapter*{\blue{List of Acronyms}}
\label{acroList}
\vspace{-0.5cm}
\begin{acronym}[MPC] 
 \acro{DTW}{Dynamic Time Warping}
 \acro{HMM}{Hidden Markov Model}
 \acro{NWTW}{Needleman-Wunsch Time Warping}
 \acro{CRF}{Conditional Random Field}
 \acro{ReLU}{Rectified Linear Unit}
 \acro{CNN}{Convolutional Neural Network}
 \acro{RNN}{Recurrent Neural Network}
 \acro{LSTM}{Long Short-Term Memory}
 \acro{OMR}{Optical Music Recognition}
 \acro{MATCH}{Music Alignment Tool CHest}
 \acro{MLP}{Multi-Layer Perceptron}
 \acro{STFT}{Short-Time Fourier Transform}
 \acro{CQT}{Constant-Q Transform}
\end{acronym}


\tableofcontents
\listoffigures
\listoftables
\chapter{Introduction}\label{ch:introduction}
Recent years have witnessed a burgeoning amount of automation in all areas of media processing across multiple data modalities. The automated processing of audio-visual and textual content has also impacted all facets of music, be it composition, education, analysis or editing. 
The development of such technologies is enabled by the fundamental research conducted in the areas of Music Information Processing (MIP), Signal Processing and Machine Learning. The heart of this thesis lies at the crossroads of these three fields of study, and develops automated processing methods for \textit{music synchronisation}.

\par The alignment of time-series based media pertaining to multiple information sources that either encode different facets of a single entity or correspond to different entities is an integral task in signal processing, with applications in a variety of scenarios such as performance analysis, video captioning and speech recognition.
Music synchronisation is one such alignment task that is aimed at computing the optimal path or mapping between multiple representations of a piece of music. This task can take on multiple forms, depending upon the input representations and the nature of the alignment computation.  Generally speaking, given a position in one representation of a piece of music, the goal of music synchronisation is to determine the corresponding position in another representation of the same musical piece. It must be noted that the individual correspondences are generally required to correspond sequentially too, albeit with room for jumps to allow for structure-aware alignment. Since the thesis primarily focuses on performance-score synchronisation, it is assumed that the performance, typically represented in the audio domain, increases linearly through time, and each position in the performance axis corresponds to a single position in the score axis. One-to-many correspondences are not typically applicable to the offline synchronisation task, wherein disambiguation of multiple positions is possible with the use of bidirectional context, unlike in the score following or online tracking scenario. The thesis assumes that there is a single alignment path that correctly maps each performance-score pair (as indicated by the ground truth data), referred to as the optimal alignment path in the upcoming chapters.
\par Depending upon the nature of the alignment computation, the task could be classified as either \textit{offline alignment}, wherein the entire information about the entities to be aligned is available beforehand; or \textit{online alignment}, wherein the algorithm does not have \textit{a priori} access to future performance events whilst aligning the current event.
Additionally, a given piece of music could be represented using different formats, for instance audio recordings, symbolic representations and sheet music images. These representations could correspond to a single entity (multi-modal representations) or multiple entities (uni-modal/cross-modal representations). Depending upon the nature of the representations, the alignment task could further be classified into audio-to-score alignment, audio-to-audio alignment, audio-to-image alignment, lyrics-to-audio alignment and so on.  
The primary focus of this PhD is the offline audio-to-score alignment task, also called \textit{performance-score synchronisation}; however, some of the proposed methods are also applicable to audio-to-audio alignment and audio-to-image alignment.
\section{Motivation}
Having a reliable alignment of a score to an acoustic realisation of the score has applications in multiple domains.
These span from applications in the entertainment domain, where alignment could be used to drive an automatic accompaniment system; the performance domain, for automatic page turning and synchronised visualisation generation; to the music education setting, for digital illustration and automatic assessment of student performances. 
Additionally, robust alignment can also aid audio editing and analysis,  wherein selecting a measure in the score could automatically select the corresponding audio, enabling convenient navigation. Research on automatic alignment is also underpinned by the industrial interest in commercial alignment applications such as MusicPlusOne \citep{raphael2001bayesian, raphael2006aligning}, Tido Music and Antescofo \citep{cuvillier2014coherent, donat2016embedding}.

\par
\begin{figure}[H]
\begin{center}
\includegraphics[width=6in]{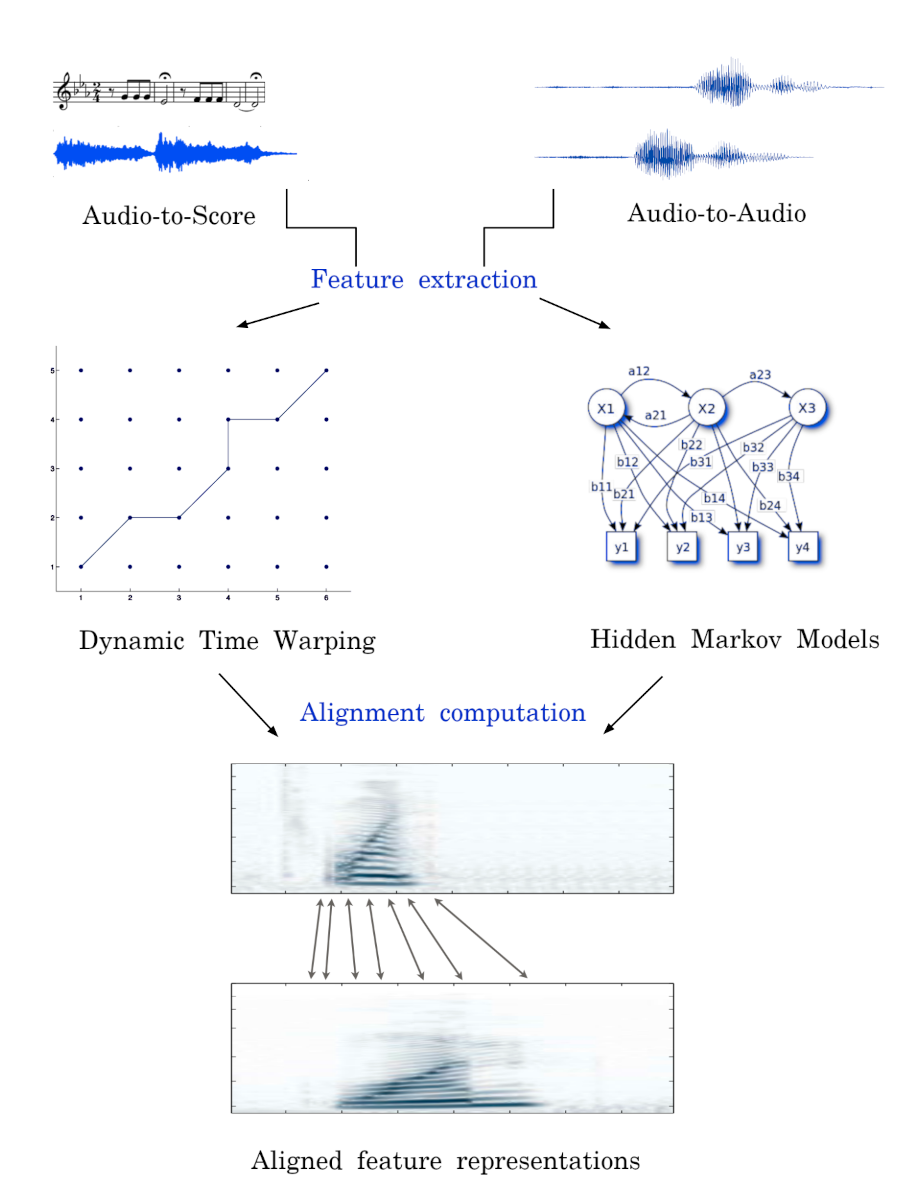}
\caption{Traditional methods for music synchronisation}
\label{fig:trad_pipeline}
\end{center}
\end{figure}

 Traditional audio synchronisation methods (as well as recent optimisations) rely on knowledge-driven approaches and stochastic approaches that are essentially based on Dynamic Time Warping or Hidden Markov Models. Figure \ref{fig:trad_pipeline} demonstrates the general pipeline employed by traditional music alignment methods. These methods bear certain limitations such as the inability to adapt to specific test settings, and the inability to capture structural differences; described in greater detail in Chapter 2.
\par Neural methods offer promise at overcoming these limitations by enabling data-driven learning. While neural networks have been around for a long time  \citep{mcculloch1943logical}, they revolutionised computation in a myriad of domains only in the recent years, equipped by a surge in the massive amounts of data generated by the expanse of the Internet and other media platforms coupled with the development of robust graphical processing units. 
Deep neural networks have demonstrated comprehensive success in a variety of fields such as computer vision, natural language processing, speech recognition, and more recently music information processing \citep{sigtia2016end, dieleman2014end, stoller2018wave, pons2017end}. While neural methods have been explored for various MIP tasks such as music transcription, music generation, genre classification and onset detection, their application to music synchronisation has been fairly limited. The alignment task entails various aspects such as multiple inputs, temporal dependencies and cross-modality that make it especially challenging to model.

\par Drawing from this motivation, this thesis proposes \textit{data-driven}, \textit{context-aware} neural methods for the synchronisation of music performances with the corresponding scores.
Some of the general benefits of this approach over traditional methods are as follows:
\begin{itemize}
     \item Support for end-to-end training and joint optimisation of all components against the same loss function
     \item Self-learning capability, no feature design and engineering required
     \item Ability to combine supervised as well as unsupervised learning approaches
     \item Multiple possible extensions and adaptations, such as measure-based alignment or domain-specific features
 \end{itemize} 
\section{Contributions of the thesis}
The primary contributions of the thesis are laid out into three fronts of data-driven performance synchronisation, corresponding to Chapters 3, 4, and 5 respectively. These are summarised below:\\ \\
\textbf{Metric Learning for Audio-to-Score Alignment}
\begin{itemize}
    \item I present a novel approach for data-driven performance-score synchronisation using learnt spectral similarity at the frame level.
    \item I propose Siamese CNNs to learn the frame similarity and demonstrate that it outperforms  traditional feature representations for various test settings.
    \item I conduct experimentation for various acoustic settings and demonstrate that the proposed approach offers high domain coverage as well as adaptability to a particular test setting.
    \item I present a study to analyse the data requirements of the model, and demonstrate that the Siamese CNNs are able to learn meaningful representations even in the presence of limited data.
    \item I discuss various optimisation methods such as deep salience representations and data augmentation to improve the performance of the proposed method in data-scare settings.
    \\
\end{itemize}
\textbf{Structure-aware Performance Synchronisation}
\begin{itemize}
    \item I propose a method for the alignment of performances to scores or other performances in the presence of structural differences.
    \item I present a data-driven method that employs a progressively dilated CNN architecture to detect inflection points, coupled with DTW to generate fine alignments.
    \item I present experimentation  conducted with varying dilation rates at different layers of the network and demonstrate that progressively increasing dilation optimally captures both short-term and long-term context.
    \item I present various ablative analyses to assess model performance for structure-aware as well as monotonic alignment and demonstrate that the dilated CNN models outperform previously proposed structure-aware methods without requiring manually annotated data.
    \item I demonstrate that the proposed method is also compatible with learnt similarity presented in Chapter 3, and can capture various kinds of structural differences regardless of the source and the type of the jumps.\\
\end{itemize}
\textbf{Towards End-to-End Neural Synchronisation}
\begin{itemize}
    \item I present a method for learnt alignment in uni-modal and multi-modal settings in a fully data-driven manner and present a way to eschew the reliance on Dynamic Time Warping and instead learn alignments using a completely neural framework.
    \item I propose a convolutional-attentional architecture trained with a custom loss based on time-series divergence for the audio-to-MIDI and audio-to-image alignment tasks pertaining to different score modalities.  
    \item I present experiments conducted for multiple test settings and comparisons with state-of-the-art data-driven approaches, and demonstrate that the proposed method outperforms contemporary methods for a variety of test settings across score modalities and acoustic conditions.
    \item I demonstrate that the proposed neural method is robust to structural differences between the performances and scores without explicitly modelling them as in Chapter 4.
    \item I present ablative analyses to demonstrate improvements offered by the convolutional-attentional framework and the custom loss and conduct significance testing to validate the effectiveness of the models.
\end{itemize}


\section{Associated publications}
\label{sec:publications}
Most of the work presented in this thesis has been published in journal
articles or international peer-reviewed conference proceedings. The peer-reviewed publications associated with this thesis are listed below:
\vspace{0.5cm}
\begin{itemize}
    \item \textbf{A Hybrid Approach to Audio-to-Score Alignment}\\ Ruchit Agrawal and Simon Dixon \\ Proceedings of the Machine Learning for Music Discovery Workshop at the International Conference on Machine Learning (ICML 2019), California, USA, June 10-15, 2019
    \item \textbf{Learning Frame Similarity using Siamese Networks for Audio-to-Score Alignment}\\ Ruchit Agrawal and Simon Dixon \\ Proceedings of the 28th European Signal Processing Conference (EUSIPCO 2020), Amsterdam, The Netherlands, January 18-21, 2021
    \item \textbf{Structure-Aware Audio-to-Score Alignment using Progressively Dilated Convolutional Neural Networks}\\ Ruchit Agrawal, Daniel Wolff and Simon Dixon \\ Proceedings of the IEEE International Conference on Acoustics, Speech, and Signal Processing (ICASSP 2021), Barcelona, Spain, June 6-11, 2021
    \item \textbf{A Convolutional-Attentional Neural Framework for Structure-Aware Performance-Score Synchronization}\\ Ruchit Agrawal, Daniel Wolff and Simon Dixon \\  IEEE Signal Processing Letters (IEEE SPL), Volume 29, December 2021
\end{itemize}
\vspace{2cm}
\par The author of the thesis is the primary contributor to all the publications listed above. This includes the development and implementation of the models, the experimentation and comparison with state-of-the-art approaches, the generation and analysis of the results and writing and editing of the papers. The supervisors Daniel Wolff and Simon Dixon contributed to all the papers in an advisory role. This entailed sharing their opinions on the research questions during the development of the methods, discussing the results of the experiments, and reviewing and suggesting changes to the drafts of these papers.
\section{Organisation of the thesis}
This thesis is laid out to demonstrate the incremental development of neural alignment methods, starting from the feature-learning level, eventually moving up to the alignment-computation level. The first phase of the research entails the development of neural methods as precursors for DTW-based alignment (Chapters 3 and 4), and the second phase consists of replacing DTW using a completely neural architecure (Chapter 5). The chapters progress as follows: \\
\begin{itemize}
    \item Chapter 2 provides a comprehensive overview of related work and describes their limitations in detail, thereby highlighting the key contributions of the research presented in the further chapters. This chapter also summarises the theoretical background (such as Dynamic Time Warping) that facilitates a better understanding of future chapters.
    \item  Chapter 3 presents a novel approach for data-driven performance-score synchronisation using learnt spectral similarity at the frame level. This chapter is based on the papers \citep{agrawal2020hybrid} and \citep{agrawal2021learning}.
    \item Chapter 4 presents a method that overcomes a major limitation of DTW-based methods and proposes a neural method for structure-aware synchronisation, applicable to both performance-performance synchronisation and performance-score synchronisation. This chapter is based on the paper \citep{agrawal2021structure}.
    \item Chapter 5 builds upon the research presented in the previous chapters and presents a \textit{learnt} alignment system that eschews the reliance on Dynamic Time Warping and enables end-to-end learning in a completely data-driven manner. This chapter is based on the paper \citep{agrawal2021convolutional}.
    \item The thesis concludes with Chapter 6, which provides the reader with a summary of the key takeaways from the thesis, and a description of directions for future research.
\end{itemize}

\chapter{Literature Review}\label{ch:literature}
\begin{sidewaysfigure}[htbp]
  \centering
  \includegraphics[width=8in]{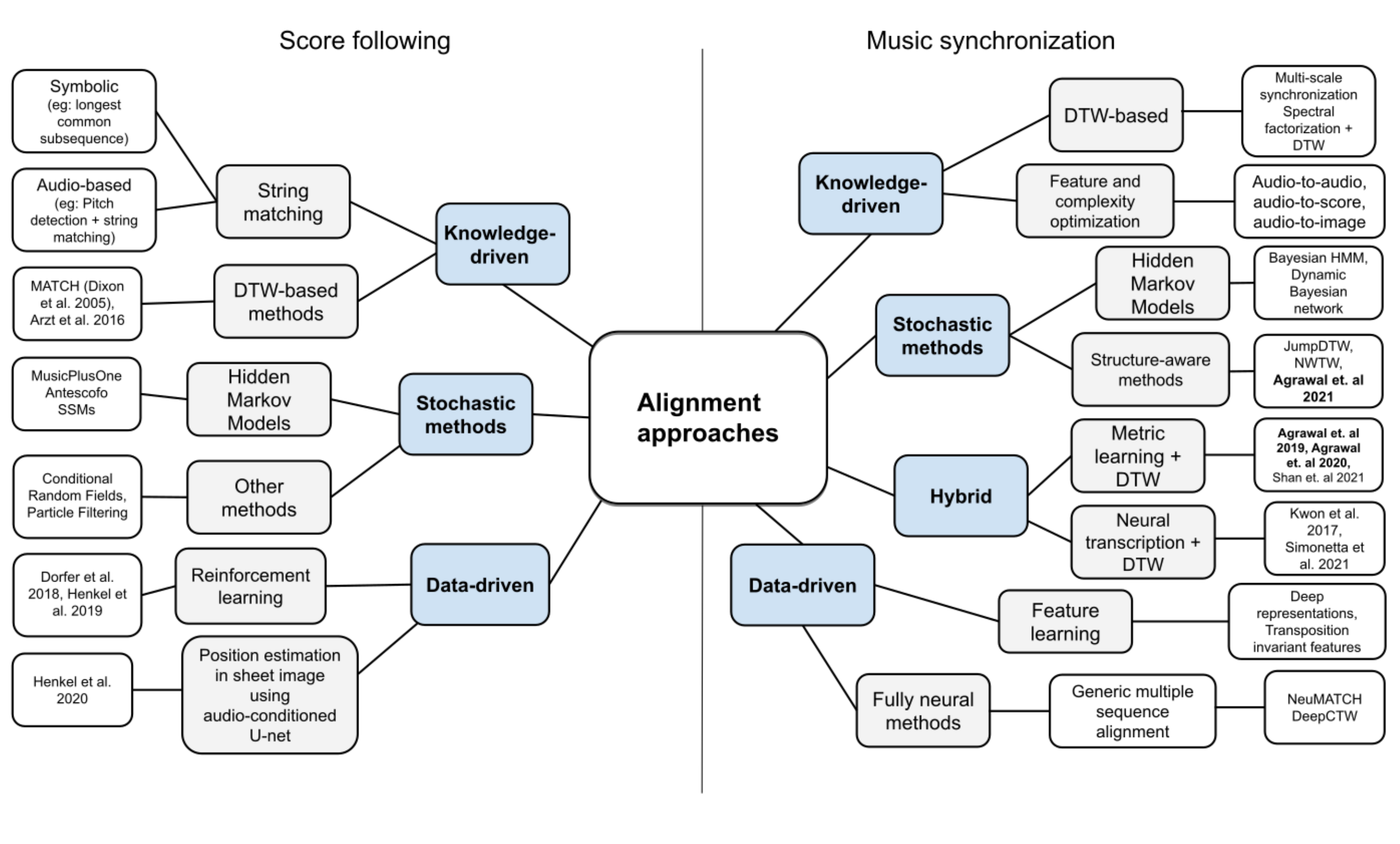}
  \caption{An overview of related work}
  \label{fig:litReview}
\end{sidewaysfigure}

\blue{I} describe a comprehensive summary of the research related to this PhD in this chapter. \blue{I} divide the relevant literature into different categories depending upon the modality of the inputs and the primary approach employed by each method. 
Depending upon the task, the alignment computation is either carried out online, known as \emph{score following}, or offline, known as \blue{\emph{music synchronisation} or \emph{performance-score synchronisation}}. The latter can also further be categorized into \emph{audio-to-score alignment}, \emph{audio-to-audio alignment} and \blue{\emph{audio-to-image alignment}}, depending upon the input modalities. While this thesis is primarily concerned with offline performance-score \blue{synchronisation}, some of \blue{my} methods are also applicable to the online setting, as will be discovered in the later chapters. The various methods for music alignment could broadly be categorised into knowledge-driven, stochastic, and neural approaches. Figure \ref{fig:litReview} presents an overview of the relevant literature. \blue{I} discuss the important alignment algorithms proposed over the years using these approaches and their applications to music processing for both these tasks, i.e. \emph{score following} and \blue{\emph{performance-score synchronisation}}. \blue{I} finally highlight the limitations of the major alignment approaches and motivate \blue{my} research for this PhD.

\section{Knowledge-driven and stochastic approaches}
Traditional methods for music \blue{synchronisation} are typically based on either knowledge-driven approaches, employing various heuristics to compute the alignment, such as the dynamic programming based \blue{Dynamic Time Warping (DTW) method \citep{sakoe1978dynamic}, or stochastic approaches, such as those based on Hidden Markov Models (HMMs) \citep{baum1966statistical} and Particle Filtering \citep{liu1998sequential}. While both DTW and HMM are \blue{optimisation} methods employing algorithms such as the Viterbi algorithm \citep{forney1973viterbi}} and dynamic programming, their mathematical formalisations differ and \blue{I} treat them separately in subsequent discussions. \blue{I} begin by a brief overview of symbolic alignment methods and move on to a more detailed overview of DTW and HMM based approaches.
\subsection{Symbolic alignment methods}
Primitive approaches for tracking a music performance relied on symbolic data formats, and employed string matching to align the two symbolic streams corresponding to the performance and the score \citep{dannenberg1984line}. As research progressed, music tracking saw the advent of the inclusion of the audio signal in 1992, wherein the audio was first converted to a symbolic format via pitch detection, and the alignment was thereafter carried out via string matching \citep{puckette1992score}. Symbolic alignment techniques matured thereafter to employ stochastic approaches, and Hidden Markov Models demonstrated promise for the task \citep{orio2003score, schwarz2004robust}. With the advent of highly effective transcription models based on deep learning \citep{hawthorne2018onsets, ycart2019comparative}, symbolic alignment has seen significant improvement in the recent years \citep{nakamura2015autoregressive, nakamura2017performance}. Since this thesis is primarily concerned with performance-score \blue{synchronisation}, with the performance being in the \textit{audio} domain, as opposed to the \textit{symbolic} domain, \blue{I} keep the discussion on symbolic methods succinct and dedicate the rest of the chapter to audio alignment methods. The reader is referred to \citep{joder2010comparative} and \citep{nakamura2017performance} for a detailed overview on symbolic alignment methods.
\vspace{0.5cm}
\par For the audio alignment task, the computation is either carried out online (\emph{score following}) or offline (\blue{\emph{music synchronisation}, \emph{performance-score synchronisation} or \emph{audio-to-score alignment}}). Figure \ref{fig:litReview} presents an overview of the relevant literature for both \emph{score following} as well as \emph{music synchronisation}. Traditional audio synchronisation methods for both these tasks rely on knowledge-driven approaches and stochastic approaches. The knowledge-driven methods are primarily based on a technique known as Dynamic Time Warping (DTW) \citep{muller2004towards,  dixon2005line, muller2006efficient, arzt2008automatic}, whereas the stochastic approaches mainly utilize Hidden Markov Models (HMMs) and similar state space models \citep{cont2005training, maezawa2011polyphonic, cuvillier2014coherent}.  
\blue{I} briefly discuss these two methods in the subsequent subsections. 

\subsection{Dynamic Time Warping}
\vspace{0.5cm}
Dynamic Time Warping (DTW) is a generic alignment method aimed at generating an optimal mapping between two time-series or sequences. While DTW was originally proposed for spoken word recognition \citep{sakoe1978dynamic}, it has developed to be a prominent technique that has been applied to a variety of tasks, including gesture recognition \citep{gavrila1995towards}, handwriting recognition \citep{vinciarelli2002survey} as well as music processing \citep{muller2015fundamentals}. 
\vspace{1.5cm}
\par DTW is based on a dynamic programming framework and generates an alignment between two sequences $A$= $(a_1,a_2,...,a_m)$ and $B$ = $(b_1,b_2,...,b_n)$ by comparing them using a local cost function, at each point, with the goal of minimizing the overall cost. The path $P$ that yields the minimum global cost is then considered to be the optimal alignment between the two sequences. Generally, $P$ is extracted by placing various constraints during the DTW computation, such as being bound by the ends of the sequences $A$ and $B$, being monotonically increasing, and being continuous. Formally, the general DTW computation can be represented as follows:

\begin{equation}
D(i, j)  = d(i, j) + min\begin{cases}
D(i, j-1) \\ D(i-1, j) \\  D(i-1,  j-1)
\end{cases}
\vspace{0.5cm}
\end{equation}
where $d(i, j)$ is the distance measure (local cost) between points $a_i$ and $b_j$; and $D(i, j)$ is the total cost for the path $P$ which generates the optimal alignment between the sequences $A_{1..i}$ and $B_{1..j}$, and $D(1, 1) = d(1, 1)$. The local cost can be computed using various distance metrics, such as the Euclidean distance, Manhattan distance or any other function as per choice and suitability. Given this formulation, the algorithm results in a quadratic time complexity via backwards recursion in the accumulated cost matrix from point $D(m, n)$, which is not scalable in real world settings. \blue{I} discuss the applications of DTW to audio alignment and the \blue{optimisations} therein in the next subsection. The introduction to DTW has been kept concise for the sake of brevity. \blue{I} refer the reader to \citep{muller2015fundamentals} for an in-depth review of DTW.

\subsection{DTW-based methods for audio alignment}
\par Dynamic Time Warping \citep{sakoe1978dynamic} was first applied to music alignment in the early 2000s \citep{orio2001alignment, hu2003polyphonic}. The former method is based on a standard DTW computation, with the performance features extracted from the peak structure distance given by the harmonic sinusoidal partials and the score features extracted using three instrument models. The latter proposes a method for aligning polyphonic audio recordings by first mapping the MIDI data to corresponding audio features and then matching these features using a standard DTW computation, thereby treating the task as an audio-to-audio alignment task. They compare three different feature representations (chromagram, pitch histogram and MFCC) and also demonstrate results on music retrieval in addition to audio alignment. While these initial approaches demonstrated the applicability of DTW to the audio alignment tasks, they had severe limitations in terms of scalability, particularly due to the long running times and the large memory requirements of the algorithm.

\par As DTW-based methods picked up traction for alignment related tasks, research progressed towards optimizing the standard DTW method for music \blue{synchronisation}, both in terms of the algorithm as well as the choice of the feature representations. 
Two notable alignment methods among these are the ones proposed by \citet{muller2004towards} and \citet{dixon2005line}. The former method \citep{muller2004towards} focuses on addressing the space and time complexity of the standard DTW algorithm for the alignment of complex, polyphonic piano music. The latter, also known as \emph{MATCH} \citep{dixon2005line}, originally proposed a DTW-based method for online music alignment. Rather than employing a standard DTW computation, the online alignment in this method is computed incrementally and has a linear time and space complexity. 

Methods were subsequently proposed to improve multiple facets of the algorithm, such as speed, memory complexity, modelling musical structure, and adaptation to the task at hand \citep{muller2006efficient, salvador2007toward, ewert2008refinement, arzt2008automatic, zhou2009canonical, arzt2010towards}. For instance, \citet{arzt2008automatic} build upon \emph{MATCH}, the online DTW method proposed by \citet{dixon2005line}, and propose various \blue{optimisations} to improve the alignment accuracy for real-time music tracking. These \blue{optimisations} include the backward-forward strategy, incorporation of musical information from the score, and maintaining multiple hypotheses. 

\par A particular track of alignment research worth mentioning here is the multi-scale approach to audio \blue{synchronisation} \citep{salvador2007toward, muller2006efficient}. This approach was initially proposed by \citet{muller2006efficient}, wherein they recursively projected the alignment path computed at a coarse resolution level to a higher level and then refined the projected path using various interpolation methods. This method yielded comparable performance to the classical DTW-based alignment method, with much lower time and space complexity. Building upon this approach, \citet{ewert2008refinement} proposed refinement strategies for music \blue{synchronisation}. They introduced novel audio features that combined onset and chroma features, and demonstrated their usage within a robust multi-scale \blue{synchronisation} framework inspired by \citep{muller2006efficient}. While the onset based features work well for music containing instrumentation with clear onsets, such as piano, these are unable to model music with smooth note transitions, such as soft violin music or orchestral music. A similar method, called FastDTW \citep{salvador2007toward}, albeit focused on generic sequence alignment, theoretically and empirically proved that a linear time and space complexity could be achieved using a multi-level \blue{synchronisation} approach. Recent alignment research has also demonstrated the applicability of multi-scale methods in memory-restricted conditions \citep{macrae2010accurate, pratzlich2016memory}.

Apart from these major earlier approaches, various other alignment methods either based mainly on DTW or utilizing DTW at some step have recently been proposed \citep{carabias2015audio, chuan2016active, wang2016robust}. Notable methods among these are the integration of active learning with DTW for audio-to-score alignment \citep{chuan2016active}, and an extension of DTW for joint alignment of multiple performances of a piece of music \citep{wang2016robust}.
While DTW is an extremely suitable method for time-series alignment, there are some inherent limitations to this approach, especially for audio-to-score alignment and audio-to-audio alignment in real world settings. \blue{I} discuss these in detail towards the end of the chapter, in Section \ref{sec:limitations}.
\subsection{Methods based on Hidden Markov Models}
In addition to DTW-based methods, several methods based on Hidden Markov Models (HMMs) have been proposed over the years for the alignment task \citep{cano1999score, orio2000score, schwarz2004robust, cont2005training, cont2006realtime, maezawa2011polyphonic, gong2015real, cuvillier2014coherent}. 
\citet{cano1999score} focus on monophonic score following using HMMs. They work with the singing voice and believe that once the score-matching  problem for the singing  voice  case is solved, they would have solved it for any other harmonic instrument. \blue{I} however disagree with that, and believe that a method optimally working for singing voice might not translate well to complex music with different instrumentation used, such as classical piano music.
Two other early methods \citep{orio2000score, schwarz2004robust}  focus on score following for polyphonic music. While the former combines spectral analysis with HMMs, the latter employs score parsing into score events and score states with the HMM proposed by the former. 
Developing on top of these methods, Cont et al. propose a novel method for real-time audio-to-score alignment that focuses on correctly choosing the music event sequence that was performed, as opposed to modelling the music signal \citep{cont2005training, cont2006realtime}. 
\par In addition to HMM-based approaches, methods based on other state space models and hybrid Markov frameworks \citep{duan2011state, cont2009coupled}, as well as methods based on Conditional Random Fields (CRF), Particle Filtering (PF) and Monte-Carlo Sampling (MCS)\citep{joder2011conditional, montecchio2011unified, yamamoto2013robust, otsuka2011real, Korzeniowski2013tracking} have also demonstrated effectiveness for score following. With the exception
of Monte Carlo methods, which are online methods, the remaining
three methods essentially operate in an offline manner (despite being sometimes applied to score following) and are thereby quite similar algorithmically.
\par A notable CRF based approach among these is by \citet{joder2011conditional}, who propose a conditional random field framework for audio-score alignment and demosntrate that it is particularly well suited to design
flexible observation functions. Among the particle filtering methods, a prominent approach is by \citet{montecchio2011unified}, who present a methodology for the real time alignment of music signals using sequential Monte Carlo inference techniques. The alignment problem is formulated as the state tracking of a dynamical system, and differs from traditional Hidden Markov Model - Dynamic Time Warping based systems in that the hidden state is continuous rather than discrete.

\par Among Bayesian approaches, a method worth mentioning here is the work proposed by \citet{maezawa2011polyphonic}, which deviates from classic methods that use ad-hoc feature design, and develops a Bayesian audio-to-score alignment method by modeling music performance using a Bayesian Hidden Markov Model. They model each state of the HMM to emit a Bayesian signal model based on Latent Harmonic Allocation. While their method works well for orchestral music, it performs poorly on solo piano and vocal music. A similar work to this is proposed by
\citet{Korzeniowski2013tracking}, who approach score following using a Dynamic Bayesian Network and employ particle filtering for inference, better modelling rests and tempo changes than other approaches.  \citet{cuvillier2014coherent} present a novel insight to the problem of duration modeling for recognition setups where events are inferred from  time-signals  using  a  probabilistic  framework. 
A similar method, albeit focused on real-time alignment of singing voice, is that proposed by \citet{gong2015real}. This HMM-based method integrates lyrics information with the observation mechanism and proposes fusion strategies to exploit information from the music as well as lyric signals.
A notable HMM-based approach that tackles the problem of handling sustain pedal effects in score following is that proposed by \citet{li2015score}. They propose modified feature representations for the performance audio in order to attenuate the effect of the sustain pedal in expressive piano performance as well as reverberation in the recording environment. 


\section{Incorporating structure in music \blue{synchronisation}}

The extraction of music information from audio has been studied to a considerable extent in recent work in Music Information Processing (MIP), however the consideration of the performance environment and the structural aspects are still areas with a significant scope for improvement \citep{paulus2010state, widmer2017getting}.
Early work on structure-aware \blue{MIP} focuses on structural segmentation of musical audio by constrained clustering \citep{levy2008structural} and music repetition detection using histogram matching \citep{tian2009histogram}. 
\citet{arzt2010towards} propose a multilevel matching and tracking algorithm to deal with issues in score following due to deviations from the score in live performance. 
A challenge faced by this approach appears when complex piano music is played with a lot of expressive freedom in terms of tempo changes. Hence, they propose methods to estimate the current tempo of a performance, which could then be used to improve online alignment \citep{arzt2010simple}. This is similar to the work proposed by \citet{muller2009towards}, wherein they develop a method for automatic extraction of tempo curves from music recordings
by comparing performances with neutral reference representations.

\par Work specifically on incorporating structural information for offline music \blue{synchronisation} includes \begin{math}\textit{JumpDTW}\end{math} \citep{Fremerey2010handling} for audio-to-score alignment and Needleman-Wunsch Time Warping (\begin{math}\textit{NWTW}\end{math}) \citep{grachten2013automatic}  for audio-to-audio alignment.
\citet{Fremerey2010handling} focus on tackling structural differences induced specifically by repeats and jumps, using a novel DTW variation called \begin{math}\textit{JumpDTW}\end{math}. This method identifies the \textit{block sequence} taken by a performer along the score, however it requires manually annotated block boundaries to yield robust performance, which are generally not readily available at test time for real world applications. Additionally, it cannot perform intra-block jumps or align deviations that are not foreseeable from the score. \blue{\begin{math}\textit{NWTW}\end{math}} \citep{grachten2013automatic}, on the other hand, is a pure dynamic programming method to align
music recordings that contain structural differences. This method is an extension of the classic Needleman-Wunsch sequence alignment algorithm  \citep{needleman1970needleman}, with added capabilities to deal with the time warping aspects of aligning music performances. A limitation of this method is that 
it cannot successfully align repeated segments owing to its waiting mechanism, which skips unmatchable parts of either sequence, and makes a clean jump when the two streams match again. 
\vspace{-0.3cm}
 \par Apart from \begin{math}\textit{JumpDTW}\end{math} and \begin{math}\textit{NWTW}\end{math}, which focus on offline alignment, work on online score following \citep{nakamura2015real} has demonstrated the effectiveness of HMMs for modeling variations from the score for real-time alignment of monophonic music.
Similar to \citet{nakamura2015real}, \citet{jiang2019offline} propose an HMM-based approach for offline score alignment in the practice scenario. They propose using pitch trees and beam search to model skips, however, their method struggles with pieces containing both backward and forward jumps.
Very recently, \citet{shan2020improved} propose Hierarchical-DTW  to automatically generate piano score following videos given an audio and a raw image of sheet music.
Their method is reliant on an automatic music transcription system \citep{hawthorne2017onsets} and a pre-trained model to extract bootleg score representations \citep{tanprasert2019midi}. It struggles when the bootleg representation is inaccurate, and also struggles on short pieces containing jumps. 

\par Apart from alignment-specific research, work on analyzing music structure in MIR is moving towards the use of machine learning based methods \citep{serra2014unsupervised, ullrich2014boundary, mcfee2014analyzing, grill2015music}.
\citet{ullrich2014boundary} apply Convolutional Neural Networks to the boundary detection task, requiring human annotated audio data for training. This approach, which is actually an adaptation of an onset detection method proposed by \citet{schluter2014improved}, proposes a CNN-based binary classifier trained directly on mel-scaled magnitude spectrograms to detect \blue{boundary-containing spectrogram} excerpts.
Based on this architecture, \citet{grill2015music, grill2015music2} present very similar methods for music boundary detection using neural networks on spectrograms and self-similarity lag matrices. A limitation of these methods is that the network cannot take advantage of structural information contained within the lag matrices over longer time contexts. 
\citet{mcfee2014analyzing} propose the application of techniques from spectral graph theory to analyze repeated patterns in musical recordings. They focus on popular music and evaluate their method on the Beatles-TUT and SALAMI datasets. Their method struggles to automatically select a single “best” segmentation without \textit{a priori} knowledge of the evaluation criteria.
\section{Data-driven/Neural Approaches}
Neural networks present a promising data-driven alternative to traditional knowledge-driven approaches. The advent of neural networks was marked with the first mathematical model tracing back to \citet{mcculloch1943logical}, and with the basic forms being employed in machine learning research until the late nineties. However, the recent surge in the amount of data being generated by the expanse of the internet and media platforms, coupled with the development of robust hardware such as graphical processing units has made it realistically possible to use deep learning architectures in a myriad of domains. 

\par The primary advantage of data-driven methods over knowledge-driven approaches is the ability to \textit{learn} from the data itself, often in an end-to-end fashion, typically achieved using neural networks. Recent advances in Music Information Processing (MIP) have demonstrated the efficacy of neural networks for tasks like music generation \citep{eck2002first}, audio classification \citep{lee2009unsupervised}, onset detection \citep{marolt2002neural}, music transcription \citep{marolt2001sonic}.
Stacking several hidden layers on top of one another makes a neural network \blue{\textit{deep}}, which is generally referred to as a deep neural network. Deep neural networks have witnessed comprehensive success in a variety of fields such as computer vision, natural language processing and speech processing, and Music Information Processing has not been an exception to this trend, evidenced by the success of deep architectures for MIP tasks \citep{sigtia2016end, dieleman2014end, stoller2018wave, pons2017end, hawthorne2018onsets}. For a detailed understanding of the architecture of \blue{deep neural networks} as well as the phenomena underlying the success of these models, the reader is referred to \citet{goodfellow2016deep}. For a detailed overview on the application of deep learning to music processing, the reader is referred to \citet{schluter2017deep}.
\blue{I} now discuss the background on neural approaches relevant to music \blue{synchronisation} on two different fronts, i.e. approaches employing feature learning or another form of neural preprocessing, and approaches having a neural component in the core alignment method itself.
\subsection{Work based on feature learning}
Feature learning or representation learning incorporates a set of methods that equips a machine learning system to automatically discover useful representations from raw data for further processing.
Early approaches for feature learning for Music Information Retrieval
(MIR) employ algorithms like Conditional Random Fields \citep{joder2013learning} or \blue{Deep Belief Networks} \citep{schmidt2012feature}, whereas 
recent work in this direction is moving towards the usage of deep neural networks \citep{thickstun2016learning}. The reader is referred to  \citet{joder2010comparative} for a comparison and evaluation of traditional feature representations for audio-to-score alignment. With the rise in deep learning, feature learning using deep neural networks has recently shown promise for a variety of MIR tasks, including note prediction \citep{thickstun2016learning}, genre classification \citep{sigtia2014improved, oramas2017multi}, chord recognition \citep{korzeniowski2016feature}, fundamental frequency estimation \citep{bittner2017deep}, and generic deep music representations \citep{kim2020one}. 

\par Work specifically on learning features for alignment includes learning the mapping for several common audio representations based on a best-fit criterion \citep{joder2011optimizing}, learning mid-level representations for Dynamic Time Warping using a Multi-Layer Perceptron model \citep{izmirli2012bridging}, and learning transposition-invariant features for alignment \citep{lattner2018learning, arzt2018audio} using gated auto-encoders. A particular limitation of the transposition-invariant features is that they are not robust to tempo variations, as opposed to chroma-based features. With the advent of data-driven approaches, recent methods have demonstrated the efficacy of learnt representations coupled with DTW-based alignment computation for music \blue{synchronisation} \citep{dorfer2017learning, tanprasert2019midi}. 
\par In addition to learning representations for DTW-based alignment models, neural networks have also been used as precursors to a standard \blue{synchronisation} procedure in other ways. A notable direction that employs neural networks to aid music \blue{synchronisation} is that using Automatic Music Transcription (AMT). \citet{kwon2017audio} proposes audio-to-score alignment of piano music using RNN-based automatic music transcription, in combination with DTW.  
A similar method which employs neural networks as a precursor for DTW-based alignment was proposed by 
\citet{waloschek2019identification}. They employ CNNs and focus on the identification and cross-document alignment of measures in music score images. It should be noted that \citet{waloschek2019identification} \blue{generate} a coarse alignment at the measure-level, as opposed to finer levels of alignment (for instance, note-level) generated by the majority of alignment algorithms. A similar line of work for cross modal retrieval is carried out by  \citet{munoz2020score}, wherein they propose a parallel score identification system based on audio-to-score alignment. Their focus is on building a real-time system targeted for handheld devices, using parallel computing techniques with ARM processors. Very recently, \citet{simonetta2021audio} propose a method to aid audio-to-score alignment using AMT using deep neural networks, followed by an HMM-based alignment computation. \blue{I} discuss the limitations of the existing feature learning methods in more detail in Section \ref{sec:limitations}. 
\subsection{Alignment methods employing neural networks}
\par Apart from methods that combine learnt representations with DTW or HMM based approaches and methods that employ other neural preprocessing (such as transcription) in combination with standard DTW or HMM based alignment, another very promising direction of work is to leverage neural networks during the alignment computation itself. Few approaches have been proposed on this front recently, albeit only for generic multiple sequence alignment. These include Deep Canonical Time Warping \citep{trigeorgis2017deep}, the first deep temporal alignment method for simultaneous feature selection and alignment; NeuMATCH \citep{Dogan18neumatch}, a neural method based on LSTM blocks, and more recently, Neural Time Warping \citep{kawano2020neural} that models multiple sequence alignment as a continuous \blue{optimisation} problem. While such learnt approaches have been explored for generic sequence alignment, the development of fully learnt methods for music \blue{synchronisation} remains limited, especially in the offline setting. The particular caveats in music \blue{synchronisation} such as instrumentation effects and structural factors make it important to build task-specific architectures motivated by musical domain knowledge. 

\par There have been a few recent approaches that explore deep learning in the performance-score analysis setting. These have mainly been proposed for the tasks of piece identification, cross-modal music retrieval, and score following, with the latter mainly concerning the image domain \citep{dorfer2018learning, dorfer2018learning2}. 
\citet{dorfer2018learning} extend their previous work \citep{dorfer2017learning}, proposing an end-to-end multi-modal \blue{Convolutional Neural Network} trained on sheet music images and audio spectrograms of the corresponding snippets for cross-modal retrieval and piece identification. \citet{dorfer2018learning2} focus on learning a score following model using reinforcement learning. They work on snippets of performance aligned to snippets of an image as proposed in their earlier method \citep{dorfer2017learning}, but using reinforcement learning to align the two streams in an online manner. Recently, \citet{henkel2019audio} propose an audio-conditioned U-net architecture for estimating the current position of the audio performance in the sheet image. These methods assume that the performer follows the score completely, and therefore treat each pair of image and audio snippet independently, and restrict large jumps by setting a threshold on the speed of the score follower. Owing to this treatment, these methods are unable to capture structural deviations from the score such as jumps and repeats. 
\par 
While neural methods have recently been explored for cross-modal retrieval and score following \citep{henkel2020learning, dorfer2018learning2}, their application to score-performance \blue{synchronisation} in the offline setting remains relatively unexplored. There are a number of differences between the offline and online alignment tasks, as well as the audio-to-score and \blue{audio-to-image alignment} tasks. This makes a comparison between corresponding methods difficult, specifically since the latter uses the absolute alignment error (distance in pixels) of the estimated alignment to its ground truth alignment for each of the sliding window positions as the evaluation measure, whereas the former uses a more musically meaningful evaluation procedure, including the offset between predicted and ground truth alignments, typically at beat or note level.  
\par  While all of these works focus on generating alignments for a snippet of audio to an image of the sheet music using multi-modal training, \blue{this} thesis is concerned with building neural \blue{synchronisation methods} in situations where the scores of the different pieces are available, generally in an offline setting, thereby making \blue{the} target application quite \blue{different.}
Some of the methods developed during this PhD are however also applicable directly to raw sheet images, and also in an online setting, as will be discovered in the later chapters.

\section{Other methods}
Apart from the aforementioned methods, various methods using customized pipelines for alignment and those using alignment for other music processing tasks have been proposed. 
Examples of application of alignment techniques for other MIP tasks include the identification of cover songs using Dynamic Time Warping with chroma binary similarity \citep{serra2008chroma}, fast identification of the piece being played along with the score position \citep{arzt2012fast}, and analysis of expressive timing in music performance through the study of alignment patterns \blue{\citep{liem2011expressive, kosta2018mazurkabl}}.
\par Various methods have also explored hybrid models via the combination of different alignment frameworks. For instance, \citet{devaney2009improving} developed a method for improving DTW-based MIDI alignment using a Hidden Markov Model that uses aperiodicity and power estimates from the signal as observations and the results of a DTW alignment as a prior. Similarly, methods that extend DTW to handle onset and offset asynchronies in polyphonic music \citep{devaney2014estimating}, as well as asynchronies between musical voices 
\citep{wang2015compensating} have demonstrated an improvement in the alignment accuracy, particularly in settings prone to having asynchronous elements. Another example of a hybrid approach is by \citet{syue2017accurate}, who propose a two-stage alignment system composed of Dynamic Time Warping, simulation of overlapped sustain notes, a background noise model, silence detection, and a refinement process. Similar to this method, \citet{chen2019effective} propose an effective method for audio-to-score alignment using onsets and modified constant Q spectra.  Their framework consists of onset detection, note matching, and dynamic programming. 
Other extensions to DTW-based methods \citep{arzt2015real, wang2016robust} focus on better alignment in the scenario where multiple performances are available, showing some promising results. The alignment of singing voice is also a related research direction \citep{gong2017audio, sharma2019automatic}, however it is not dealt with specifically in this thesis.
\par In addition to research methods for core \blue{Music Information Processing}, application oriented research has also progressed over the years, for tasks such as score following \citep{cont2009coupled}, automatic accompaniment \citep{sako2014ryry} and music analysis software \citep{herremans2017multi}.  
\citet{dannenberg2015arrangements} focus on Human-Computer music performance systems for popular music and present an interface to adapt music data in different formats, allowing the user to specify the live performance order via a re-arrangement of the material. \citet{alonso2017parallel} aim at developing online alignment software for multi-core architectures, including x86/x64 processors and ARM processors. They develop a client-server architecture employing a parallel online time warping solution for real-time audio-to-score alignment in such multi-core systems. A multi-modal platform for semantic music analysis was recently proposed by \citet{herremans2017multi} to help musicologists visualize audio-score tension. They present a web-based \textit{Interactive system for Multi-modal Music Analysis (IMMA)} that provides musicologists with an intuitive interface for joint analyses of scores and performances. In related work, albeit not specifically in the music domain, \citet{li2018read} propose a multi-modal \blue{summarisation} method for asynchronous text, image, audio and video. Another notable approach that incorporates visual information in music processing is the detection of playing/non-playing activity of musicians from symphonic music videos \citep{bazzica2016detecting}, and exploiting this information to aid score \blue{synchronisation} \citep{bazzica2014exploiting}. \blue{The reader is referred to \citet{duan2018audiovisual} for a detailed review of methods aimed at audiovisual analysis of music performances.}
\par Additionally, methods for dataset creation with regards to alignment have recently been proposed \citep{joysingh2019development, meseguer2019dali}. Among these, the former method explores the development of large annotated music datasets using HMM based Forced Viterbi Alignment \citep{joysingh2019development}, while the latter presents the \emph{DALI} dataset, which is a large dataset of synchronised audio, lyrics and notes, automatically created using a teacher-student machine learning paradigm \citep{meseguer2019dali}. Some of these research methods have been successfully employed as part of commercial applications, such as \emph{Antescofo} \citep{cont2009coupled}, \emph{MusicPlusOne} \citep{raphael2009current}, \emph{TONARA} and \emph{Tido Music}.
\section{Limitations of major approaches}\label{sec:limitations}
Having discussed a wide range of music alignment and \blue{synchronisation} techniques, I now proceed to discuss the limitations of the major alignment algorithms in this section and thereby highlight the motivation of \blue{my} research. This will also help the reader to have a better insight on the specific problems to be addressed by my research. Table 2.1 summarises the key contributions and limitations of the major alignment approaches discussed in the previous sections.


Table 2.1: Key contributions and limitations of major alignment approaches

| Category | Approach | Method | Contributions | Limitations |
|---|---|---|---|---|
| Knowledge driven | String matching | Dannenberg 1984 | Method to align two symbolic streams | Symbolic method, very basic, cannot deal with expressive variations |
| | DTW-based | Orio and Schwarz 2001, Hu 2003 | Initial approaches to use DTW for music alignment | Limited scalability, large memory requirement, unable to handle large tempo changes |
| | | Müller 2004 Müller 2006 | Multi-scale approaches to alignment | Struggles with pianissimo passages and repeated chords |
| | | Dixon 2005 | Developed the Music Alignment Tool Chest (MATCH) based on DTW | Struggles with path irregularities and does not model structural differences between performances |
| | | Arzt 2008 | Built upon MATCH by proposing various optimisations for online score following | Difficult to deal with tempo changes and onset features make it tough to model instruments such as violin |
| | | Ewert and Müller 2008 | Propose refinement strategies for music synchronisation | Struggles with capturing smooth note transitions and handling structural differences |

| | | | | |
|---|---|---|---|---|
| | | Macrae and Dixon 2010 | Proposed WTW, a linear cost variation on DTW for real-time synchronisation | Assumes that the alignment is continuous from the start to the end, no avenue to capture jumps |
| | | Prätzlich 2016 | Proposed a multi-scale, parallelizable DTW algorithm for memory restricted conditions | Predicted warping path may deviate from the optimal global warping path, happens more frequently with smaller constraint regions |
| Stochastic | HMM-based | Cont 2009 | Proposed a coupled duration-focused architecture for score following | Instability when the reference tempo is different from the expected tempo |
| | | Maezawa 2011 | Develops a Bayesian alignment method using Bayesian HMM | Performs poorly on solo piano and vocal music |
| | | Cuvillier and Cont 2014 | Proposes criteria for temporal coherency in semi-Markov alignment models | Potential confusion of the observation model in the presence of heavy sustain pedal usage |
| | | Li and Duan 2015 | Proposed HMM-based method for score following in presence of sustain pedal effects | Introduces potential new mismatch caused by the removal of notes whose keys have not been released, unable to capture structure |
| | CRF-based | Joder 2013 | Proposes CRF method that is well suited to design flexible observation functions | Disregards the possibility of large variations due to different instruments and recording conditions |

| | PF-based | Korzeniowski 2013 | Proposes a Dynamic Bayesian Network + Particle Filtering for inference | Inconsistent behaviour, the extensions have different impacts on different pieces |
|---|---|---|---|---|
| Structure aware | JumpDTW | Fremerey et al. 2010 | Extends DTW to allow for jumps and repeats specified in the score. | Reliant on *a priori* manually annotated block boundaries, cannot handle intra-block jumps, unable to capture unmarked deviations |
| | Multi-level matching | Arzt et al. 2010b | Propose a structure-aware multi-level matching and tracking algorithm | Unable to handle complex piano music with tempo changes |
| | NWTW | Grachten et al. 2013 | Propose the Needleman Wunsch Time Warping for structure-aware audio alignment | Waiting mechanism when two streams mismatch, does not allow backward jumps and thereby unable to align repetitions |
| | HMM-based | Nakamura et al. 2015b | HMMs for modeling variations from the score in score following | Works only for monophonic music |
| | | Jiang et al. 2019 | Offline alignment for practice scenario, using HMMs + pitch trees + beam search | Struggles with pieces containing both forward and backward jumps |

| | | | | |
|---|---|---|---|---|
| | Hierarchical DTW | Shan and Tsai 2020 | Method to generate piano score following videos given an audio and raw image of sheet music | Reliant on bootleg representation, struggles when said representation is inaccurate, difficult to handle short pieces containing jumps |
| Neural | Feature learning | Arzt and Lattner 2018 | Learnt transposition invariant features | Not robust to tempo changes |
| | | Tanprasert 2019 | Learnt mid-level representations | Cannot handle structural differences |
| | Reinforcement learning | Dorfer 2018b, Henkel 2019a | Score following in raw images of sheet music | Struggles with structural changes, does not generalise well to real music data |
| | Transcription + Alignment | Kwon 2017 | Alignment using RNN-based transcription + DTW | Does not capture structure, poor generalizability to other instruments |
| | | Waloschek 2019 | CNNs for cross-document measure alignment | Can only generate a coarse alignment, unable to capture structure |
| | | Simonetta 2021 | Deep neural networks + HMM | Does not work well for non-piano solo music |
| | Position estimation | Henkel 2020 | Audio-conditioned U-net for estimating current position in sheet image | Cannot capture structural differences, works well for synthetic data but struggles with real music |


The following insights regarding the limitations of major previous alignment approaches can be drawn from Table 2.1:
\begin{itemize}
    \item Approaches based on DTW assume that the alignment path relating the two streams increases monotonically, thereby rendering the model to be unable to allow jumps. This hinders the capability of DTW-based methods to handle possible structural deviations from the score, such as repeated measures and skipped sections.
    \item Standard methods based on low-order HMMs (for instance first-order Markov models) suffer from the inherent limitation of the Markov assumption, which implies that given the immediate past, future alignment decisions are not dependent on any further history. This limits contextual modelling essential for robust structure-aware music alignment.
    \item The method developed by \citet{arzt2016flexible}, contains different components, including one for feature extraction, one for modeling tempo, one for anticipating structural changes, and so on. This hinders end-to-end training and adaptability to different test settings across acoustics and instrumentation. 
    \item Transposition-invariant features \citep{arzt2018audio} work well for aligning transposed performances to their score, however they struggle in the presence of tempo differences between the performances and the corresponding scores, as opposed to chroma-based features, owing to the manner in which the features incorporate the local context.
    \item Most methods described in this chapter have a dependence on heterogeneous technologies, i.e. the components are individually estimated rather than jointly optimized as done by deep neural networks.
    \item DTW-based and HMM-based approaches typically incorporate handcrafted features. This impedes the learning of representations in a data-driven manner, which can be realized by using neural networks for this task.  
    \item JumpDTW, a method proposed to allow for jumps during DTW-based alignment requires manually labelled `block boundary' annotations that mark the jump locations in the performance audio. The method requires that annotations that are accurate at the frame level (in the audio feature representation), which are generally not available \textit{a priori} for real music data. Thus its performance is highly dependent on the quality of Optical Music Recognition (OMR) systems in the absence of manual annotations. Additionally, this method is unable to align deviations induced by a performer that are not foreseeable from the score, and cannot capture intra-block jumps, i.e. deviations within the annotated block boundaries.
    \item The Needleman Wunsch Time Warping (NWTW) technique, another prominent method proposed for structure-aware audio alignment, is unable to align repeated segments or measures since there is no provision for backward jumps. This method waits until the two streams can be aligned again, thereby being unable to align repeats.
    \item The score following approach based on reinforcement learning \citep{dorfer2018learning2} assumes that the performer follows the score completely, and hence is not robust to structural changes. Additionally, this approach treats each pair of image and audio snippets independently, and restricts large jumps by setting a threshold on the speed of the score follower. This method thus discards any structural information available from the score which would otherwise bear potential to improve alignment performance.
\end{itemize}

\par The remainder of the thesis expands on my research that aims at developing alignment methods that overcome some of the limitations listed above. Broadly speaking, I propose various novel methods based on neural network architectures for \textit{context-aware} performance-score synchronisation of real music performances with the corresponding scores.
\par \blue{The methods proposed in the upcoming chapters tackle these limitations and incrementally develop towards end-to-end data-driven alignment. In the next chapter, I describe metric learning approaches using Machine Learning methods to aid audio-to-score alignment. This chapter addresses the \textit{learning from data} and \textit{adaptability} limitations of the existing approaches. Thereafter, Chapter 4 proposes a novel method for structure-aware alignment and therefore addresses the previous methods' limitations around \textit{monotonicity} and \textit{contextual incorporation} in alignment prediction. Finally Chapter 5 develops a completely neural framework with a custom loss function that removes the dependence on DTW for alignment generation, thereby enabling \textit{end-to-end} training along with structural incorporation.}

\chapter{Metric Learning for Audio-to-Score Alignment}\label{ch:frameSimilarity}
\vspace{1cm}
This chapter describes the exploration of using neural networks as precursors for DTW-based audio-to-score alignment. It begins with a discussion of representation learning for the alignment task and highlights the challenges and limitations faced by this approach. It then presents a novel approach to performance-score synchronisation that yields robust performance on multiple music datasets. It further describes experiments on metric learning for alignment and demonstrates the applicability of this method in various scenarios encompassing different instrumentation and acoustic settings. The chapter concludes with a discussion on the advantages as well as the limitations of the method, which motivates further research presented in the upcoming chapters.
\vspace{3cm}
\section{Introduction} 
\vspace{0.5cm}
Audio-to-score alignment or performance-score \blue{synchronisation} aims at generating an accurate mapping between an audio recording of a performance and the corresponding score of a given piece of music.  
The alignment computation in performance-score \blue{synchronisation} is traditionally carried out via algorithms based on Dynamic Time Warping (DTW) \citep{sakoe1978dynamic} or Hidden Markov Models (HMM) \citep{baum1966statistical}, which operate on handcrafted feature representations of the two inputs, such as standard spectrograms, log-frequency spectrograms and chromagrams \citep{dixon2005line, ewert2009high, arzt2012adaptive}. 
\vspace{0.5cm}
\par The most popular feature representation for music alignment is the chromagram, which is a time-chroma representation generated from the log-frequency spectrogram \citep{bartsch2005audio}. A chromagram comprises a time-series of chroma vectors, which represent harmonic content at a specific time in the audio as \begin{math} c \in \mathbb{R}^{12} \end{math}. Each $c_i$ stands for a pitch class, and its value indicates the current saliency of the corresponding pitch class. The central idea of using this representation is to aggregate all spectral information relating to a
specific pitch class into one coefficient. 

\vspace{0.5cm}
\par While chromagrams have demonstrated promising performance for a variety of MIR tasks, it must be noted that they are a reductive choice for audio-to-score alignment. Since chromagrams are contrived representations, achieved via a simple projection based on pitch classes, they are not necessarily the optimal representation for the alignment task. For instance, octave information that is discarded by chromagrams might be useful for the disambiguation between two possible alignment locations. Additionally, these representations are not trainable and hence cannot be adapted to a particular type of data. 
\vspace{0.5cm}
\par These limitations can be overcome by employing a data-driven approach that \textit{learns} the representations from the data itself, which is a promising research direction given the recently increased availability of music data.  Additionally, a data-driven approach opens up the possibility of adapting the feature representations to specific settings, due to their trainable nature. An example of this can be observed in the study carried out by \citet{ewert2016piano}, wherein they demonstrate that adapting an automatic transcription system to a specific piano yields a significant improvement ($\approx 10\%$) in transcription accuracy. 
\vspace{0.5cm}
\par To this end, this chapter explores machine learning strategies based on a data-driven approach to learn feature representations that are optimal for the alignment task.  The goal is to come up with a feature representation which is robust to aspects like tempo and rhythm changes whilst being adaptable to factors like timbre, tuning and acoustic conditions. The chapter focuses on offline audio-to-score alignment of piano music and describes experiments using representation learning methods as well as metric learning methods to improve DTW-based alignment. The next section briefly describes an exploration of representation learning for audio-to-score alignment. The description of this method has been kept brief since it did not result in significant improvements over standard feature representations. The subsequent section proposes a metric learning based strategy and discusses the method in greater detail, since it demonstrates promising results across various settings.

\vspace{0.5cm}
\section{Representation learning for alignment}\label{sec:representation}
Driven by the motivation described in the previous section, this section explores representation learning for the alignment task. It presents an endeavour to learn the features from the audio and score inputs at the frame level, in an unsupervised manner. It must be noted that frame-level features must be learnt for DTW-based alignment, rather than learning \textit{shapelets}, which discover discriminative feature segments from the entire time-series \citep{zhang2016unsupervised}, and are mainly useful for classification tasks. \blue{The alignment task, on the other hand, would require a \textit{sequence} of feature vectors}, which could then be employed by a DTW/HMM computation to generate the fine alignment. 
\begin{figure}[ht]
\vspace{1cm}
  \centering
  \includegraphics[width=0.9\columnwidth]{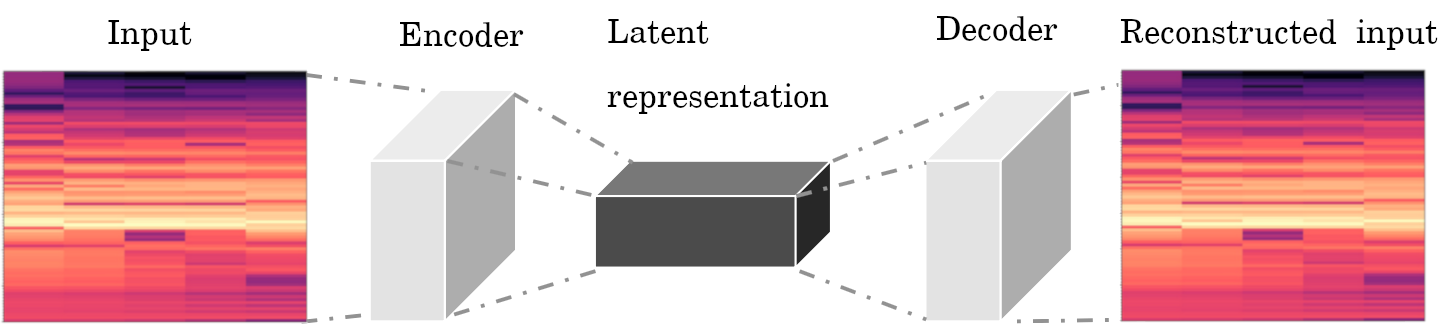}
  \caption{The \blue{autoencoder} architecture to learn features for the alignment task}
  \label{fig:auto}
\end{figure}
\vspace{1cm}
\par \blue{To this end, an exploration was carried out to learn features  from the audio and score inputs using an autoencoder model. Autoencoders are able to convert a high-dimensional input to a low-dimensional representation, which can then be employed to reconstruct the original high-dimensional input.} 
\par \blue{For the experimentation on representation learning, I employed the
convolutional autoencoder containing an encoder-decoder architecture. The architecture of the convolutional autoencoder is shown in Figure \ref{fig:auto}. The encoder comprises two convolutional (conv2D) and max-pooling (MaxPooling2D) layers, and converts the input into a latent representation. The decoder comprises two transposed convolutional layers (Conv2DTranspose) followed by a convolutional layer (conv2D) with a sigmoid activation. Each transposed convolutional layer is passed through a nearest neighbor upsampling layer before being passed as input to the next layer. The convolutional kernels used are of dimension $3 \times 3$, the input and output dimensionality is $128 \times 128$. The model was implemented using the deep learning API Keras and trained on the GeForce RTX 2080 Ti GPU card containing 11 GB GPU memory.} 
\begin{table*}[th]
\vspace{1cm}

\caption{Results (alignment accuracy in \%) of the autoencoder model. \\$\mathit{CAE}$: convolutional \blue{autoencoder}}

\centering
\begin{tabular}{cccc} \toprule
\hline 
\multirow{2}{*}{\textbf{Model}} & \multicolumn{3}{c}{\textit{Error Margin}} 
\tabularnewline
  &  \textbf{$<$50ms} & \textbf{$<$100ms} & \textbf{$<$200ms} \\
\midrule 
 $\mathit{MATCH}$ \blue{\citep{dixon2005match}} & 72.1 & 77.6 & 83.7  \\
\midrule
 $\mathit{DTW_{Chroma}}$ & 70.5 & 76.3 & 82.4   \\
\midrule 
 $\mathit{DTW_{CAE}}$ & 70.2 & 75.8 & 84.5 \\
\midrule
\midrule 
\bottomrule
\end{tabular}
\label{tab:autoencoders}
\end{table*}

\vspace{0.5cm}
\par The autoencoder model was trained on the MIDI Aligned Piano Sounds (MAPS) dataset \citep{emiya2010maps} in an unsupervised manner. \blue{From the original MAPS database, which contains \blue{synthesised} MIDI-aligned audio for a range of acoustic settings, we select the subset \emph{MUS} containing complete pieces of piano music.} It must be noted that the alignment labels were not employed to train the models, rather the \blue{autoencoder} models essentially learn a feature representation from the input while trying to reconstruct the input as the output. To this end, the MIDI files corresponding to the score are first converted to audio through FluidSynth \citep{henningsson2011fluidsynth} using piano soundfonts. The two audio inputs to be aligned are then represented as sequences of analysis frames, using a low-level spectral representation computed via \blue{the} Short Term Fourier Transform of the signal. 
The model is then used to generate the frame-level representations at test time for the audio and score sequences, and these feature sequences are passed on to a DTW computation to yield the alignment. 
\par \blue{The model was trained using the binary cross-entropy loss for 30 epochs.} The model was tested on the Mazurka dataset \citep{sapp2007comparative}, which contains piano music across a range of acoustic settings. The alignment error is computed as $e_i$ = $| {t_i}^e$ - ${t_i}^r |$, encoding the time difference between the alignment positions of corresponding events in the reference ${t_i}^r$ and the estimated alignment time ${t_i}^e$ for score event $i$. The results are shown in terms of alignment accuracy, denoting the percentage of events which are aligned within an error of up to 50ms, 100ms and 200ms respectively. \blue{The results obtained were compared with those obtained by the methods employing standard feature representations.} The results are given in Table \ref{tab:autoencoders}.
\par As can be seen from Table \ref{tab:autoencoders}, this method did not yield significant improvements over traditional feature representations, especially for fine-grained alignment. This could be attributed to the difference in the acoustic settings between the training and test set. The learnt representations do outperform traditional feature representations ($DTW_{Chroma})$ for coarse alignments, however only slight improvements can be observed in the alignment accuracy. Owing to the negligible improvements over the chromagram representation, the detailed architecture and experimental setup of the autoencoder experiments are not described. Rather, an improved \textit{alignment-specific} approach is explored and described in detail in the remainder of the chapter. 
\section{Metric learning for DTW-based alignment}
The previous section explored methods for learning feature representations from audio and score inputs using autoencoders. While the methods provided a slight improvement over the chromagram representation, they bear a number of inherent limitations:
\begin{itemize}
    \item The representations of the audio and score inputs are not learnt jointly.
    \item Fine-grained alignment performance drops compared to standard representations.
    \item While the representations are learnt directly from data, and are thereby adaptable, they are not necessarily optimal for the alignment task.
\end{itemize}
\par This section explores a more intuitive method for learning the features \textit{suitable} for the DTW-based alignment computation. I pose the question:
\begin{quote}
    \textit{How can we jointly learn a representation from the performance and score sequences that is aimed at optimising DTW-based alignment performance?}
\end{quote}
The remainder of the chapter presents the following answer to the above question:
\begin{quote}
\textit{Learn the similarity matrix that is operated on by DTW from the data directly.}
\end{quote}
This will ensure that the data-driven approach to learning the features will operate jointly on both the performance and score sequences, and at the same time, will be geared towards improving alignment accuracy. 
\par Driven by this motivation, the rest of the chapter explores a novel method that leverages neural networks to learn performance-score similarity at the frame level. This provides a method to extract pertinent information helpful for performing alignment, and discard extraneous information such as percussive noise and timbral variations that could potentially hinder effective alignment.
\par The feature engineering step of standard alignment methods is overriden and the focus is on learning frame similarity using Siamese Convolutional Neural Networks (CNNs), since they can jointly optimise the representation of input data conditioned on the similarity measure being used.


 \begin{sidewaysfigure}[htbp]
  \centering
  \includegraphics[width=8in]{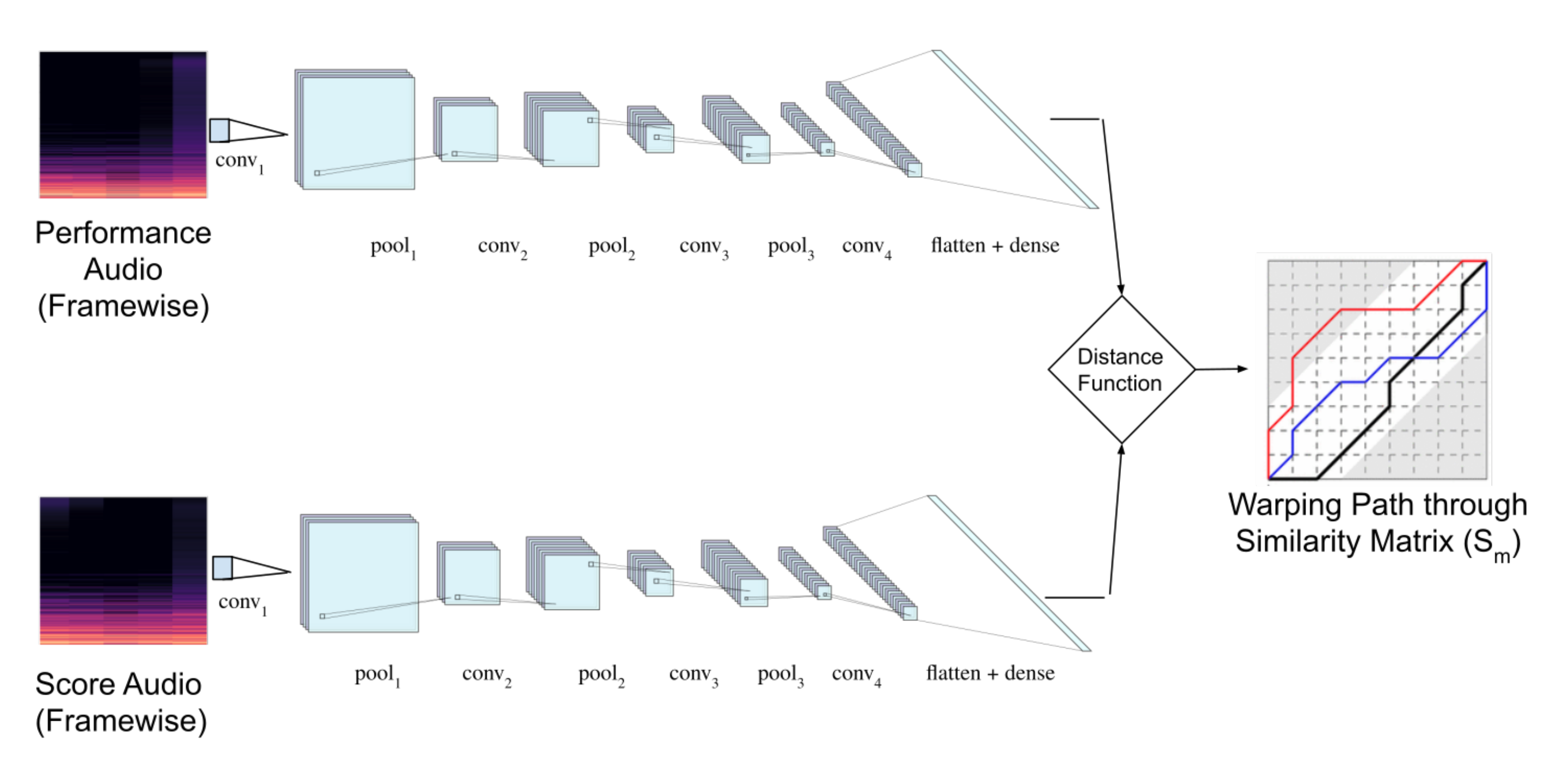}
  \caption{Model Pipeline. The model is trained to classify the middle column of the spectrogram patches.}
  
\begin{tabular}{r@{ : }l r@{ : }l}
$conv_i$ & \textit{$i_{th}$ convolution layer} & $pool_j$ & \textit{$j_{th}$ pooling layer}\\
$flatten$& \textit{flatten layer} & $dense$& \textit{fully connected layer} 
\end{tabular}
  \label{fig:siamesePipeline}
\end{sidewaysfigure}
To elaborate the intuition behind this method further, rather than extracting a feature representation from the separate inputs using autoencoders (\textit{What is the best way to represent this input sequence?}), the endeavour is to determine how similar two frames corresponding to the performance and score respectively are (\textit{Do these two corresponding frames contain similar spectral content?}). The latter question pertains to the metric learning task, and is specifically geared towards aiding DTW-based alignment.
More specifically, this method aims to \textit{learn} the frame similarity matrix using a neural approach that looks at both inputs together, rather than learning representations individually and thereafter comparing them. 
The learnt similarity matrix is then passed on to a DTW-based algorithm to generate the alignments. 
\par The next section presents the proposed method to answer the aforementioned question, \textit{How similar is the spectral content contained in these two frames?}

\section{Proposed Methodology}\label{method}

The proposed metric learning based approach employs a Siamese Convolutional Neural Network, a class of neural network architectures that contains two or more identical subnetworks \citep{bromley1994signature} for this task. This framework has shown promising results for similarity estimation in the field of computer vision \citep{zagoruyko2015learning} as well as  natural language processing \citep{mueller2016siamese}. The motivation to employ Siamese CNNs for similarity learning is that they can jointly optimise the representation of the input data conditioned on the similarity measure being used. This learnt similarity matrix then serves as the input to a DTW algorithm to generate the fine-level alignments via a computation of the optimal warping path through the matrix.

The method is described in detail in the subsequent subsections.
\subsection{Method pipeline}\label{sub:pipeline}
\begin{figure}[ht]
  \centering
  \includegraphics[width=\columnwidth]{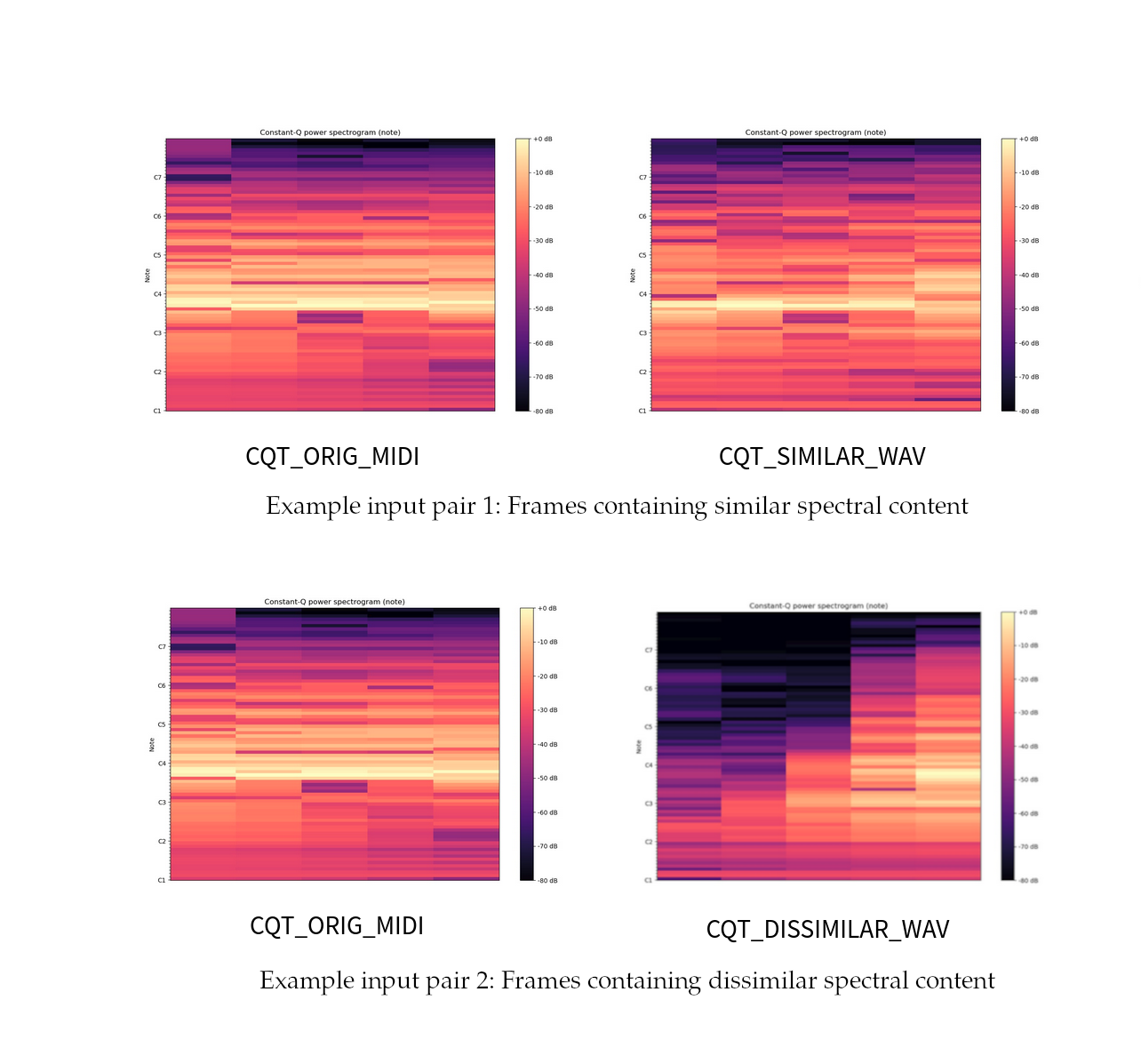}
  \caption{\blue{Sample inputs to our Siamese model}}
  \label{fig:sampleInputs}
\end{figure}
An overview of the model pipeline is presented in Figure \ref{fig:siamesePipeline}. A Siamese CNN, akin to that prototyped in \citep{agrawal2020hybrid}, is trained to compute a frame similarity matrix $S_m$ to be fed to DTW to generate alignment. The network comprises two identical branches (twin subnetworks), and takes two inputs, one for each subnetwork. Each subnetwork computes an embedding at the final layer which is compared using a distance function that returns the final output similarity. The arrangement of convolutional and pooling layers within each subnetwork is similar to that of a standard Convolutional Neural Network (CNN), however, it does not contain a softmax layer, rather it has a fully connected layer as the final layer. \\

\par The network thus takes two inputs, corresponding to the audio and score respectively, and each ends up with an embedding generated by the fully connected layer in the respective subnetwork. The difference between these embeddings is computed using the distance function, and the result is passed to a single neuron with the sigmoid activation function, which outputs either 0 or 1, denoting a \textit{matching} or \textit{non-matching} frame pair. The training data for the Siamese models is thus formatted as a list of frame pairs, from the audio and score sequences, with a corresponding binary label encoding the ground truth.
\par The MIDI files corresponding to the score are first converted to audio through FluidSynth \citep{henningsson2011fluidsynth} using piano soundfonts. The two audio inputs to be aligned are then represented as sequences of analysis frames, using a low-level spectral representation computed via \blue{the} Short Term Fourier Transform of the signal. 
The training data (described in a further section) contains synchronised audio and MIDI files, so it is straightforward to extract matching frame pairs. For each matching pair, a non-matching pair is randomly selected (using MIDI-information) in order to have a balanced training set. 
The inputs to the Siamese network are labelled frame pairs from the performance audio and the \blue{synthesised} MIDI respectively. \blue{Figure \ref{fig:sampleInputs} shows an example of the input pairs operated on by the Siamese models. Given two sequences containing five frames corresponding to the audio and score inputs, the Siamese CNN model is trained to determine if the central frame in these \textit{match} or not.}
Further details about input representations and training are provided in the Experimental Setup section (Section \ref{sec:ch3_expSetup}).

\subsection{Loss function}\label{sub:loss}
Task-specific loss functions have shown promising results in the fields of image processing and natural language processing \citep{qi2017contrastive, amirhossein2018multi}. Various loss functions could be employed for the metric learning task. The proposed method employs the contrastive loss function \citep{hadsell2006dimensionality} for training the models. This formulation is chosen over a standard classification loss function like cross entropy since the objective is to differentiate between two audio frames, rather than classifying a single frame. Let $X = (X_1, X_2)$ be the pair of inputs $X_1$ and $X_2$, $W$ be the set of parameters to be learnt and $Y$ be the target binary label ($Y$ = 0 if they match and 1 if otherwise). The contrastive loss function 
for each tuple is computed as follows:
\begin{equation}
    L(W, X, Y) = (1-Y)\frac{1}{2}(D_W)^2 + (Y)\frac{1}{2}\{max(0, m - D_W)\}^2
\end{equation}
where $m$ is the margin for dissimilarity and $D_W$ is the Euclidean Distance between the outputs of the subnetworks. \blue{The margin $m$ is set as a hyperparameter and its value is determined using grid search.} Pairs with dissimilarity greater than $m$ do not contribute to the loss function. \blue{It can be noted that $\hat{Y}$ is not part of the loss, since  the goal of the proposed model is not to \textit{classify} the input image pairs per se, but instead to \textit{differentiate} between them. Hence, the loss function relies on $D_W$ to evaluate how well the model distinguishes the input image pairs, given the target labels.}
\par $D_W$ can be formally expressed as follows:
\begin{equation}
    D_W(X) = \sqrt{\{G_W(X_1) - G_W(X_2)\}^2}
\end{equation}
where $G_W$ is the output of each twin subnetwork for the inputs $X_1$ and $X_2$. Since it is a distance-based loss, it tries to ensure that semantically similar examples are embedded close to each other, which is a desirable trait for extracting alignments. 
\subsection{Generating fine alignments}
The Siamese network thus learns to classify the sample pairs as similar or dissimilar. 
This is done for each audio frame pair and the similarity matrix thus generated is then passed on to a 
DTW-based algorithm to generate the alignment path. Given the similarity values along the performance axis $A$= $(a_1,a_2,...,a_m)$ and those along the score axis $B$ = $(b_1,b_2,...,b_n)$, the alignment path that optimises the overall cost of aligning the two sequences is computed as follows:

\begin{equation}
D(i, j)  = d(i, j) + min\begin{cases}
D(i, j-1) \\ D(i-1, j) \\  D(i-1,  j-1)
\end{cases}
\end{equation}
where $d(i, j)$ is the distance measure (local cost) between points $a_i$ and $b_j$; and $D(i, j)$ is the total cost for the path which generates the optimal alignment between the sequences $A_{1..i}$ and $B_{1..j}$. The  Euclidean distance is employed as the distance measure and the DTW framework of \citet{giorgino2009computing} is used to compute the warping paths.

\section{Experimental Setup}\label{sec:ch3_expSetup}
\subsection{Datasets}

\begin{table}[ht]
\vspace{1cm}
   \caption{Datasets used for metric learning experiments on piano music}
   \centering
   \begin{tabular}{ c c c c} \toprule
      \textbf{Name} & \textbf{Annotations} & \textbf{Recordings}  & \textbf{Stage} \\ \midrule
      MAPS \citep{emiya2009multipitch} &  MIDI-audio alignment  &  238 & Train\\ \midrule
      Saarland \citep{muller2011saarland} & MIDI-audio alignment  & 50  & Train \\ \midrule
      Mazurka \citep{sapp2007comparative} & Beat level annotations & 239 & Test \\ \midrule
      \bottomrule
\end{tabular}
\label{tab:datasets_ch3}
\end{table} 
The datasets employed in the experiments on metric learning for alignment are summarised in Table \ref{tab:datasets_ch3}.
The experiments utilise the MAPS database \citep{emiya2009multipitch}, the Saarland database \citep{muller2011saarland} and the Mazurka dataset \citep{sapp2007comparative}. From the original MAPS database, which contains a combination of acoustic and synthetic MIDI-aligned audio for a range of settings, the subset \emph{MUS} containing complete pieces of piano music is extracted, and appended to the Saarland database. The resultant database comprising 288 recordings is randomly divided into sets of 230 and 58 recordings. These sets form the training and validation sets respectively.
The performance of the models is tested on the Mazurka dataset \citep{sapp2007comparative}, which contains recordings of Chopin's Mazurkas dating from 1902 to the early 2000s, thereby spanning across various acoustic settings. This dataset contains annotations of beat times for five Mazurka pieces. 


\subsection{Model Architecture}
 \begin{table}[ht]
 \vspace{1cm}
   \caption{Architecture of the Siamese model}
   \centering
   \begin{tabular}{ c c c c c } \toprule
       \textbf{Type of layer} & \textbf{Input size} & \textbf{Kernels} & \textbf{Kernel size}    \\ \midrule
       Convolution  & $128 * 128 * 3$ & 64 &   $5 * 5$ \\ \midrule
       Max-Pooling & $128 * 128 * 64$  & 1 &  $2 * 2$   \\ \midrule
       Convolution  & $64 * 64 * 64$ & 128 & $5 * 5$ \\ \midrule
       Max-Pooling & $64 * 64 * 128 $ & 1 & $2 * 2$   \\ \midrule
       Convolution  & $32 * 32 * 128$ & 256  & $3 * 3$  \\ \midrule
       Max-Pooling & $32 * 32 * 256$ & 1 & $2 * 2$   \\ \midrule
       Convolution  & $16 * 16 * 256$ & 512 & $3 * 3$  \\ \midrule
       Flatten & $16 * 16 * 512$ & - & -  \\ \midrule
       Fully Connected  & $131072$  & - & - \\ \midrule
      \bottomrule
\end{tabular}
\label{tab:arch}
\end{table} 
The proposed Siamese model has four convolutional layers of varying dimensionality with different convolutional kernels. The outputs of each layer are passed through a Rectified Linear Unit in order to add non-linearity. These are then passed on to a batch \blue{normalisation} operation, the outputs of which are fed as the inputs for the next layer. The fourth convolutional layer is finally followed by a fully connected layer, which generates the similarity output. The detailed architecture of the model is given in Table \ref{tab:arch}.

\par In order to keep the modality consistent, the MIDI files are first converted to audio \blue{using piano soundfonts as described earlier in Section \ref{sub:pipeline}.} 
The two audio inputs are converted to a low-level spectral representation using a Short Time Fourier Transform, with a hop size of 23 ms and a hamming window of size 46 ms. 
\blue{The Short-time Fourier transform (STFT) is a Fourier-related transform used to determine the sinusoidal frequency and phase content of local sections of a signal as it changes over time. The constant-Q transform (CQT), very closely related to the Fourier transform, transforms a data series to the frequency domain using a bank of filters that are equally spaced in log-frequency. } Experiments are conducted for \blue{these two} spectral representations, i.e. the STFT as well as the CQT representations. For the latter, a CQT is employed with 24 bins per octave, with the first bin corresponding to frequency 65.4 Hz (midi note C2). 





\subsection{Evaluation Methodology}

The results obtained using the Siamese models are compared with various alignment approaches, including \emph{MATCH} \citep{dixon2005match}, and three other DTW-based frameworks, one using the Chroma representation \citep{bartsch2005audio} ($DTW_{Chroma}$), and the other two using representations learnt using a multi-layer perceptron \citep{izmirli2010understanding} model and  ($DTW_{MLP}$), the convolutional autoencoder model ($\mathit{DTW_{CAE}}$), which was proposed in Section \ref{sec:representation}. The comparative experimentation is carried out using two different mechanisms for computing the similarity matrix $S_m$:
\begin{itemize}
    \item Using binary labels: For these experiments, the outputs of the Siamese CNN are directly used to populate the similarity matrix $S_m$, wherein 0 and 1 correspond to similar and dissimilar pairs respectively. 
    \item Using distances: For these experiments, the distance $D_W$ yielded by the contrastive loss computation is employed to populate the similarity matrix $S_m$. This distance directly corresponds to the dissimilarity between the two inputs, thereby adding more information in the similarity matrix than the binary counterpart. 
\end{itemize}
\par  The similarity matrix $S_m$ yielded by the Siamese networks is passed on to a DTW computation, as detailed in Section \ref{sub:pipeline}.
 The output warping path through $S_m$ predicted by DTW is then compared with the reference alignments to generate the alignment scores for the test set using the method proposed by \citet{cont2007evaluation}.
\par The alignment error is computed as $e_i$ = ${t_i}^e$ - ${t_i}^r$, measuring the time difference between the alignment positions of corresponding events in the reference ${t_i}^r$ and the estimated alignment time ${t_i}^e$ for score event $i$. The results are given in  accuracy and denote the percentage of events which are aligned within an error of up to 25 ms, 50ms, 100ms and 200ms respectively. \blue{I also conduct significance testing using the Diebold-Mariano test \citep{harvey1997testing} to examine the statistical significance of the results. To this end, I conduct pairwise comparisons of all model predictions with the predictions of the best performing model for each error margin and for each experimental setup described in the upcoming subsections. All the Siamese CNN models were trained on a GeForce RTX 2080 Ti GPU card containing 11 GB GPU memory, with the NVIDIA driver version 418.87.00 and CUDA version 10.1. The training for all models was completed in less than 8 hours when trained on a single GPU core.}
\section{Results and Discussion}
The results obtained by the methods are given in Table \ref{tab:siamese_binary} for various error margins. Results are reported for the methods using binary as well as distance labels in the similarity matrix.
\subsection{Improvements over standard feature representations}
\begin{table*}[ht]
\caption{Results of Siamese networks trained using binary matrix\\ \blue{$*$: significant differences from \begin{math}\mathit{SCNN_{CQT}}\end{math} (Distance), $p < 0.05$}}
   \centering
\begin{tabular}{ccccc} \toprule
\hline 
\multirow{2}{*}{\textbf{Model}} & \multicolumn{4}{c}{\textit{Error Margin}} 
\tabularnewline
  & \textbf{$<$25ms}& \textbf{$<$50ms} & \textbf{$<$100ms} & \textbf{$<$200ms} \\
\midrule 
 $\mathit{MATCH}$ \blue{\citep{dixon2005match}}  & 64.8* & 72.1* & 77.6* & 83.7*  \\
\midrule
 $\mathit{DTW_{Chroma}}$ & 62.9* & 70.5* & 76.3* & 82.4*   \\
\midrule 
 $\mathit{DTW_{MLP}}$  & 63.8* & 68.7* & 76.9* & 83.1* \\
\midrule 
 $\mathit{DTW_{CAE}}$ & 63.5* & 70.2* & 75.8* & 84.5* \\
\midrule
 \begin{math}\mathit{SCNN_{STFT}}\end{math} (Binary) & 65.6* & 71.9* & 78.1* & 84.8*   \\
\midrule 
 \begin{math}\mathit{SCNN_{CQT}}\end{math} (Binary) & 66.4* & 73.1* & 78.7* & 85.3* \\
\midrule 
 \begin{math}\mathit{SCNN_{STFT}}\end{math} (Distance) & 67.2* & 73.4*  & 78.7* & 85.6* \\
\midrule 
 \begin{math}\mathit{SCNN_{CQT}}\end{math} (Distance) & \textbf{68.1} & \textbf{74.8} & \textbf{80.1} & \textbf{86.7}\\
\midrule 
\bottomrule
\end{tabular}
\label{tab:siamese_binary}
\end{table*}

The Siamese models  \begin{math}\mathit{SCNN_{STFT}}\end{math} and \begin{math}\mathit{SCNN_{CQT}}\end{math} outperform the DTW-based methods using the standard chroma representation ($DTW_{Chroma}$) as well as those using the representations learnt using multi-layer perceptron ($DTW_{MLP}$) and \blue{convolutional autoencoder} ($\mathit{DTW_{CAE}}$) (Table \ref{tab:siamese_binary}, rows 1-5). This corroborates that frame similarity learnt from real data is effective for generating robust alignments.
 The CQT representation (\begin{math}\mathit{SCNN_{CQT}}\end{math}) yields better results than the STFT representation (\begin{math}\mathit{SCNN_{STFT}}\end{math}). This could be attributed to the more musically condensed representation offered by CQT over STFT. 
\par Additionally, one can observe the trend that the models trained using a non-binary distance matrix outperform those trained on binary matrices. This suggests that thresholding the similarity into binary labels discards potentially useful information and the distances facilitate the DTW algorithm to take better long-term decisions. This emphasises the advantage of our method over the $DTW_{MLP}$ representation, which employs a multi-layer perceptron model to classify input semigram pairs into binary labels as opposed to measuring dissimilarity via distances.
\par The next section inspects the confusion matrix generated by the Siamese model to aid a more detailed analysis. The confusion matrix is generated using the classification results of the Siamese model  \begin{math}\mathit{SCNN_{CQT}}\end{math}.
\subsection{Optimisations}
\subsubsection{Confusion Matrix}
In addition to the overall alignment accuracy for different error margins, this section reports the precision and recall of the methods for the matching/non-matching classification. In order to test this, all the matching frame pairs from the test set are extracted using the time correspondences and labelled 0, with the remaining frames labelled 1. The binary output of the Siamese model \begin{math}\mathit{SCNN_{CQT}}\end{math} is then compared with this frame-level ground truth (rather than comparing the output of the DTW computation with the aligned ground truth). The resulting confusion matrix is shown below. \blue{The values are listed in percentage.}
\begin{center}
\begin{tabular}{c >{\bfseries}r @{\hspace{0.7em}}c @{\hspace{0.4em}}c @{\hspace{0.7em}}l}
  \multirow{10}{*}{\rotatebox{90}{\parbox{1.1cm}{\bfseries\centering Actual\\ value}}} & 
    & \multicolumn{2}{c}{\bfseries Prediction outcome} & \\
  & & \bfseries p$'$ & \bfseries n$'$ & \bfseries 
  \\
  & p & \MyBox{TP}{84} & \MyBox{FN}{16} & 
  \\[2.4em]
  & n & \MyBox{FP}{23} & \MyBox{TN}{77} & 
  \\
  & & Confusion matrix &  \begin{math}\mathit{SCNN_{CQT}}\end{math} &
\end{tabular}\label{tab:confusion1}
\end{center}
\vspace{0.5cm}
\par As it can be observed from the confusion matrix, the number of false positives yielded by \begin{math}\mathit{SCNN_{CQT}}\end{math} was greater than the number of false negatives. It could be speculated that this is due to the sparsity in terms of the representations, i.e. presence of noise / unpitched sound activity in the spectrograms. 
\vspace{0.5cm}
\begin{figure}[ht]
  \centering
  \includegraphics[width=0.8\columnwidth]{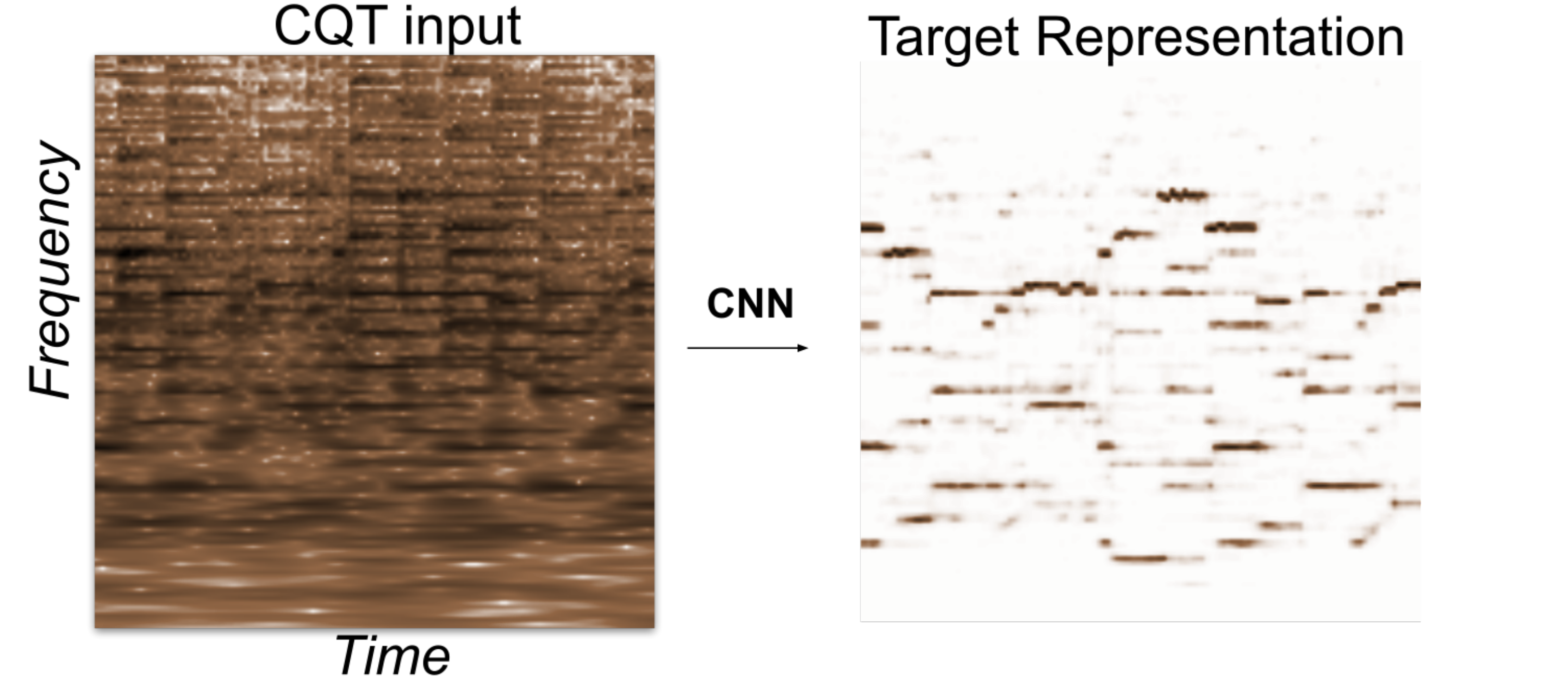}
  \vspace{0.3cm}
  \caption{Deep salience representation to address data scarcity}
  \label{fig:salience}
\end{figure}
\par In order to overcome the limitation mentioned above, experiments were conducted using deep salience representations \citep{bittner2017deep} for effective training of the Siamese models in these conditions. These are time-frequency representations aimed at estimating the likelihood of a pitch being present in the audio. Figure \ref{fig:salience} shows an example of a pitch salience representation. 

\par The primary motivation behind using such a representation is that it de-emphasises non-pitched content and emphasises harmonic content, thereby aiding training in data-scarce conditions. To compute the salience representations, a CNN is trained to learn a series of convolutional filters, constraining the target salience representation to have values between 0 and 1, with larger values corresponding to time-frequency bins where fundamental frequencies are present \citep{bittner2017deep}. The model is trained to minimize the cross entropy loss as follows:
\begin{equation}
    L(y, \hat{y}) = -y log(\hat{y}) - (1 - y)  log(1 - \hat{y})
\end{equation}
where both $y$ and $\hat{y}$ are continuous values between 0 and 1. 
\par The performance obtained using the Deep Salience representation is compared with that obtained using the \blue{Constant-Q Transform (CQT) of the raw audio files, since the CQT representation outperforms the STFT representation as demonstrated in the previous experiments.} A CQT  with 24 bins per octave is employed, with the first bin corresponding to a frequency of 65.4 Hz (MIDI note C2).

\begin{table*}[th]
\vspace{0.5cm}
\caption{Results of Siamese models trained using various optimisations (binary matrix)\\ \blue{$*$: significant differences from \begin{math}\mathit{SCNN_{Sal+DA}}\end{math}, $p < 0.05$ }}
   \centering
\begin{tabular}{ccccc} \toprule
\hline 
\multirow{2}{*}{\textbf{Model}} & \multicolumn{4}{c}{\textit{Binary Matrix}}
\tabularnewline
&\textbf{$<$25ms}& \textbf{$<$50ms} & \textbf{$<$100ms} & \textbf{$<$200ms}  \\
\midrule 
 \begin{math}\mathit{SCNN_{CQT}}\end{math} & 66.4* & 73.1* & 78.7* & 85.3* \\
\midrule 
 \begin{math}\mathit{SCNN_{Sal}}\end{math} & 68.2* & 75.3* & \textbf{81.4} & \textbf{87.8}* \\
\midrule 
 \begin{math}\mathit{SCNN_{CQT+DA}}\end{math} & 67.9* & 74.4* & 80.8* & 86.7*  \\
\midrule 
 \begin{math}\mathit{SCNN_{Sal+DA}}\end{math} & \textbf{69.4} & \textbf{76.4} & 81.2 & 87.5   \\
\midrule 
\bottomrule
\end{tabular}
\label{tab:optimisationsBinary}
\end{table*}

\begin{table*}[th]
\vspace{1cm}
\caption{Results of Siamese models trained using various optimisations (distance matrix)\\$*$: significant differences from \begin{math}\mathit{SCNN_{Sal+DA}}\end{math}, $p < 0.05$ }
   \centering
\begin{tabular}{ccccccccc} \toprule
\hline 
\multirow{2}{*}{\textbf{Model}}  & \multicolumn{4}{c}{\textit{Distance Matrix}}
\tabularnewline
   & \textbf{$<$25ms} &\textbf{$<$50ms} & \textbf{$<$100ms} & \textbf{$<$200ms}\\
\midrule 
 \begin{math}\mathit{SCNN_{CQT}}\end{math}  & 68.1* & 74.8* & 80.1* & 86.7*\\
\midrule 
 \begin{math}\mathit{SCNN_{Sal}}\end{math}  & 70.3* & 76.7*  & 82.1* & 88.4* \\
\midrule 
 \begin{math}\mathit{SCNN_{CQT+DA}}\end{math}  & 69.6* & 75.4* & 81.6* & 87.9*\\
\midrule 
 \begin{math}\mathit{SCNN_{Sal+DA}}\end{math}   & \textbf{71.7} & \textbf{78.2} & \textbf{83.3} & \textbf{90.1} \\
\midrule 
\bottomrule
\end{tabular}
\label{tab:optimisationsDistance}
\end{table*}
\par In addition to the Deep Salience representation, data augmentation (DA) was also explored to increase the available training data for our experiments. To this end,  20\% additional training samples were generated by employing a random pitch shift of up to \SI{\pm 30} cents, using librosa \citep{mcfee2015librosa}.
For the Siamese models trained without data augmentation, the naming convention employed is \begin{math} \mathit{SCNN_{x}}\end{math}, where $x$ is the feature representation used during training. The models trained using data augmentation are named \begin{math} \mathit{SCNN_{CQT+DA}}\end{math} and \begin{math} \mathit{SCNN_{Sal+DA}}\end{math} for the CQT  and the salience representations respectively.
\par As Tables \ref{tab:optimisationsBinary} and \ref{tab:optimisationsDistance} demonstrate, both the pitch salience representation (\begin{math}\mathit{SCNN_{Sal}}\end{math}) and data augmentation (\begin{math}\mathit{SCNN_{DA}}\end{math}) prove to be effective in improving the performance of our model over \begin{math}\mathit{SCNN_{CQT}}\end{math}, with the salience representation contributing greater improvements. It could be inferred that using salience representations makes it easier for the model to learn meaningful features from the input representations, since it emphasises pitched content. Improvements using data augmentation can be attributed to the fact that pianos are not always tuned to $A = 440$ Hz in the real world, and often the relative intervals are also not tuned perfectly, hence comparison with MIDI files might lead to false negatives in such cases. Data augmentation ensures that the disparity between our training and test conditions is \blue{minimised} by simulating more realistic conditions in our training data.  A combination of distance matrix, salience representation and data augmentation (\begin{math}\mathit{SCNN_{Sal+DA}}\end{math}) yields the best results, as can be seen from Table \ref{tab:optimisationsDistance}.
\subsubsection{Confusion Matrix}
\begin{center}
\begin{tabular}{c >{\bfseries}r @{\hspace{0.7em}}c @{\hspace{0.4em}}c @{\hspace{0.7em}}l}
  \multirow{10}{*}{\rotatebox{90}{\parbox{1.1cm}{\bfseries\centering Actual\\ value}}} & 
    & \multicolumn{2}{c}{\bfseries Prediction outcome} & \\
  & & \bfseries p$'$ & \bfseries n$'$ & \bfseries %
  \\
  & p & \MyBox{TP}{82} & \MyBox{FN}{18} & 
  \\[2.4em]
  & n & \MyBox{FP}{17} & \MyBox{TN}{83} & 
  \\
  &  & Confusion Matrix & ($\mathit{SCNN_{Sal}}$) &
\end{tabular}\label{tab:confusion2}
\end{center}
\par The confusion matrix \blue{(values in percentage)} using the optimisations suggests that the salience representation \textit{evens out} the disparity between false positives and false negatives, evident from the comparison of $\mathit{SCNN_{Sal}}$ with $\mathit{SCNN_{CQT}}$. This corroborates that using the optimisations prescribed in this section helps improve model performance, especially in data scarce conditions. This also raises the question: \textit{Given enough data, would the salience representation become redundant?} This shall be answered in an upcoming section analysing performance of the $\mathit{SCNN_{CQT}}$ and $\mathit{SCNN_{Sal}}$ across different data settings.

\par The experimentation until now focused on the alignment of piano music. The next section presents a study on the capacity of the Siamese models to generalise to different instrumentation settings containing non-piano music. Since the similarity matrix containing distance labels yields better results, further studies on the generalisation capability of the models to different instrumentation settings only report results using the distance labels, in the upcoming section.
\subsection{Generalisation to other instruments}\label{ssec:gen}
 A number of feature representations have been explored over the years for alignment-related tasks. These include the plain spectrogram and its reductive derivatives, the semigram, the chromagram (the most popular choice), onset features and mel-frequency cepstral coefficients (MFCCs). Different feature choices have different advantages and are hence suitable in different conditions. For instance, onset features work well for piano music but not so well for violin/cello music. MFCCs work better on similar signals (for instance for two spoken utterances), whereas chroma features are able to generalise better to audio-to-score alignment. The question arises: 
 \begin{quote}
     Can we create a representation that is the \textit{Jack of all trades}, but with the potential to be the \textit{Master of one?}
 \end{quote}
 \par The previous sections compared the performance of the Siamese representations and its optimisations with the chromagram representation for classical piano music. A key advantage offered by the data-driven approach described in this chapter is the ability to \textit{learn} the similarity from the data itself, hence improving domain coverage and generalisation capacity, i.e. making the method applicable to various target domains. This section explores the applicability of the Siamese models to audio-to-score alignment for \textit{non-piano music}, and thereby attempts to answer the aforementioned question.

\begin{table}[ht]
\vspace{0.5cm}
   \caption{Datasets used for metric learning experiments on non-piano music}
   \vspace{0.2cm}
   \centering
   \begin{tabular}{ c c c c} \toprule
      \textbf{Name} & \textbf{Instrument} & \textbf{Recordings}  & \textbf{Stage} \\ \midrule
      MAPS \small{\citep{emiya2009multipitch}} &  Piano  &  238 &  Train\\ \midrule
      RWC \small{\citep{goto2002rwc}} & Various  & 61  & Train \\ \midrule
      \small{SCREAM-MAC-EMT \citep{li2015analysis}} & Violin & 60 & Test \\ \midrule
      \small{Traditional Flute Dataset \citep{brum2018traditional}} & Flute & 30 & Test \\ \midrule
      \bottomrule
\end{tabular}
\label{tab:generalisationDatasets}
\end{table} 
\par It is desirable to test the performance of these models on unseen data containing non-piano music whilst being trained on data combining various instrumentation settings. This will ensure that the model is actually \textit{learning} meaningful relationships \blue{that can generalise well to different test settings.}
To this end, the classical part of the Real World Computing (RWC) dataset \citep{goto2002rwc} is employed in addition to the MAPS dataset to train the Siamese models. This dataset consists of audio-MIDI alignment data for 61 classical pieces, for various instruments including the piano, violin, clarinet and string quartet. These pieces are appended to the MAPS dataset to yield the entire training dataset.

\par The capability of the models for generalisation on other instruments is tested for flute music and violin music respectively. The performance for violin music is tested on the SCREAM-MAC-EMT dataset \citep{li2015analysis} and the performance for flute music is tested on the Traditional Flute Dataset for Score Alignment \citep{brum2018traditional}. The contents of the training and testing datasets are summarised in Table \ref{tab:generalisationDatasets}. The results on violin music and flute music are provided in Table \ref{tab:violin} and Table \ref{tab:flute} respectively. For the experimentation on violin music, in addition to $\mathit{MATCH}$ and $DTW_{Chroma}$, the performance of the $\mathit{SCNN}$ models is also compared with $\mathit{SCNN}$ models trained only on piano music. These are reported to demonstrate the difference made by incorporating the RWC data during training. 
 
\begin{table*}[th]
\vspace{0.5cm}
\caption{Results of Siamese models for violin music\\ \blue{$*$: significant differences from \begin{math}\mathit{SCNN_{Sal+DA}}\end{math}, $p < 0.05$ }}
   \centering
\begin{tabular}{ccccc} \toprule
\hline 
\multirow{2}{*}{\textbf{Model}} & \multicolumn{4}{c}{\textit{Error Margins}}
\tabularnewline 
& \textbf{$<$25ms}& \textbf{$<$50ms} & \textbf{$<$100ms} & \textbf{$<$200ms} \\
\midrule 
$\mathit{MATCH}$ \blue{\citep{dixon2005match}}  & 60.7* & 68.9* & 74.8* & 80.4* \\
\midrule
 $DTW_{Chroma}$  & 57.3* & 67.5* & 72.4* & 78.6* \\
\midrule 
\begin{math}\mathit{SCNN_{CQT}} \end{math} (Piano)  & 63.1* & 70.5* & 76.7* & 81.5*\\
\midrule 
 \begin{math}\mathit{SCNN_{Sal}} \end{math} (Piano)  & 64.8* & 71.6*  & 77.4* & 82.4* \\
\midrule 
 \begin{math}\mathit{SCNN_{CQT}}\end{math}  & 65.4* & 72.6* & 78.9* & 84.8*\\
\midrule 
 \begin{math}\mathit{SCNN_{Sal}}\end{math}  & 67.1* & 73.7*  & 79.5* & 86.9* \\
\midrule 
 \begin{math}\mathit{SCNN_{CQT+DA}}\end{math} & 68.3* & 74.8* & 80.4* & 87.5*\\
\midrule 
 \begin{math}\mathit{SCNN_{Sal+DA}}\end{math}  & \textbf{69.5} & \textbf{76.1} & \textbf{81.2} & \textbf{89.2} \\
\midrule 
\bottomrule
\end{tabular}
\label{tab:violin}
\end{table*}

\begin{table*}[th]
\vspace{0.4cm}

\caption{Results of Siamese models for flute music\\ \blue{$*$: significant differences from \begin{math}\mathit{SCNN_{Sal+DA}}\end{math}, $p < 0.05$ }}
   \centering
\begin{tabular}{ccccc} \toprule
\hline 
\multirow{2}{*}{\textbf{Model}} & \multicolumn{4}{c}{\textit{Error Margins}}
\tabularnewline 
& \textbf{$<$25ms}& \textbf{$<$50ms} & \textbf{$<$100ms} & \textbf{$<$200ms} \\
\midrule 
$\mathit{MATCH}$ \blue{\citep{dixon2005match}}  & 61.6* & 69.2* & 76.4* & 81.2* \\
\midrule
 $DTW_{Chroma}$  & 58.1* & 67.9*  & 73.0* & 79.3* \\
\midrule 
 \begin{math}\mathit{SCNN_{CQT}}\end{math}  & 66.3* & 73.1* & 77.1* & 85.4*\\
\midrule 
 \begin{math}\mathit{SCNN_{Sal}}\end{math}  & 67.9* & 74.6*  & 79.2* & 87.3* \\
\midrule 
 \begin{math}\mathit{SCNN_{CQT+DA}}\end{math} & 69.1* & 75.2* & 80.9* & 88.2*\\
\midrule 
 \begin{math}\mathit{SCNN_{Sal+DA}}\end{math}  & \textbf{70.4} & \textbf{76.8} & \textbf{81.7} & \textbf{89.8} \\
\midrule 
\bottomrule
\end{tabular}
\label{tab:flute}
\end{table*}

\par The results on both instruments suggest that this method is a promising approach to alignment and is able to generalise well to different test settings with limited training data of the target domain. This is possible due to the trainable nature of the Siamese Network, unlike the manually handcrafted features that are used by standard DTW-based algorithms. Additionally, Siamese models are typically less data-hungry than other neural architectures based on CNNs. \blue{This is because for $n$ samples in a dataset, the Siamese network can be trained on $(n^2 - n)/2$ unique input pairs: $n^2$ possible pairings between each input, minus the $n$ pairings between two of the same samples, and divided by two to account for two permutations being counted as separate combinations.} This enhances the generalisation capacity in the presence of limited data. In order to boost the performance for a target domain even further, model adaptation \citep{li2020model, agrawal2017three} could be employed, which is left to be explored as part of future work. 



\subsection{Further analysis}
\subsubsection{In which scenarios are learnt frame similarities most useful?}
\begin{figure*}[ht]
  \centering
  \includegraphics[width=\columnwidth]{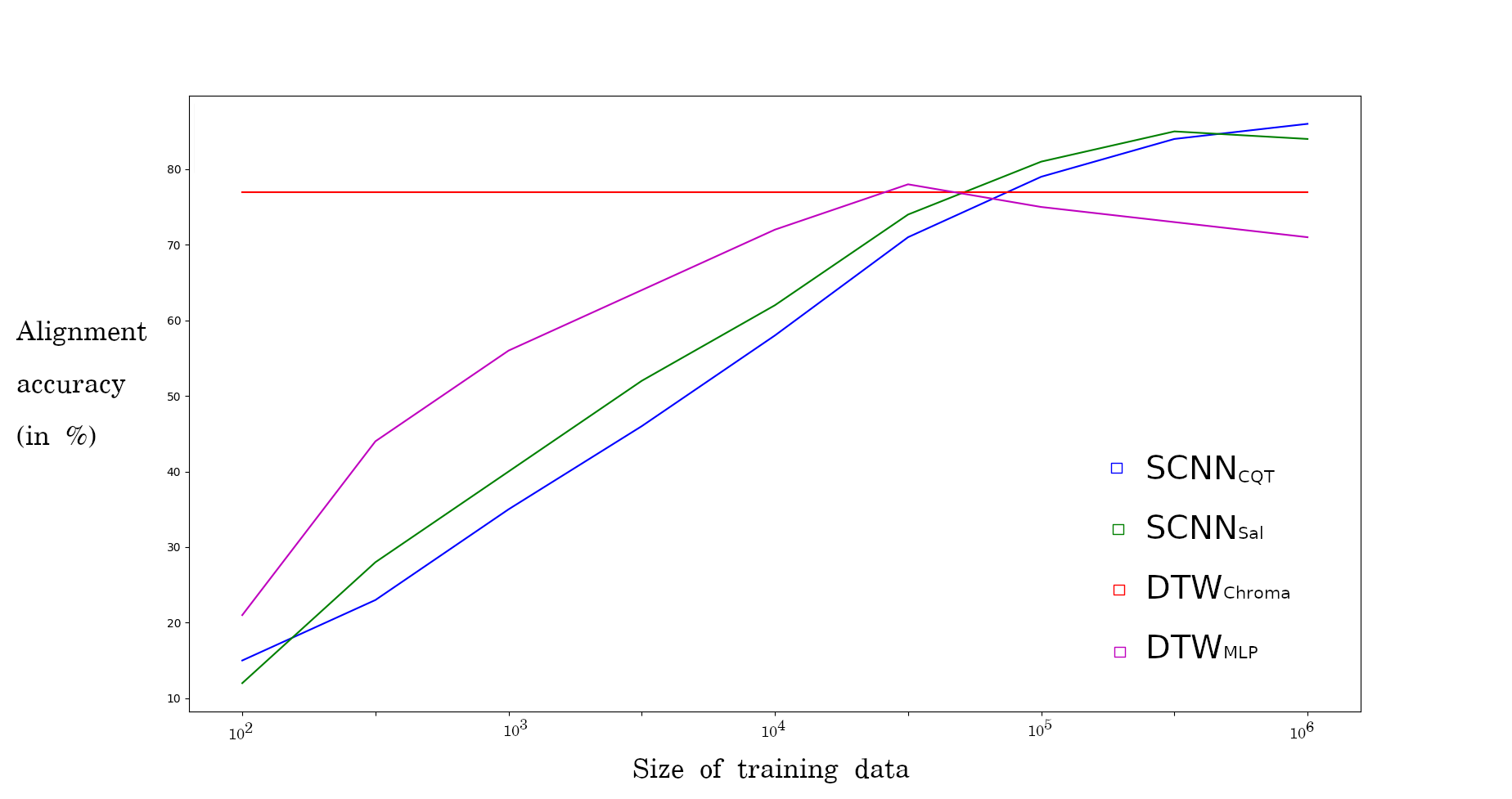}
  \vspace{0.2cm}
  \caption{Model performance according to available training data}
  \label{fig:ablationData}
\end{figure*}
\par The previous section highlights the applicability of the metric learning method across various acoustic and instrumentation settings. However, obtaining sufficient data for the target domain remains a challenging step for training Machine Learning models, especially for obscure test conditions. To determine the data needs of our models and delineate the optimal data settings where the metric learning approach results in better performance over traditional methods, this section compares the performance of the models across various \textit{data settings}.
\par An ablative analysis was conducted by training the Siamese models on varying amounts of data, in order to determine the data settings that optimise model performance. To this end, a comparison of the models \begin{math}\mathit{SCNN_{CQT}}\end{math}, \begin{math}\mathit{SCNN_{Sal}}\end{math} and \begin{math}\mathit{DTW_{MLP}}\end{math} trained on various subsets of the entire training set discussed earlier in the chapter is carried out. To obtain the learning curve, incremental amounts of training data are employed, containing 100, 500, 1000, 5000, 10000, 50000, 100000, 500000, and 1000000 data samples respectively. 
The results of this study are shown in Figure \ref{fig:ablationData}.

 The graph demonstrates that the Siamese models begin outperforming the\\ \begin{math}\mathit{DTW_{Chroma}}\end{math} and \begin{math}\mathit{DTW_{MLP}}\end{math} models at around 75000 data points, and are especially promising in scenarios with $10^5$ or higher number of data samples (frame pairs) available. This can readily be provided by a dataset containing approximately 200 pieces or more. This study suggests that the proposed metric learning based method would be best employed when there are 250 or more audio-score pairs available. Additionally, since the method is trained on matching/non-matching pairs of frames, synthetic data mimicking the test conditions can be readily employed during training, by generating the audio using different soundfonts and adding various acoustic perturbations. This can significantly increase the amount of training data without requiring annotated alignments, which are often not readily available for real-world settings. \blue{An interesting observation from the graph is that the performance of the \begin{math}\mathit{DTW_{MLP}}\end{math} goes down with increasingly large training data. On inspection of the data, it was found that the drop in performance was caused due to the disparity in the acoustic conditions between the additional training data and the test dataset.}
\begin{figure*}%
    \centering
   \vspace{-0.5cm}
      \subfloat[Example piece with no structural changes]
  {\includegraphics[width=7cm, height=6cm]{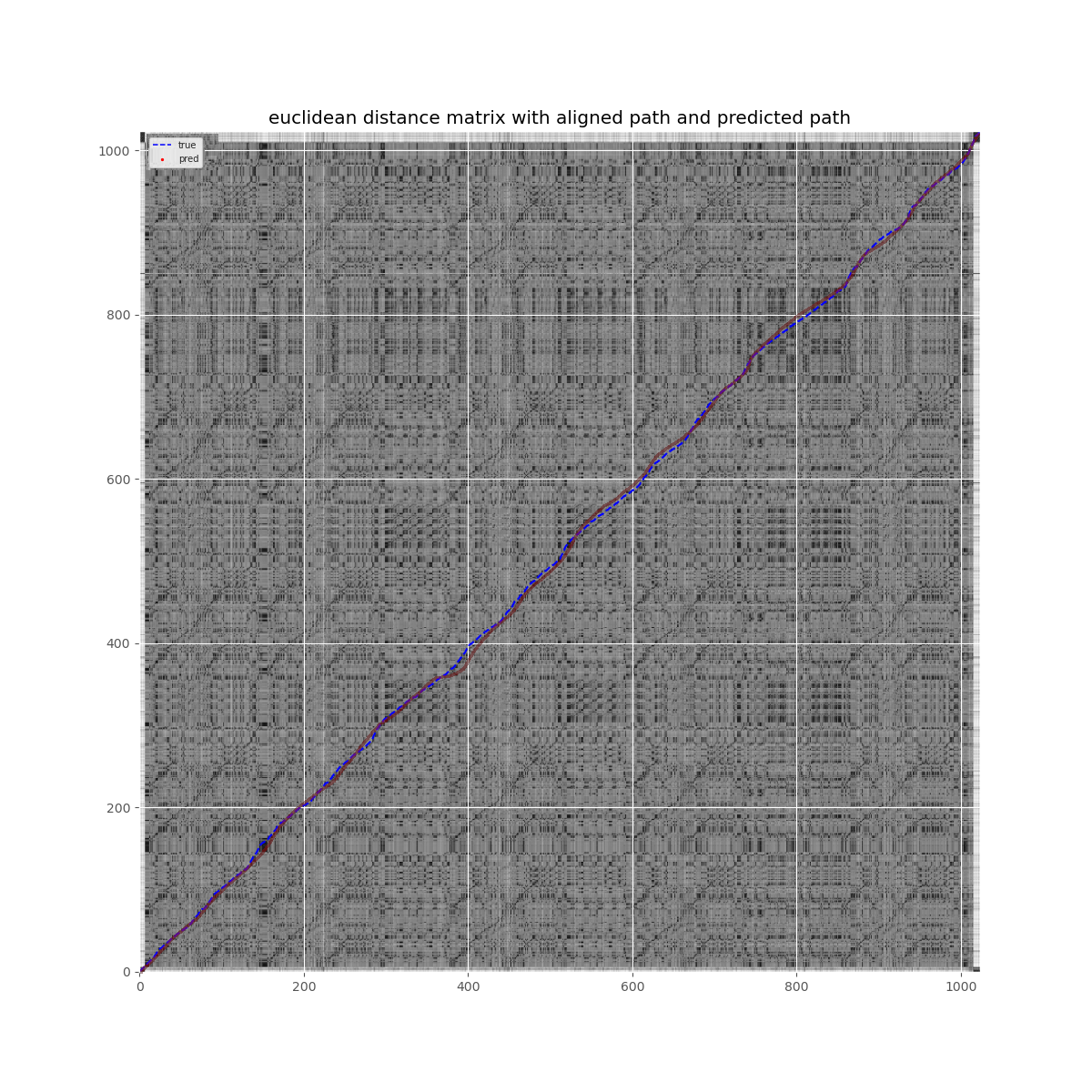} }

 \subfloat[Example piece with a short repeat]
  {\includegraphics[width=7cm, height=6cm]{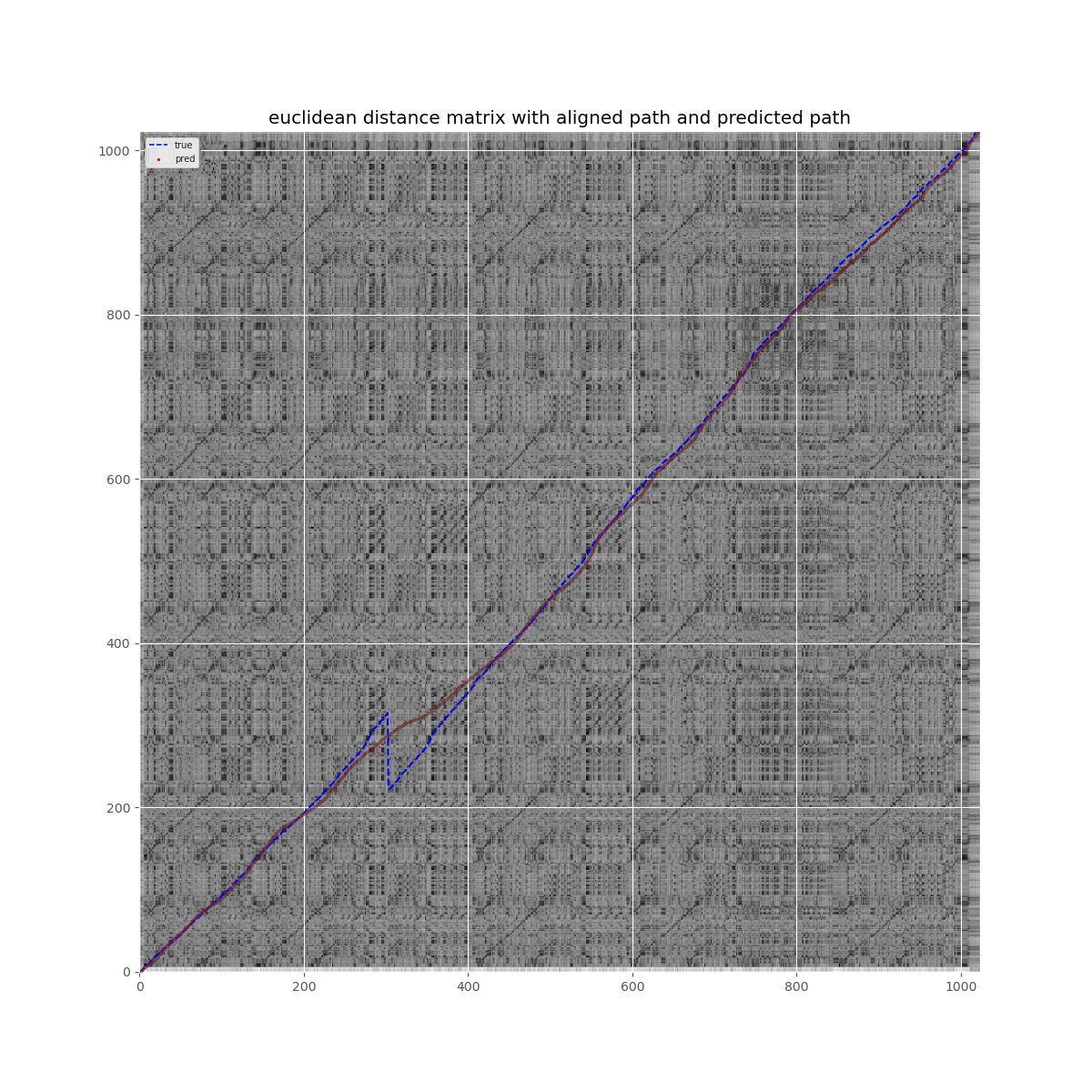} }
  
    \subfloat[Example piece with a long repeat]
    {\includegraphics[width=7cm, height=6cm]{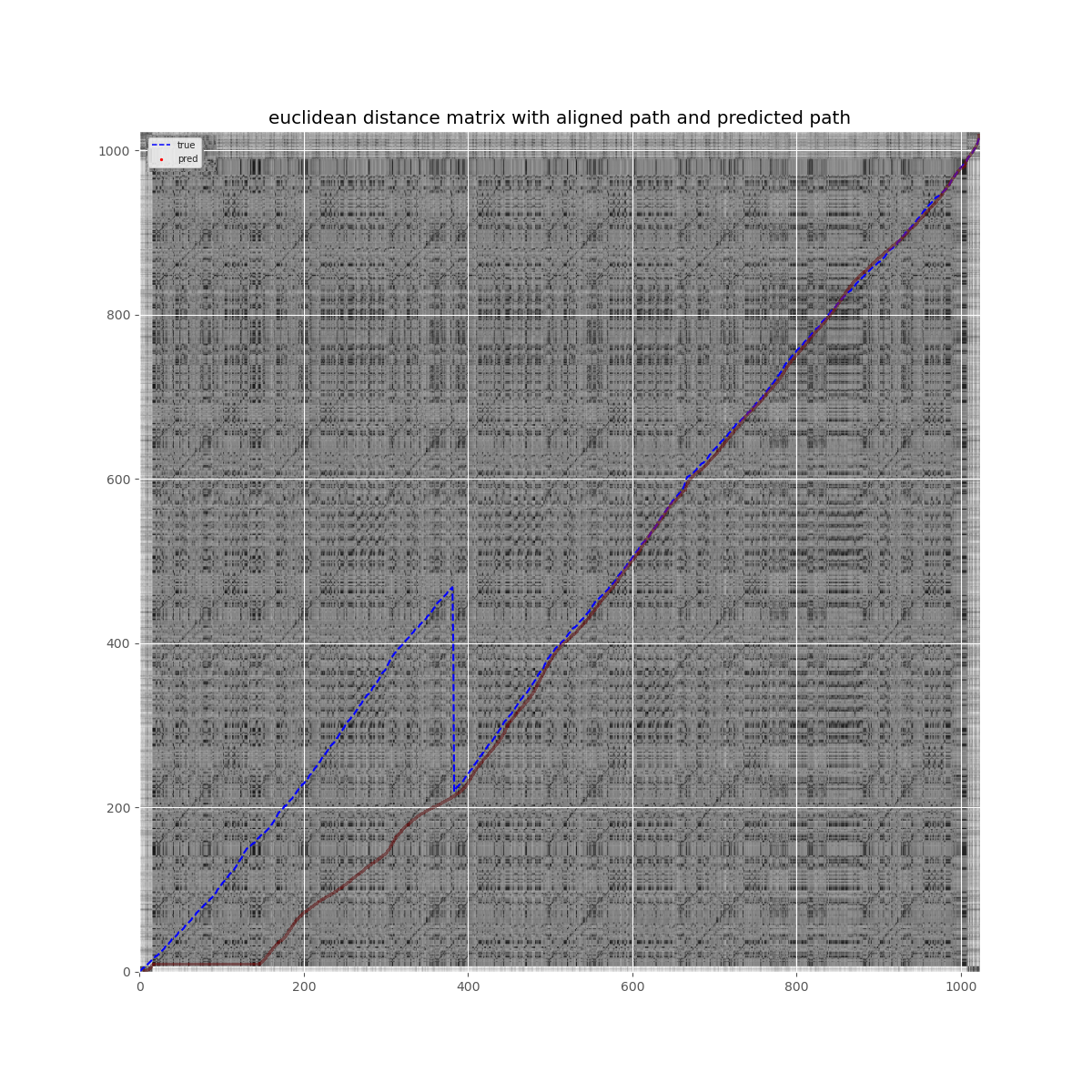} }

    \caption{Performance of the Siamese model on pieces containing structural deviations from the score.\\
    Predicted path in red, ground truth in blue.\\
    X-axis: Frame index (performance), Y-axis: Frame index (score)}%
    \label{fig:robustness}
\end{figure*}
\subsubsection{A note on salience representations}
The figure also suggests that \begin{math}\mathit{SCNN_{CQT}}\end{math} starts outperforming \begin{math}\mathit{SCNN_{Sal}}\end{math} beyond ~500000 data points. The trend continued beyond 1000000 data points, albeit with little difference between subsequent orders of magnitude. While the salience representation improves performance and reduces the false positive rate over the CQT representation, the ablative study on data settings suggests that given ample training data, the Siamese model could be able to differentiate between meaningful and unnecessary information from the CQT representation itself. The salience representation, therefore, could be eschewed in the presence of data-abundant settings.
\subsubsection{Robustness to structural changes}


In addition to quantitative analysis, some alignment plots are provided in this subsection in order to aid qualitative analysis, and to specifically analyse model performance depending upon structural agreement between the performance and score. 
Figure \ref{fig:robustness} depicts the performance of the Siamese model on some example pieces. Example (a) contains complete structural agreement between the score and the performance. Examples (b) and (c) contain structural differences between the score and the performance via the presence of a short and long repeated segment respectively.
\par Since DTW assumes structural agreement between the score and the performance, it does not allow for jumps and hence the method is unable to align pieces containing major structural deviations from the score. The next chapter provides a method to handle such structural differences between the performance and score and presents a novel method for structure-aware alignment.

\section{Conclusion and further developments}\label{sec:ch3_conc}
Audio-to-score alignment is the task of finding the optimal mapping between a performance and the score for a given piece of music. Dynamic Time Warping (DTW)  has been the de facto standard for this task, typically incorporating handcrafted features \citep{dixon2005line, ewert2009high, arzt2012adaptive}. The primary limitation of handcrafted features lies in their inability to adapt to different acoustic settings and thereby model real world data in a robust manner, in addition to not being \blue{optimised} for the task at hand. 
This chapter presented novel applications of representation learning and metric learning for audio-to-score alignment. 
\par While the representation learning method demonstrated slight improvements over standard feature representations, the metric learning method provided much better improvements and an intuitive approach to aid DTW-based alignment via \textit{learnt frame similarity}. This approach employed Siamese Convolutional Neural Networks to learn the frame similarity matrix, which was then used in the DTW computation to generate the fine alignment. The proposed metric learning based method is efficiently able to learn representations for DTW directly from data. 
\par The learnt similarity matrix provides a way to improve the domain coverage of the representations and enables the models to generate robust alignments across various acoustic and instrumentation conditions. In principle, the method is also adaptable to different acoustic settings using domain adaptation techniques, unlike traditional DTW-based methods that employ handcrafted features. Additionally, the proposed method could also be applied to other synchronisation tasks, such as audio-to-audio alignment of multiple audio recordings of the same piece. 
  \par The experimentation on music data from different acoustic conditions demonstrates that the proposed method of learning frame similarity using Siamese neural networks combined with the DTW computation is a promising method for audio-to-score alignment. The principal advantage of this approach over standard methods with traditional feature choices (like chroma features or MFCCs) is the ability to learn directly from data, which provides higher relevance, coverage and adaptability. In other words, this method is able to yield accurate alignments across different acoustic settings, at the same time offering the capability to be adapted to a particular setting.
This chapter also demonstrated that salience representations and data augmentation are effective techniques to improve alignment accuracy, in the presence of data-scarce conditions. 
\par Potential further developments to this method include the incorporation of attention into the convolutional models to aid training and improve performance. The exploration of transfer learning using pretrained models trained for a different task but on a large-scale dataset such as AudioSet \citep{gemmeke2017audio}, and the adaptation of the Siamese models to a particular target domain using various domain adaptation techniques could also serve to be a promising exploration.
  A limitation of the proposed method is the inability to handle structural differences between the scores and performances. This limitation stems from the reliance on the DTW computation, which assumes structural agreement and only allows for a monotonically increasing warping path. 
  The next chapter presents a novel method that addresses this issue and generates robust structure-aware alignments.

\chapter{Structure-Aware Performance Synchronisation}\label{ch:repeatDetection}
The previous chapter presented a novel approach for data-driven performance-score synchronisation using learnt spectral similarity at the frame level. The method proposed therein progresses the development of DTW-based methods by incorporating a learnt similarity matrix, rather than employing traditional handcrafted feature representations. A limitation of standard DTW-based alignment techniques is the inability to capture structural differences between multiple representations of a piece of music, stemming from the assumption of complete structural agreement between the input representations. This chapter focuses on overcoming this limitation of DTW-based methods, and presents a novel method for the offline structure-aware alignment task. The proposed method is applicable to both performance-score synchronisation as well as performance-performance synchronisation, unlike previously proposed structure-aware approaches that generally cater to one of these tasks. The next section introduces the structure-aware alignment task and presents the motivation behind the method proposed in this chapter. 

\section{Introduction}


\par The analysis of 
music performance is a challenging area of \blue{Music Information Processing (MIP)}, owing to multiple factors such as suboptimal recording conditions, structural differences from the score, and subjective interpretations. 
Deviations from the structure and/or tempo prescribed by the score are some methods through which music performers add expressiveness to their music. Such deviations are 
common in several genres of music, particularly classical music \citep{widmer2016getting}. The identification and handling of structural differences between a music performance and the score is therefore a challenging yet integral component of a robust synchronisation method. The particular path through the score that the musician is going to take is difficult to predict, and although there have been several methods proposed in the recent past to handle these impromptu changes;  it still remains a problem that is not fully solved \citep{arzt2016flexible}.  
\par The majority of the prominent approaches for audio-to-score alignment are based on Dynamic Time Warping (DTW) \citep{dixon2005line}, or Hidden Markov Models (HMM) \citep{muller2015fundamentals}. The primary reason that such approaches do not work well for structure-aware alignment is that these methods typically assume that the musician follows the score from the beginning to the end without inducing any structural modifications, which is often not the case in real world scenarios. The alignments computed using DTW-based methods are constrained to progress monotonically through the piece, and are thereby unable to model structural deviations such as repeats and jumps. Similarly, the Markov assumption inherent in HMMs assumes that the probability of each event depends only on the state attained in the previous event, thereby limiting the incorporation of context that is necessary for structure-aware alignment. 
\par A few approaches have been proposed over the years for handling structural changes during alignment. These methods are either reliant on Optical Music Recognition (OMR) to detect repeat and jump directives \citep{Fremerey2010handling}; or on frameworks fundamentally different from DTW, such as the \blue{Needleman-Wunsch Time Warping} method \citep{grachten2013automatic}. The former method, called \begin{math}\textit{JumpDTW}\end{math}, requires manually annotated frame positions for block boundaries, initially provided by an OMR system, and is unable to model impromptu jumps and deviations that are not foreseeable from the score. The latter is unable to align repeated segments, since it does not introduce backward jumps, but is rather based on a waiting mechanism upon mismatch of the two streams. 
 \par This chapter presents a novel method for structure-aware alignment of a performance and the corresponding score, or two performances of a given piece of music. We propose a custom Convolutional Neural Network (CNN) based architecture for modeling the structural differences between the input representations, coupled with a flexible DTW framework to generate the fine-grained alignments. 
 Our method does not require a large corpus of hand-annotated data and can also be trained exclusively using synthetic data if hand-annotated data is unavailable. While the method is described with a focus on the performance-score synchronisation task, which is a more common application of alignment methods, it is also applicable to the performance-performance synchronisation task, i.e.\ the alignment of two performances or audio recordings, with potential unavailability of the reference scores. 
 \par The remainder of the chapter is structured as follows: The next section discusses the types of structural differences present in (classical) music and their possible sources. Section \ref{sec:related} summarises the prominent approaches that were previously proposed for the task and highlights their limitations. The proposed approach to overcome these limitations is presented in Section \ref{sec:method}, followed by a description of the experimental setup in Section \ref{sec:setup} and a discussion of the results in Section \ref{sec:results}. The chapter then concludes with the key takeaways and briefly discusses potential further extensions of the proposed method in Section \ref{sec:conc}. 
\section{Types of structural differences and their sources}\label{sec:types}
\begin{figure}[htbp]
  \centering
  \includegraphics[width=6.5in]{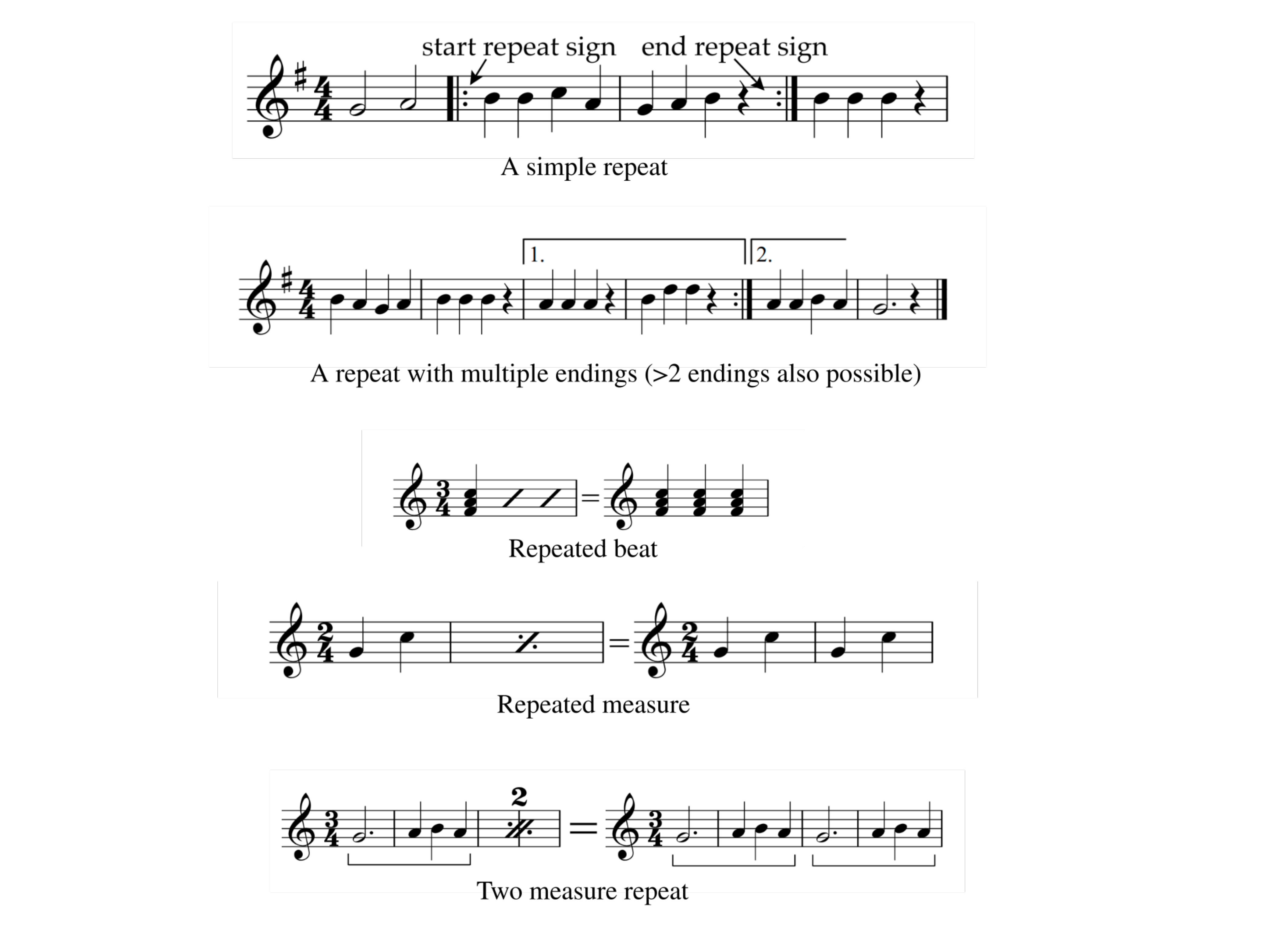}
  \caption{Types of repeats in classical music.\\Some examples from \textit{learnmusictheory.net}}
  \label{fig:repeats}
\end{figure}

\begin{figure}[htbp]
  \centering
  \includegraphics[width=6.5in]{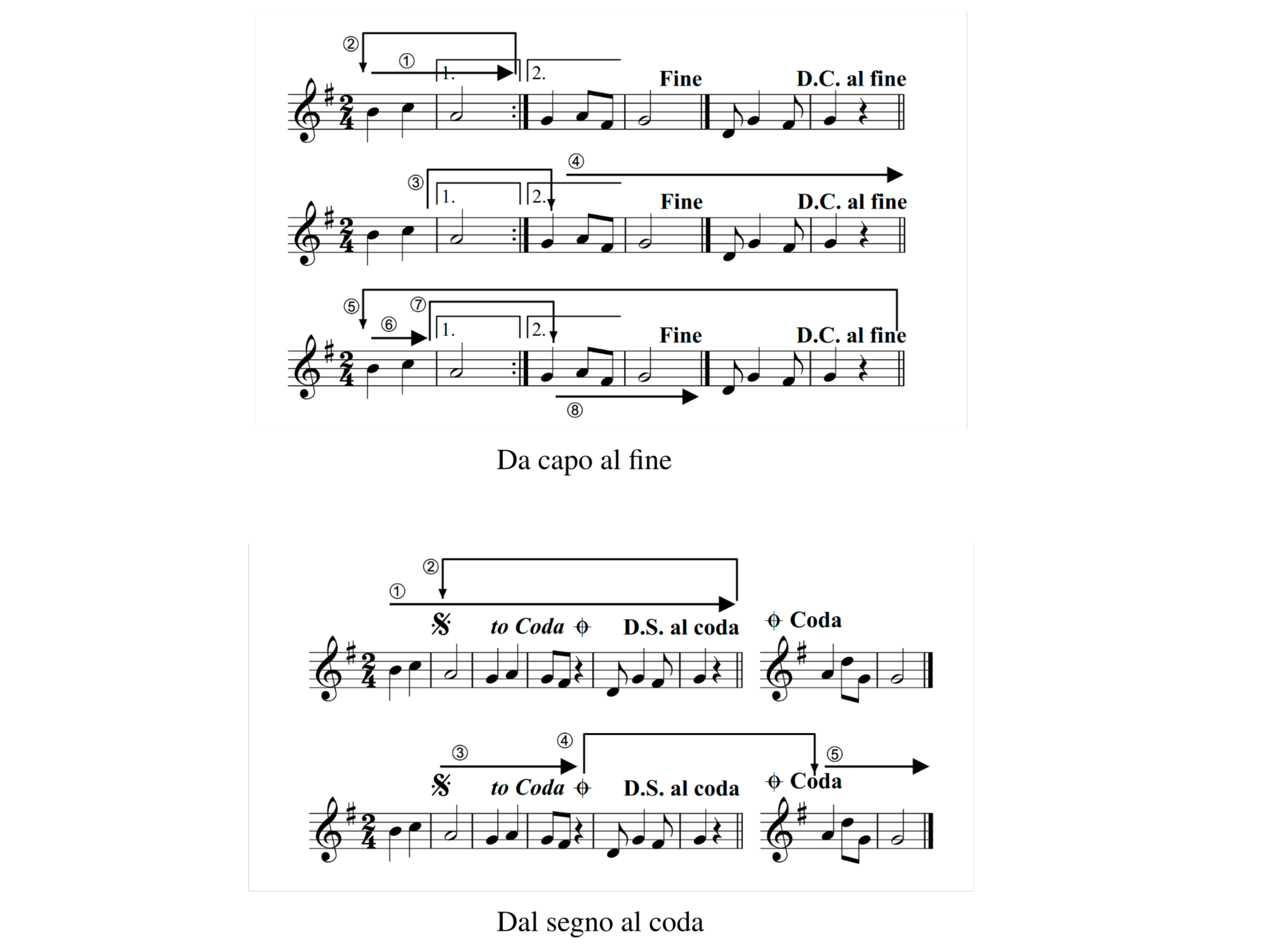}
  \caption{Examples of other section order markers.\\Some examples from \textit{learnmusictheory.net}}
  \label{fig:dc_ds}
\end{figure}

\blue{A structural difference between a given score and performance audio is defined as a deviation exhibited by the performance from the global structure prescribed by the score. Since this typically occurs at the bar level, in most cases these differences correspond to the discrepancy between the notated bar sequence and the performed bar sequence. However, the proposed method in this chapter offers greater flexibility and allows for \textit{intra-measure} jumps, thereby also considering structural differences in cases where the score and performance have the same bar sequence but a different frame sequence, i.e. cases where the frame sequence does not increase monotonically along the score axis.} 
The majority of alignment approaches assume complete structural agreement between a performance and the score or the sheet notation. However, in practice, a performer does not always completely follow the structure prescribed by the score from the beginning to the end without any modifications. Rather, depending upon the scenario (stage/informal/practice), the performance could assume a specific structure, with some changes introduced in a spontaneous manner by the performer.\\
\\
\textit{What could be the causes of structural differences in such cases?}
\begin{itemize}
    \item Performers often induce structural modifications by means of ignoring or annexing repeats indicated in the score, in order to add expressivity to their music or to alter the length of a programme.
    \item A performer might even skip certain sections of the music, depending upon the performance scenario. This is more common during long performances.
    \item There could be an \textit{on the fly} addition of sections not included in the score, such as cadenzas or improvisational passages. These are the most challenging to model for an automatic alignment method. 
\end{itemize}
\vspace{0.5cm}
\textit{What could be the types of structural differences induced due to the above reasons?}
\begin{itemize}
    \item Backward jumps: A backward jump occurs when a performance jumps \textit{back} to a previous position in the score. For example, repeats, a prominent component of classical music, could contribute to such jumps depending upon their realisation. Figure \ref{fig:repeats} depicts the types of repeats that are commonly encountered in classical music. A section repeated more often than prescribed could lead to such jumps. Additionally, a section repeated without any repeat marker, or certain mistakes could also contribute to backward jumps.
    \item Forward jumps: A forward jump occurs when a performance jumps \textit{ahead} to a further position in the score. For example, a repeat sign that is ignored or certain sections of the sheet music that are skipped could contribute to such jumps. 
\end{itemize}
\vspace{0.5cm}
These forward and backward jumps could further be categorized into \textit{inter-block} and \textit{intra-block} jumps, as follows:
\begin{itemize}
    \item Inter-block jumps: Inter-block jumps primarily occur at the boundaries of \textit{sections} of the sheet music. Note that these sections are loosely defined in many cases, however, certain markers such as \textit{D.S. al Coda, D.C. al Coda, D.S. al Fine} and double bar lines often mark such sections and thus notate inter-block jumps. Figure \ref{fig:dc_ds} provides examples of these markers.
    \item Intra-block jumps: These are jumps that occur within \textit{sections} in the sheet music. These are more common in the improvisational and practice scenarios.
\end{itemize}
This results in four possible jump types in total:  the forward inter-block jump, the backward inter-block jump, the forward intra-block jump and the backward intra-block jump.
\vspace{0.2cm}
\par The section so far discussed the causes and types of structural differences between a specific performance and a specific score for a given piece of music, such as the presence of structural markers in sheet music that can result in such differences depending upon their realisation and interpretation by the performer, as well as changes added in an impromptu manner. In addition to these cases where the structural differences are induced \textit{by the performer}, there might be scenarios where a particular performance could be following a different version or edition of the sheet music from the one being used at analysis time. This discrepancy in the sheet music versions could also result in inadvertent deviations from the structure prescribed by the  score that is being employed at analysis time.

\par Structural incorporation is also essential for performance-performance synchronisation, wherein multiple (audio) performances vary among each other in terms of the structure. 
The method proposed in this chapter addresses all these scenarios, regardless of the input modalities (audio-to-audio or audio-to-score), the sources of the structural differences, as well as the types of the induced jumps. 
I now discuss previously proposed approaches for structure-aware alignment to facilitate a better understanding of the context as well as the motivation behind the proposed method.

\section{Previous approaches and their limitations}\label{sec:related}
The alignment of performances containing structural deviations from the score has been studied mainly using rule-based and stochastic approaches \citep{Fremerey2010handling, grachten2013automatic, nakamura2015real, jiang2019offline}. The prominent methods among these for offline audio-to-score and audio-to-audio alignment include  \begin{math}\textit{JumpDTW}\end{math} \citep{Fremerey2010handling} and Needleman-Wunsch Time Warping (\begin{math}\textit{NWTW}\end{math}) \citep{grachten2013automatic}  respectively.
\par \citet{Fremerey2010handling} focus on tackling structural differences induced specifically by repeats and other section markers in the sheet music using a novel DTW variation called \begin{math}\textit{JumpDTW}\end{math}. This method identifies the \textit{block sequence} taken by a performer along the score, with a block being a musically relevant section in the score. This method has the following limitations:
\begin{itemize}
    \item Since it is a score alignment method, it assumes that the block markers are available \textit{a priori} during test time. However, such markers are not always available at test time for real world applications. In such scenarios, this method is reliant on block boundaries automatically detected using OMR, and hence is affected by commonly occurring OMR-induced errors. These boundaries often need to be manually corrected to ensure robust performance.
    \item The approach cannot align deviations that are not foreseeable from the score.
    \item It is unable to capture intra-block jumps (defined in Section \ref{sec:types}).
    \item It is not applicable to performance-performance synchronisation without the availability of scores. 
\end{itemize}
\begin{math}\textit{NWTW}\end{math} \citep{grachten2013automatic}, on the other hand, is a pure dynamic programming method for the audio-to-audio alignment of music recordings that contain structural differences. This method is an extension of the classic Needleman-Wunsch sequence alignment algorithm  \citep{likic2008needleman}, with added capabilities to deal with the time warping aspects of aligning music performances. 
A caveat of this method is that it cannot skip certain parts of the score, thereby being unable to effectively model forward jumps.  Additionally, 
it does not successfully align repeated segments owing to its waiting mechanism. This mechanism makes the model wait during the non-matching parts of either sequence, and then introduces a clean jump (i.e. a horizontal transition with the repeated portions unaligned) when the two streams match again.
\par Apart from \begin{math}\textit{JumpDTW}\end{math} and \begin{math}\textit{NWTW}\end{math}, which focus on offline structure-aware alignment using time warping based methods, a few methods have explored Hidden Markov Models (HMMs) for modelling variations from the score, in both the online \citep{nakamura2015real} and offline \citep{jiang2019offline} settings. While the former method \citep{nakamura2015real} focuses on real-time score following for monophonic music, the method proposed in this chapter deals with offline audio-to-score and audio-to-audio alignment for polyphonic music performance. 
The latter method \citep{jiang2019offline} focuses on offline score alignment for the practice scenario. Similar to \citet{nakamura2015real}, their approach is also based on HMMs; but they propose using pitch trees and beam search to model skips. However, their method struggles with pieces containing both backward and forward jumps, which is an important challenge that is tackled by the method proposed in this chapter. 
\par Recent work on audio-to-score alignment has moved towards machine learning based methods and has demonstrated the efficacy of multimodal embeddings \citep{dorfer2018learning}, reinforcement learning \citep{dorfer2018learning2, henkel2019score} and learnt frame similarities \citep{agrawal2021learning}, albeit these are not structure-aware methods. \citet{shan2020improved} propose Hierarchical-DTW to automatically generate piano score following videos given an audio and a raw image of sheet music. Their method is reliant on an automatic music transcription system \citep{hawthorne2017onsets} and a pre-trained model to extract bootleg score representations \citep{tanprasert2019midi}. It struggles when this representation is not accurate, and also on short pieces containing jumps. 
While they work with raw images of sheet music and generate score following videos; the method proposed in this chapter is focused on offline alignment, is not reliant on other pre-trained models, and performs well on both short and long pieces. 
\section{Proposed Methodology}\label{sec:method}

Having discussed the previously proposed approaches for structure-aware alignment and their limitations, it is evident that since the particular path through the score to be taken by the performer is difficult to predict beforehand, structural changes  are challenging to model using rule-based approaches, and machine learning methods offer promise at effectively addressing these challenges.

\par A time-series analysis based approach could be applicable in this scenario, utilising sequence models such as RNNs or LSTMs. However, modelling the two separate inputs using a sequence-to-sequence model such as an RNN would typically require multiple encoders, with each encoder representing a token (at each timestamp) from the performance and score sequences respectively. This formulation would thereby result in a complex model with a reasonably high number of parameters, and would require large annotated training data to yield accurate results. In addition to the data-hungry nature of RNNs, these models also suffer from the problem of vanishing gradient for very long inputs \citep{hochreiter2001gradient, agrawal2017building}. This makes these models unsuitable for the alignment task (in their native form), since the task deals with inputs that could be thousands of frames long. 

\par In order to circumvent these limitations of recurrent models, I cast the structure-aware alignment task as a prediction problem that works on a joint representation of the two inputs. I pose the question:\\ \textit{How can we detect the synchronous subpaths, whether they correspond to a musically meaningful segment or not, from a single joint representation of the two inputs (audio and score OR audio and audio)?}
\par The remainder of the chapter presents an answer to the above question. I propose a novel neural approach for structure-aware alignment that detects the structural differences between the two input representations of a given piece of music (two performances or a performance and a score) via predicting the \blue{\textit{inflection points}} that mark the locations of the transitions between the various synchronous subpaths that overlap between the two inputs. These inflection points are then passed on to an extended DTW framework to yield the fine-grained alignments.
\par I propose a progressively dilated \blue{Convolutional Neural Network} architecture to perform the inflection point detection. This is implemented as a multi-label prediction task, with the model outputting the synchronous subpaths between the performance and score sequences via the prediction of the inflection points. A key advantage of this approach is that it can be employed successfully even in the presence of limited annotated data. The proposed architecture employs progressively dilated convolutions in addition to standard convolutions, with varying amounts of dilation applied at different layers of the network. The primary motivation behind our architecture is that it allows us to effectively capture both short-term and long-term context, 
using much fewer parameters than sequential models such as recurrent neural networks, and without facing the vanishing gradient problem. 

\begin{figure}[H]
\vspace{1cm}
\begin{center}
\includegraphics[width=4in]{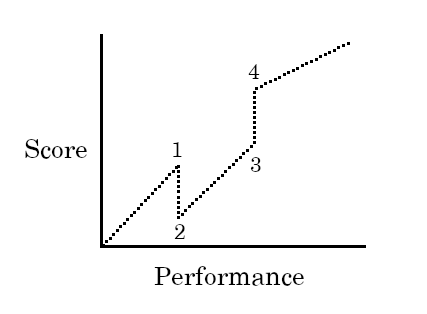}
\caption{An illustration of the inflection points that mark structural changes}
\label{fig:ch4_inflection}
\end{center}
\end{figure}

 \begin{sidewaysfigure}[htbp]
  \centering
  \includegraphics[width=8in]{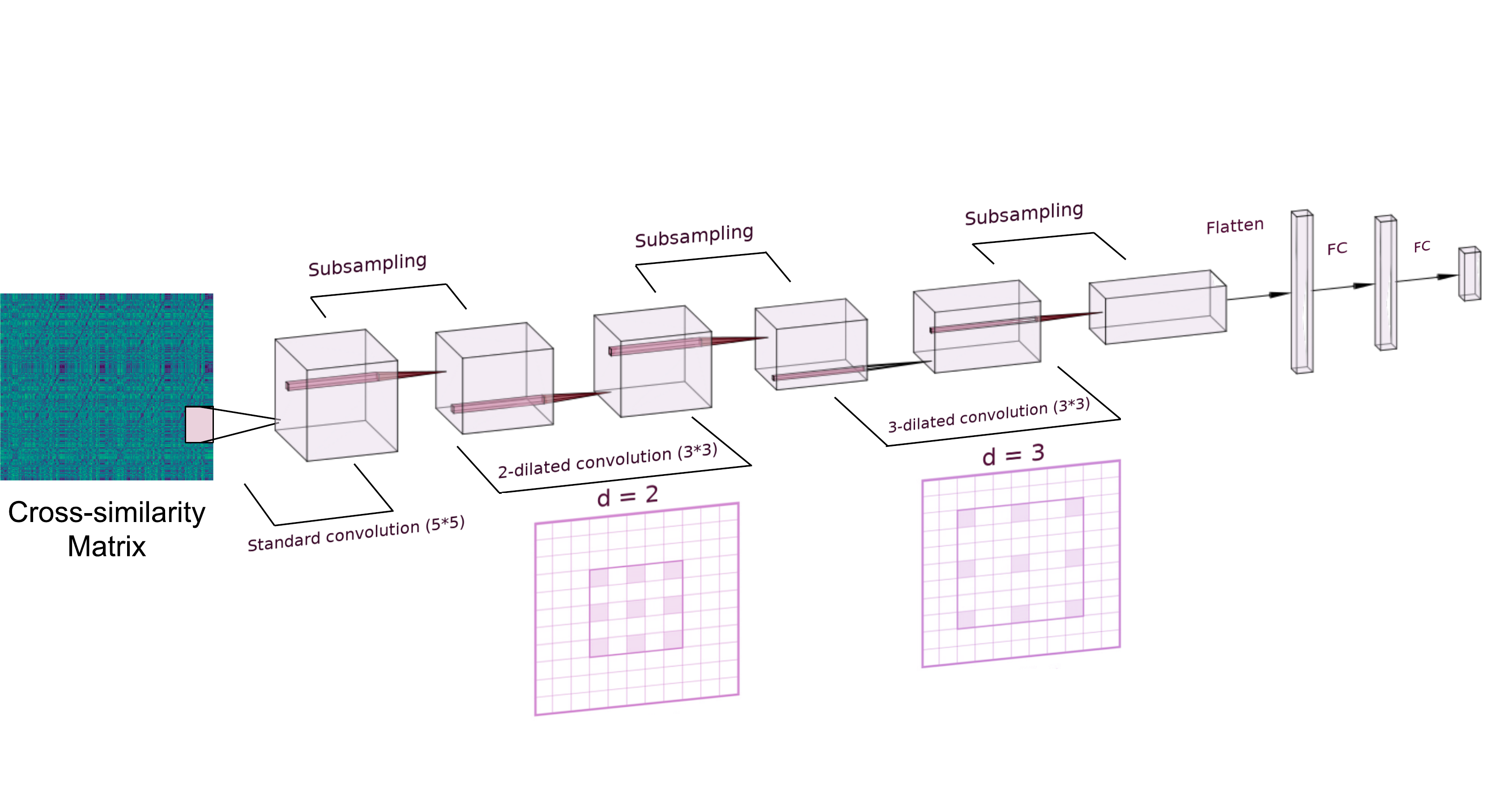}
  \caption{Schematic diagram illustrating the general architecture of our models.\\ \textit{d: Dilation rate, FC: Fully connected layer}}
  \label{fig:dilatedCNN}
\end{sidewaysfigure}
\par The general architecture of the proposed method is illustrated in Figure \ref{fig:dilatedCNN}. Formally, let $X$= $(x_1, x_2,..., x_p)$ and $Y$ = $(y_1, y_2,..., y_q)$ be the feature sequences corresponding to the performance and score respectively. The goal is to obtain the sequence of frame indices $\hat{Y}_p$= $(\hat{y}_1, \hat{y}_2,..., \hat{y}_p)$, denoting the path taken by the performance $X$ through the score $Y$. This is achieved via first predicting the set of inflection points $I = \{(a_1, b_1), (a_2, b_2), .., (a_N, b_N) \mid a_m \in [1, p], b_m \in [1, q]\}$ marking the transition locations between the synchronous subpaths between the performance and the score (or between the two performances), followed by the generation of fine-grained alignments using Dynamic Time Warping. Figure \ref{fig:ch4_inflection} illustrates the inflection points present in the ground-truth alignment of a performance-score pair containing two structural changes. The next subsection presents the method proposed for the prediction of the inflection points.

\subsection{Progressively dilated convolutions for inflection point detection}

\begin{figure}[H]
\vspace{1cm}
\begin{center}
\includegraphics[width=5in]{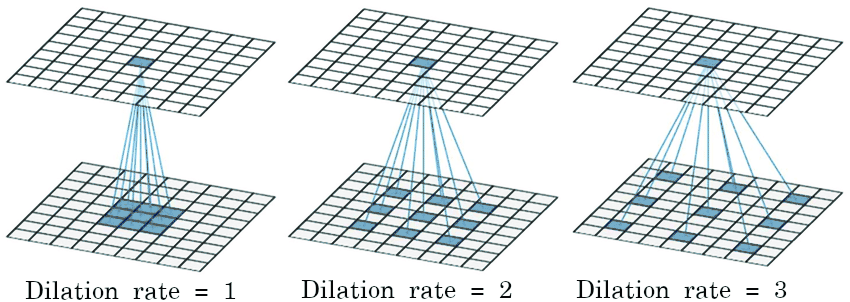}
\vspace{1cm}
\caption{The dilated convolution operation - A 3 $\times$ 3 convolution dilated using different rates}
\label{fig:dilatedConv}
\end{center}
\end{figure}
I propose a progressively dilated CNN architecture to predict the set of inflection points that encode the jump locations that mark the structural differences. The dilated convolution operation was initially proposed by \citet{yu2015multi} for the semantic segmentation task as a means to achieve accurate dense prediction without losing resolution or coverage. Figure \ref{fig:dilatedConv} illustrates the dilated convolution operation. The concept of dilation involves introducing ``holes'' or ``gaps'' in the convolutional kernel, thereby increasing the receptive field while keeping the same number of kernel weights. The dilated convolution of a discrete function $F$ with a discrete filter $f$ on an element $\textbf{p}$ is defined as follows:
\begin{equation}
(F*_df)(\textbf{p}) = \sum\limits_{\textbf{t}}F(\textbf{p}-d\textbf{t})f(\textbf{t})
\end{equation}\label{eq:dilatedConv}
where $d$ is the factor by which the kernel is inflated, referred to as the dilation rate. Note that setting d = 1 renders the dilated convolution operation as the normal convolution operation:
\begin{equation}
(F*f)(\textbf{p}) = \sum\limits_{\textbf{t}}F(\textbf{p}-\textbf{t})f(\textbf{t})
\end{equation}\label{eq:Conv}
\par The proposed architecture combines both standard convolutions and \blue{dilated} convolutions in a strategic manner. The motivation behind using varying amounts of dilation is to incorporate both short-term and long-term context to model structure, 
thereby effectively modeling different types of structural deviations from the score such as extra repeated sections, skipped measures, and arbitrary jumps. The incorporation of multi-scale contextual information has proven to be useful in computer vision as well as natural language processing tasks \citep{lee2017going, agrawal2018no, agrawal2018contextual}.
\par The first layer of our proposed network employs standard convolutions in order to capture low-level relationships essential to understand the immediate context of a region in the similarity matrix. The subsequent layers introduce dilation at different rates in order to substantially expand the receptive field, thereby capturing high-level information better as we move deeper into the network. Inflating the kernel using dilation allows us to incorporate larger context without increasing the number of parameters, which is a desirable property for the neural models that approach the structure-aware alignment task in the presence of limited hand-annotated data. With progressively dilated convolutions, the receptive field is exponentially increased, and for a dilation rate $d$, a kernel of size $m$ effectively works as a kernel of size $m'$  as follows:
\begin{equation} m' = m + (d-1) \times (m-1) 
\end{equation} This facilitates the incorporation of context better than standard convolutions, which can only offer linear growth of the effective receptive field as we move deeper into the network. Put simply, our architecture offers a wider field of view at the same computational cost.

\par Experiments are conducted using varying amounts of dilation to determine the optimal dilation rate to be applied at each layer of the network. The results of the experiments suggest that progressively increasing dilation as we move deeper into the network produces the best results for detecting the structural differences, as will be discovered in Section \ref{sec:results}. In order to test specific improvements using the dilated convolutions, experiments are also conducted using a baseline CNN model trained without any dilation to generate the inflection points. The next section describes the method for the generation of fine-grained alignments after the inflection points have been detected.
\vspace{0.1cm}
 \subsection{Generation of fine-grained alignment}
 \par In order to generate the fine alignments, the inflection points predicted by our dilated convolutional models are employed as potential jump positions to assist a DTW-based alignment algorithm. We implement such an extended DTW framework, inspired by \begin{math}\textit{JumpDTW}\end{math}  \citep{Fremerey2010handling}, to allow for jumps between the synchronous subpaths.  
Let us assume $X$= $(x_1, x_2,..., x_p)$ to be the feature sequence corresponding to the performance and $Y$ = $(y_1, y_2,..., y_q)$ to be the feature sequence corresponding to the score. Furthermore, let ($a_i$, $b_i$) denote the ($x$, $y$) co-ordinates of the $i_{th}$ inflection point, and $N$ denote the total number of inflection points. The odd numbered inflection points correspond to the end of the synchronous subpaths and the even numbered points correspond to the beginning of the subpaths. We modify the classical DTW framework to extend the set of possible predecessor cells for the cell $(a_i, b_i)$ for all \begin{math} i \in \{2, 4, 6, .., N\} \end{math}, as follows:
\vspace{-0.2cm}
\begin{equation}
D(m, n)  = e(m, n) + min\begin{cases}
D(m, n-1) \\ D(m-1, n) \\  D(m-1,  n-1) \\
D(a_{i-1}, b_{i-1}) \hspace{0.1cm}  \iff  (m, n) \in [(a_i - 5, b_i - 5), (a_i + 5, b_i + 5)]\\ 
\hspace{2.2cm} i \in \{2, 4, ..., N\}
\end{cases}
\end{equation}
where $e(m, n)$ is the Euclidean distance between points $x_m$ and $y_n$, and $D(m, n)$ is the total cost to be minimised for the path until the cell $(m, n)$. The path which yields the minimum value for $D(p, q)$ is taken to be the optimal alignment path between the performance and score sequences.

\section{Experimental Setup}\label{sec:setup}
The detection of synchronous subpaths is implemented as a \blue{multivariate regression} task using progressively dilated CNNs, with each output label predicting an inflection point that encodes a deviation in the performance from the score.  
A key challenge in modelling structural changes for alignment is the lack of \blue{hand-annotated} data, marked for repeats and jumps accurately at the frame level. This is also one of the caveats of \begin{math}\textit{JumpDTW}\end{math}, which is reliant on the accuracy of the OMR system to detect jump and repeat directives in the absense of manually annotated boundaries.
To overcome the lack of annotated training data, we generated synthetic samples containing jumps and repeats using the audio from the MSMD \citep{dorfer2018learning} dataset. 

\begin{table}[ht]
\vspace{0.5cm}
   \caption{Datasets used for our experiments on structure-aware alignment}
   \centering
   \begin{tabular}{ c c c c} \toprule
      \textbf{Name} & \textbf{Instrument} & \textbf{Recordings}  & \textbf{Stage} \\ \midrule
      MSMD \small{\citep{emiya2009multipitch}} &  Piano  &  495 &  Train\\ \midrule
      Tido  & Piano  & 150  & Train and Test \\ \midrule
      Mazurka \small{\citep{sapp2007comparative}} & Piano & 239 & Test \\ \midrule
      JAAH \small{\citep{eremenko2018audio}} & Various & 113 & Train and Test \\ \midrule
      \bottomrule
\end{tabular}
\label{tab:datasets}
\end{table} 
\par The datasets employed for the experimentation in this chapter are listed in Table \ref{tab:datasets}. The training dataset for the experiments on performance-score synchronisation contains 2625 pairs of audio recordings, corresponding to the MIDI score and performance respectively. 2475 of these are obtained from the MSMD dataset, which originally contains 497 pieces of performance-score pairs (generated by LilyPond) with fine-grained alignment annotations. Two of these recordings were found to have insufficient annotations, and were discarded. Each of the remaining 495 pieces was utilised to generate four structurally different performance-score pairs, such that these pairs contain varying numbers of structural differences, generated randomly by splitting and joining the performance audios at various locations. This was done using pydub (\textit{https://pypi.org/project/pydub/}). In addition to these four pairs, each pair was also employed once as it is, i.e. without any repetition. \blue{It must be noted that while the MSMD dataset contains only one Chopin Mazurka (Op.~6 No.~1), there is no direct replication in the test set, since the MSMD dataset contains synthetically generated audio.} In addition to synthetic data, a small amount of \blue{hand-annotated} data from a private dataset procured from Tido UK Ltd. is employed, referred to as the Tido dataset further in the chapter. The training set taken from the Tido dataset comprises audio pairs for 150 pieces, 80 of which contain structural differences. Experiments are also conducted for the performance-performance synchronisation task, and these are presented after the primary experimentation for performance-score synchronisation. 
\vspace{0.2cm}
\par The networks operate on the cross-similarity matrix between the score and performance and predict the ($x$, $y$) co-ordinates of the inflection points as the outputs.
The cross-similarity matrix for each performance-score pair is computed using the Euclidean distance between the chromagrams corresponding to the score and performance respectively. We employ librosa \citep{mcfee2015librosa} to compute the chromagrams as well as the cross-similarity. It must be noted that any other similarity metric could also be employed, depending on the test setting, as will be shown in the experiment described in Section \ref{sec:learnt}.
\par The proposed dilated CNN models consist of three convolutional and subsampling layers, with standard convolutions at the first layer and dilated convolutions with varying dilation rates at the second and third layer respectively. The outputs of each layer are passed through a Rectified Linear Unit to add non-linearity, followed by batch normalisation before being passed as inputs to the next layer. The output of the third convolutional and subsampling layer is sent through a flatten layer, following which it is passed through two fully connected layers of size 4096 and 1024 respectively to predict the $(x, y)$ co-ordinates of the inflection points. The output of the final layer is a one-dimensional tensor of size 64, signifying that the model can predict up to 32 inflection points, with their $(x, y)$ co-ordinates in chronological order. \blue{These suffice for our test data, since the number of inflection points is less than 32 in all cases. To deal with the varying number of inflection points pertinent to different training examples, the target vector is padded using \textit{post-padding} with a value of 4096 passed as the \textit{value parameter}. After a few epochs of training, the network eventually learns to predict the correct number of inflection points, with the remaining ones predicted as 4096 and therefore not contributing to the loss. It must be noted that masking could also be employed (to inform the model that the remaining part of the vector is actually padded and should be ignored), however, I do not employ it since the number of inflection points is unknown at test time. Rather, the network is trained to correctly predict the number of inflection points along with their positions.}
\par \blue{The architecture could also be modified to allow for more inflection points if faced with a scenario containing very long performances  (such as operas) with a high number of jumps or skips. In such cases, the modification would primarily involve adjusting the input-output dimensionality. It must be noted that the computational overhead in such cases would primarily arise from the higher input dimensionality, however it could be mitigated (at the cost of losing granularity) by limiting the input dimensionality and modifying the DTW formulation to consider the model predictions as neighbourhoods containing the jump locations.} 
\par The output of the final layer is compared with the ground truth using the \blue{mean squared error} loss, since we want to capture the distance of the predicted inflection points from the ground truth inflection points in time.  We employ a dropout of 0.5 for the fully connected layers to avoid overfitting. Our batch size is 64 and the models are trained for 40 epochs, with early stopping, \blue{i.e. stopping the training when the model performance on the validation set does not improve any further}. The dilated CNN models are denoted as  \emph{DCNN}${}_{m+n}$, where $m$ and $n$ correspond to the dilation rates employed at the second and third layer respectively.

\par We test the performance of our models on three different datasets, all containing recordings of real performances. The results for the performance-score synchronisation task are demonstrated on the publicly available Mazurka dataset \citep{sapp2007comparative}
In order to analyze specific improvements for structurally different pieces, results 
are also demonstrated on subsets of the Tido dataset \emph{with} and \emph{without} structural differences. These are detailed in Section \ref{sec:ablationStructure}.
\par The results obtained by the dilated CNN models are compared with \begin{math} \textit{JumpDTW}\end{math} \citep{Fremerey2010handling}, \begin{math}\textit{NWTW}\end{math} \citep{grachten2013automatic}, \begin{math}\textit{MATCH}\end{math} \citep{dixon2005line} and a vanilla CNN model without dilation     \emph{CNN}${}_{1+1}$. The number of parameters of the \emph{DCNN}${}_{m+n}$ networks is comparable with that of the baseline \emph{CNN}${}_{1+1}$ network.
For comparison with \begin{math}\textit{JumpDTW}\end{math}, the SharpEye OMR engine is employed to extract frame predictions for block boundaries from the  sheet images \citep{Fremerey2010handling}. These are then passed on to our implementation of \begin{math}\textit{JumpDTW}\end{math} to generate the alignment path. 
Similarly, the proposed method is also compared with the \begin{math}\textit{NWTW}\end{math} method, with the optimal gap penalty parameter $\gamma$  \citep{grachten2013automatic} being estimated on the training datasets presented in Table \ref{tab:datasets}.
\par The next four subsections present experimental studies that analyse the performance of the proposed method in various application scenarios as well as conduct ablative analyses that delineate the improvements obtained using various components of the proposed method. \blue{As in Chapter 3, significance testing is also carried out using the Diebold-Mariano test \citep{harvey1997testing} in order to examine the statistical significance of the results. To this end, I conduct pairwise comparisons of all model predictions with the predictions of the best performing model for each error margin and for each experimental setup described in the upcoming subsections. All the dilated CNN models were trained on a GeForce RTX 2080 Ti GPU card containing 11 GB GPU memory, with the NVIDIA driver version 418.87.00 and CUDA version 10.1. The training for all models was completed in less than 10 hours when trained on a single GPU core.}
\section{Results and Discussion}\label{sec:results}
\subsection{Results on performance-score synchronisation\\
Study 1: Overall accuracy on the Mazurka dataset and the effect of different dilation rates}

\begin{table*}[ht]
\vspace{1cm}
\centering
  \begin{adjustbox}{max width=\textwidth}
\begin{tabular}{ccccc}
\toprule
\hline 
\multirow{2}{*}{\textbf{Model}} & 
\multicolumn{4}{c}{\textit{On Mazurka dataset}}  

\tabularnewline
  & \textbf{$<$25ms}& \textbf{$<$50ms} & \textbf{$<$100ms} & \textbf{$<$200ms} 
  \\
\midrule 
 \begin{math}\textit{MATCH}\end{math} & 64.8* & 72.1* & 77.6* & 83.7*   \\
\midrule
 \begin{math}\textit{JumpDTW}\end{math}  & 65.8* & 75.2* & 79.8* & 85.7* \\
\midrule 
 \begin{math}\textit{NWTW}\end{math}  & 67.6* & 75.5* & 80.1* & 86.2*  \\
\midrule 
     \emph{CNN}${}_{1+1}$ & 68.2* & 75.7* & 80.5* & 87.1*   \\
\midrule 
      \emph{DCNN}${}_{2+2}$ & \textbf{69.9} & 76.4* & 81.6* & 88.9*  \\
\midrule 
    \emph{DCNN}${}_{2+3}$ & 69.7 & \textbf{77.2} & \textbf{82.4} & \textbf{89.8} \\
\midrule 
     \emph{DCNN}${}_{3+3}$ & 69.2* & 76.1* & 81.2* & 88.7* \\
  \midrule
\bottomrule
\end{tabular}
\end{adjustbox}
\vspace{0.3cm}
\caption{Alignment accuracy in \% on  the Mazurka dataset.\\ \emph{DCNN}${}_{m+n}$: Dilated CNN model with dilation rates of $m$ and $n$ at the second and third layer respectively.\\
\blue{$*$: significant differences from \emph{DCNN}${}_{2+3}$, $p < 0.05$ }}

\vspace{1cm}
\label{resultsMazurka}
\end{table*}
The results obtained by our models in terms of overall accuracy for performance-score synchronisation on the Mazurka dataset are given in Table \ref{resultsMazurka}. We report alignment accuracy in \%, 
where each value denotes the percentage of beats aligned correctly within the corresponding time durations of 25, 50, 100 and 200 ms respectively. In order to determine the optimal dilation rates, experimentation is conducted using four different dilation settings; two of which contain equal levels of dilation applied at the second and third layer, one comprises increasing dilation applied with network depth, and one comprises a baseline CNN, i.e. the models are trained without any dilation.

\par The experimentation with different dilation rates reveals that progressively increasing dilation as we move deeper (\emph{DCNN}${}_{2+3}$) yields better results than models trained using equal amounts of dilation (\emph{DCNN}${}_{2+2}$, \emph{DCNN}${}_{3+3}$). \blue{This could be attributed to better multi-scale contextual incorporation, with local context being captured earlier in the network and global context being captured further down the network.} Models trained with dilation at the first layer and those trained using dilation rates of 4 and higher did not yield improvement over the vanilla CNN model \emph{CNN}${}_{1+1}$ and hence are not reported. This suggests that progressively increasing dilation helps the model learn higher level features better further down the network. 
Overall accuracy on the Mazurka dataset suggests that the dilated CNN models perform better than \begin{math}\textit{MATCH}\end{math} by 4-6\% as well as the \begin{math}\textit{JumpDTW}\end{math} and \begin{math}\textit{NWTW}\end{math} frameworks by 1-4\% (Table \ref{resultsMazurka}, columns 1-4) for all error margins. \blue{The accuracy of inflection point detection using the dilated CNN models is also computed to analyse how it affects the alignment accuracy. The accuracy results for the inflection point detection are 88.9\%, 92.8\%, 95.4\% and 91.5\% for the \emph{CNN}${}_{1+1}$, \emph{DCNN}${}_{2+2}$, \emph{DCNN}${}_{2+3}$, and \emph{DCNN}${}_{3+3}$ models respectively. This suggests that the inflection point detection accuracy is positively correlated with the alignment accuracy, with the progressively dilated model (\emph{DCNN}${}_{2+3}$) yielding the best performance.}

\subsection{Study 2 - Model performance when trained without hand-annotated data}

\begin{table*}[ht]
  \vspace{1cm}

   \centering
  \begin{adjustbox}{max width=\textwidth}
\begin{tabular}{ccccc}
\toprule
\hline 
\multirow{2}{*}{\textbf{Model}} & 
\multicolumn{4}{c}{\textit{On Mazurka dataset}}  

\tabularnewline
  & \textbf{$<$25ms}& \textbf{$<$50ms} & \textbf{$<$100ms} & \textbf{$<$200ms} 
  \\
\midrule 
 \begin{math}\textit{MATCH}\end{math} & 64.8* & 72.1* & 77.6* & 83.7*   \\
\midrule
 \begin{math}\textit{JumpDTW}\end{math}  & 65.8* & 75.2* & 79.8* & 85.7* \\
\midrule 
 \begin{math}\textit{NWTW}\end{math}  & 67.6* & 75.5* & 80.1* & 86.2*  \\
\midrule 
     \emph{CNNsyn}${}_{1+1}$ & 66.8* & 73.4* & 78.2* & 85.9*   \\
\midrule 
      \emph{DCNNsyn}${}_{2+2}$ & 68.1* & \textbf{77.1} & 80.6* & 87.4*  \\
\midrule 
    \emph{DCNNsyn}${}_{2+3}$ & \textbf{68.9} & 76.4 & \textbf{81.2} & \textbf{88.6} \\
\midrule 
     \emph{DCNNsyn}${}_{3+3}$ & 67.8*  & 75.1* & 80.4* & 87.1* \\
  \midrule
\bottomrule
\end{tabular}
\end{adjustbox}
\vspace{0.1cm}
\caption{Alignment accuracy in \% on the the Mazurka dataset.\\ \emph{DCNNsyn}${}_{m+n}$: Dilated CNN model with dilation rates of $m$ and $n$ at the second and third layer respectively, trained only on synthetic data.\\ $*$: significant differences from \emph{DCNNsyn}${}_{2+3}$, $p < 0.05$ }

\label{resultsSynthetic}
\end{table*}
 A key advantage of the proposed method is the ability to perform well in the presence of limited hand-annotated data. It would be desirable to examine the ability of the model to generate alignments without \textit{any} hand-annotated data. To test the model performance on this front, I also conduct experimentation wherein the models are trained using different dilation rates as described earlier. However, rather than employing the entire training set, these models are trained exclusively on the subset of the training set containing only synthetic data. These models are notated as (\emph{DCNNsyn}${}_{i+j}$), with $i$ and $j$ denoting the dilation rates employed at the second and third layer of the network respectively. The results obtained by these models are given in Table \ref{resultsSynthetic}. 
\par The performance of the models follow a similar trend to the previous experiments in terms of dilation. The \emph{DCNNsyn} model trained with progressively increasing dilation yields the best performance among the dilated CNN models. When compared with previous structure-aware approaches, the progressively dilated CNN model trained exclusively on synthetic data (\emph{DCNNsyn}${}_{2+3}$) yields better alignment accuracy than \emph{NWTW} as well as \begin{math}\textit{JumpDTW}\end{math}, which requires manually labelled block boundaries to handle repeats and jumps \citep{Fremerey2010handling}. This emphasises the applicability of our method in real-world scenarios, where hand-annotated data marked with structure annotations is not readily available at test time. Additionally, the dilated CNN models noticeably outperform all methods when a limited amount of real data is added to the synthetic data during training (Table \ref{resultsMazurka}, rows 5-7).

\subsection{Study 3 - Specific improvements on pieces with and without structural differences}\label{sec:ablationStructure}

\begin{table*}[ht]
\vspace{1cm}
\centering
  \begin{adjustbox}{max width=\textwidth}
\begin{tabular}{ccccc}
\toprule
\hline 
\multirow{2}{*}{\textbf{Model}} &  \multicolumn{4}{c}{\textit{With structural differences (Tido)}}
\tabularnewline
  & \textbf{$<$25ms}& \textbf{$<$50ms} & \textbf{$<$100ms} & \textbf{$<$200ms} 
  \\
\midrule 
 \begin{math}\textit{MATCH}\end{math} & 61.5* & 70.4* & 74.6* & 80.7* \\
\midrule
 \begin{math}\textit{JumpDTW}\end{math} & 69.1* & 77.2* & 82.0* & 88.4*   \\
\midrule 
 \begin{math}\textit{NWTW}\end{math}  & 68.6* & 75.8* & 80.7* & 87.5*   \\
\midrule 
     \emph{CNN}${}_{1+1}$  & 70.4* & 78.3* & 83.4* & 90.1* \\
\midrule 
      \emph{DCNN}${}_{2+2}$  & 72.7* & 80.1*  & 84.5* & 91.4*  \\
\midrule 
    \emph{DCNN}${}_{2+3}$  & \textbf{73.9} & \textbf{81.3} & \textbf{85.6} & \textbf{92.8}\\
\midrule 
     \emph{DCNN}${}_{3+3}$   & 72.3* & 79.5* & 84.2* & 90.4* \\
  \midrule
  \emph{DCNNsyn}${}_{2+3}$  & 70.5* & 78.6* & 83.8* & 90.5*  \\
\midrule 
\bottomrule
\end{tabular}
\end{adjustbox}
\vspace{0.3cm}
\caption{Alignment accuracy in \% on the subset of the Tido dataset containing structural differences.\\ \emph{DCNN}${}_{m+n}$: Dilated CNN model with dilation rates of $m$ and $n$ at the second and third layer respectively.\\
\blue{$*$: significant differences from \emph{DCNN}${}_{2+3}$, $p < 0.05$ }}

\vspace{1cm}
\label{resultsTidoWith}
\end{table*}

\begin{table*}[ht]
  \vspace{1cm}

   \centering
  \begin{adjustbox}{max width=\textwidth}
\begin{tabular}{ccccccccccccc}
\toprule
\hline 
\multirow{2}{*}{\textbf{Model}} &  \multicolumn{4}{c}{\textit{Without structural differences (Tido)}}
\tabularnewline
  & \textbf{$<$25ms}& \textbf{$<$50ms} & \textbf{$<$100ms} & \textbf{$<$200ms}
  \\
\midrule 
 \begin{math}\textit{MATCH}\end{math}& 70.2* & 78.4* & 84.7* & 90.3*  \\
\midrule
 \begin{math}\textit{JumpDTW}\end{math}  &  68.7* & 77.5* & 82.1* & 88.9*  \\
\midrule 
 \begin{math}\textit{NWTW}\end{math}   & 68.4* & 77.1* & 82.8* & 89.4*  \\
\midrule 
     \emph{CNN}${}_{1+1}$  & 69.3* & 78.0* & 84.1* & 89.3* \\
\midrule 
      \emph{DCNN}${}_{2+2}$ & \textbf{71.4} & 79.5* & 85.3* & 90.5* \\
\midrule 
    \emph{DCNN}${}_{2+3}$  & 71.0 & \textbf{80.3} & \textbf{85.8} & \textbf{91.8} \\
\midrule 
     \emph{DCNN}${}_{3+3}$ & 70.6* & 78.8* & 84.9* & 91.2*\\
  \midrule
  \emph{DCNNsyn}${}_{2+3}$ & 69.2* & 78.3* & 84.6* & 89.8*  \\
\midrule 
\bottomrule
\end{tabular}
\end{adjustbox}
\vspace{0.1cm}
\caption{Alignment accuracy in \% on the subset of the Tido dataset without structural differences.\\  \emph{DCNN}${}_{m+n}$: Dilated CNN model with dilation rates of $m$ and $n$ at the second and third layer respectively.\\
\blue{$*$: significant differences from \emph{DCNN}${}_{2+3}$, $p < 0.05$ }}

\label{resultsTidoWithout}
\end{table*}
The previous subsection demonstrates the performance of the dilated CNN models in terms of overall accuracy on the Mazurka dataset, regardless of the structural agreement between the individual performances and the corresponding scores. This subsection analyses the performance of the proposed method (compared with previous approaches) in terms of specific improvements for \textit{structure-aware alignment} and \textit{monotonic alignment}, i.e. the alignment of pieces with complete structural agreement. 
\par To this end, the models are tested on two different subsets of the Tido dataset, such that the first subset comprises performance-score pairs that contain structural differences, and the second subset comprises pairs without any structural differences between the performances and the corresponding scores. 
Both the subsets contain 75 pieces each. The results obtained by the models for these subsets are given in Tables \ref{resultsTidoWith} and \ref{resultsTidoWithout} respectively. 

The tables demonstrate that the dilated CNN models yield an increase of 2-5\% in alignment accuracy over \begin{math}\textit{JumpDTW}\end{math} and \begin{math}\textit{NWTW}\end{math} on the test subset containing structural differences and an increase of 1-3\% on the test subset not containing any structural differences. Compared with \begin{math}\textit{MATCH}\end{math}, our models show an increase of 9-10\% on the subset with structural differences, and an increase of 1-2\% on the subset without structural differences.
\par These results suggest that the proposed method yields noticeable improvements over previous methods for structure-aware alignment. At the same time, the method does not hinder the capacity of the models to align pairs with complete structural agreement, making the method suitable to be adopted in various test conditions, regardless of the structural aspects.

\subsection{Study 4 - Effect of learnt similarity on model performance}\label{sec:learnt}
\begin{table*}[ht]
\vspace{1cm}
\centering
  \begin{adjustbox}{max width=\textwidth}
\begin{tabular}{ccccc}
\toprule
\hline 
\multirow{2}{*}{\textbf{Model}} & 
\multicolumn{4}{c}{\textit{On Mazurka dataset}}  

\tabularnewline
  & \textbf{$<$25ms}& \textbf{$<$50ms} & \textbf{$<$100ms} & \textbf{$<$200ms} 
  \\
\midrule 
 \begin{math}\textit{MATCH}\end{math} & 64.8* & 72.1* & 77.6* & 83.7*   \\
\midrule
 \begin{math}\textit{JumpDTW}\end{math}  & 65.8* & 75.2* & 79.8* & 85.7* \\
\midrule 
 \begin{math}\textit{NWTW}\end{math}  & 67.6* & 75.5* & 80.1* & 86.2*  \\
\midrule 
     \emph{CNNsiam}${}_{1+1}$ & 68.2* & 75.7* & 80.5* & 87.1*  \\
\midrule 
      \emph{DCNNsiam}${}_{2+2}$ & 70.5* & 77.8* & 82.2* & 89.7*  \\
\midrule 
    \emph{DCNNsiam}${}_{2+3}$ & \textbf{71.3} & \textbf{78.4} & \textbf{83.5} & \textbf{91.2} \\
\midrule 
     \emph{DCNNsiam}${}_{3+3}$ & 70.2* & 77.5* & 82.4* & 89.5* \\
  \midrule
\bottomrule
\end{tabular}
\end{adjustbox}
\vspace{0.3cm}
\caption{Alignment accuracy in \% on  the Mazurka dataset.\\ \emph{DCNNsiam}${}_{m+n}$: Dilated CNN model trained with learnt similarity proposed in Chapter 3, with dilation rates of $m$ and $n$ at the second and third layer respectively.\\
\blue{$*$: significant differences from \emph{DCNNsiam}${}_{2+3}$, $p < 0.05$ }}

\vspace{1cm}
\label{resultsMazurkaLearnt}
\end{table*}
While the primary experimentation is carried out using the chromagram representation since it is a commonly used and readily available representation, a particularly relevant experiment on this front would be analysing the effect of employing the learnt similarity proposed in Chapter 3 on the performance of the proposed structure-aware alignment models. To this end, experiments are conducted on the same train and test splits, but using the cross similarity matrix generated using the method proposed in Chapter 3. The results of these experiments are provided in Table \ref{resultsMazurkaLearnt}.

\par The results suggest that employing frame similarity that is learnt using the method proposed in Chapter 3 boosts the model performance even further. I recommend the usage of learnt frame similarity in the presence of non-standard acoustic conditions, and especially if the application domain meets the data needs presented in Section 3.7.4.

\subsection{Qualitative analysis}

\begin{figure*}%
    \centering
       \subfloat[Input]
  {{\includegraphics[width=5cm, height=5cm]{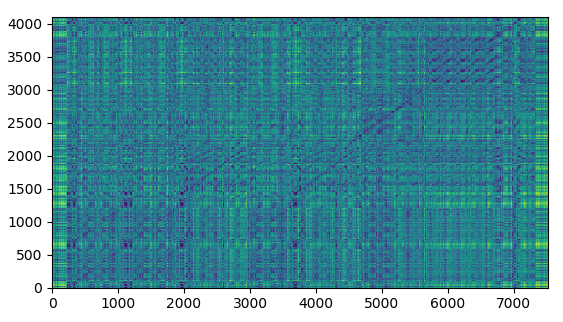} }}%
    \qquad
        \subfloat[\begin{math}\textit{JumpDTW}\end{math}]
    {{\includegraphics[width=5cm, height=5cm]{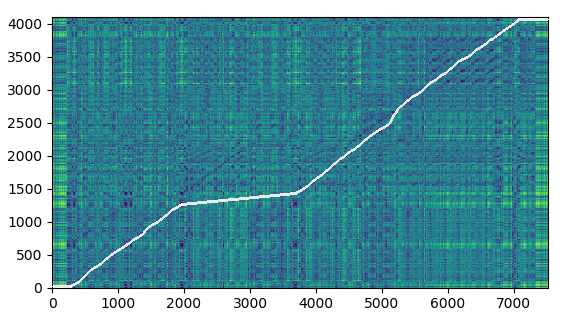} }}%
    \\
        \subfloat[\begin{math}\textit{NWTW}\end{math}]
{{\includegraphics[width=5cm, height=5cm]{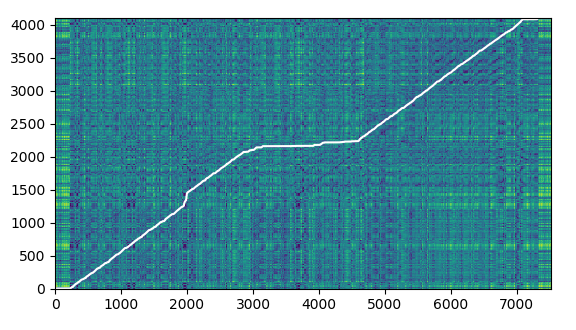} }}%
    \qquad
    \subfloat[\emph{DCNN}${}_{2+3}$]
    {{\includegraphics[width=5cm, height=5cm]{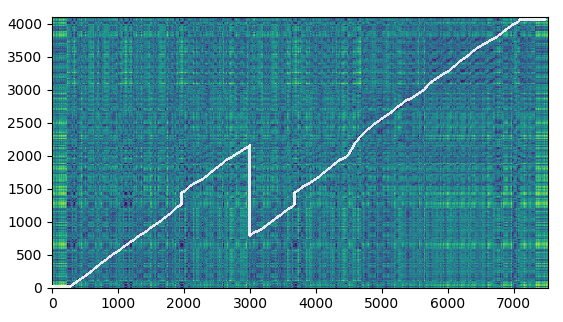} }}%
    \qquad
        \subfloat[Ground Truth]
    {{\includegraphics[width=5cm, height=5cm]{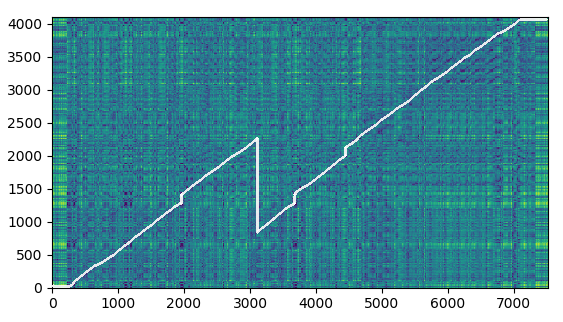} }}%
    \\
        \caption{Comparison of our alignment path with standard method predictions for a piece containing a backward jump and a few (short) forward jumps. \\Input: Cross-similarity matrix between score and performance.\\
    X-axis: Frame index (performance); Y-axis: Frame index (score).}%
    \label{fig:comparison1}%
\end{figure*}

\begin{figure*}%
    \centering
           \subfloat[Input]
{{\includegraphics[width=5cm, height=5cm]{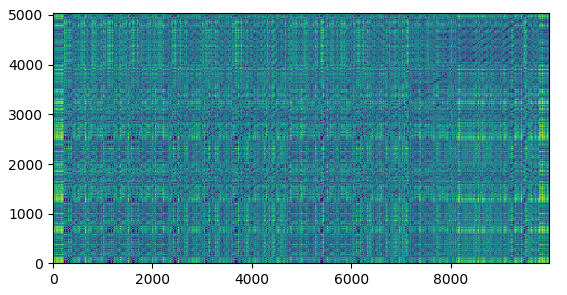} }}%
    \qquad
        \subfloat[\begin{math}\textit{JumpDTW}\end{math}]
    {{\includegraphics[width=5cm, height=5cm]{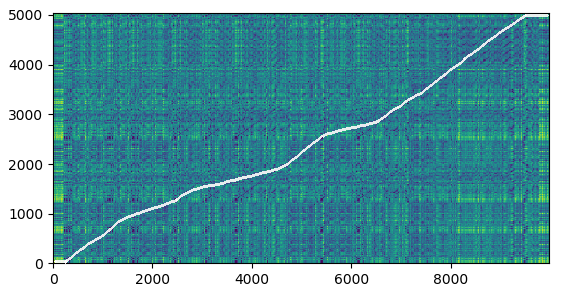} }}%
    \\
        \subfloat[\begin{math}\textit{NWTW}\end{math}]
    {{\includegraphics[width=5cm, height=5cm]{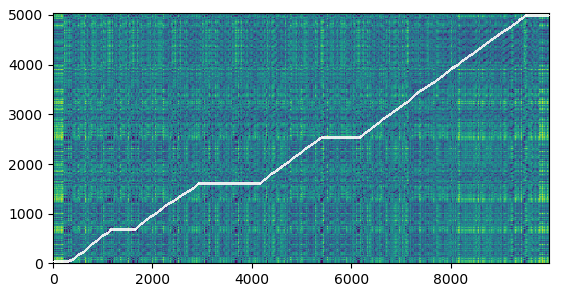} }}%
    \qquad
    \subfloat[\emph{DCNN}${}_{2+3}$]
    {{\includegraphics[width=5cm, height=5cm]{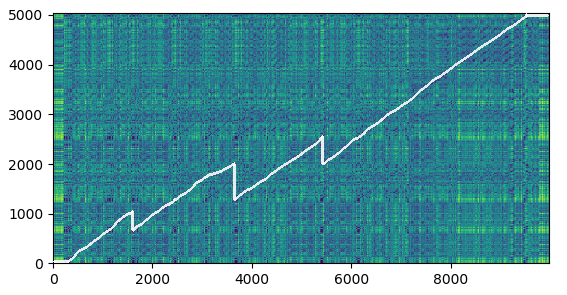} }}%
    \qquad
        \subfloat[Ground Truth]
    {{\includegraphics[width=5cm, height=5cm]{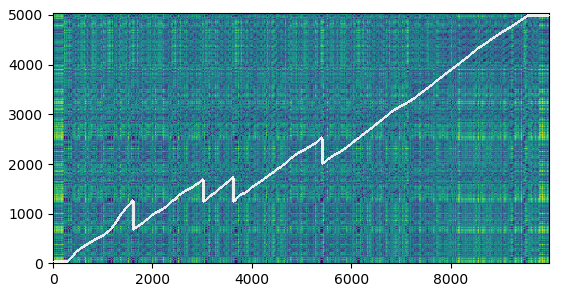} }}%
    \\
        \caption{Comparison of our alignment path with standard method predictions for a piece containing multiple (short) backward jumps. \\Input: Cross-similarity matrix between score and performance.\\
    X-axis: Frame index (performance); Y-axis: Frame index (score).}%
    \label{fig:comparison2}%
\end{figure*}

\begin{figure*}%
    \centering
    
    \subfloat[Input]
    {{\includegraphics[width=5cm, height=5cm]{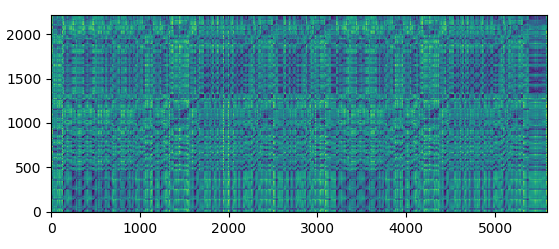} }}%
    \qquad
    \subfloat[\begin{math}\textit{JumpDTW}\end{math}]
    {{\includegraphics[width=5cm, height=5cm]{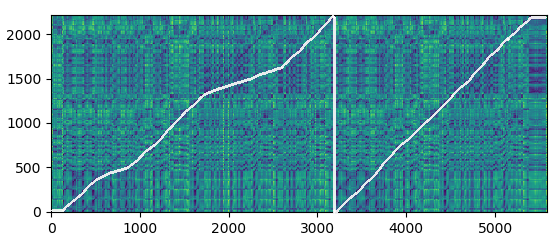} }}%
    \\
    \subfloat[\begin{math}\textit{NWTW}\end{math}]
    {{\includegraphics[width=5cm, height=5cm]{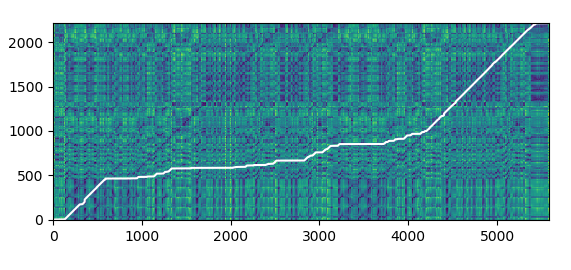} }}%
    \qquad
    \subfloat[\emph{DCNN}${}_{2+3}$]
    {{\includegraphics[width=5cm, height=5cm]{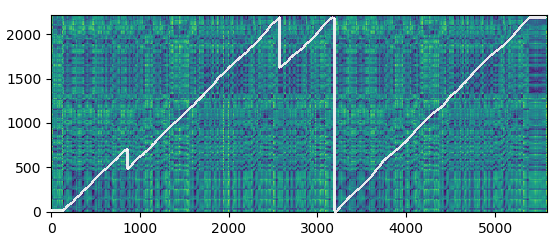} }}%
    \qquad
    \subfloat[Ground Truth]
    {{\includegraphics[width=5cm, height=5cm]{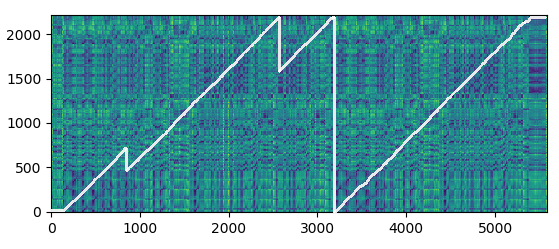} }}%
    \\
        \caption{Comparison of our alignment path with standard methods for a piece with backward jumps and a long repeated section. \\Input: Cross-similarity matrix between score and performance.\\
    X-axis: Frame index (performance); Y-axis: Frame index (score).}%
    \label{fig:comparison3}%
\end{figure*}

In addition to the quantitative analysis described in the previous subsections, a qualitative analysis of the model predictions is performed to enable visualisation of the model performance and thereby aid better comparison with previously proposed approaches. We examined the alignment paths generated by the comparative methods for various performance-score pairs to facilitate qualitative understanding of our results. Figures \ref{fig:comparison1}, \ref{fig:comparison2} and \ref{fig:comparison3} provide three such examples, depicting the alignment paths generated by the progressively dilated CNN models, compared with those generated by previous structure-aware approaches, for pieces containing varied types of structural differences.

\vspace{0.1cm}
\par The manual inspection of the alignment paths presented in these figures confirms that long-term context is better captured by the progressively dilated CNNs than other models, and these are able to detect larger deviations in addition to short ones.  The alignment plots illustrate that \begin{math}\textit{JumpDTW}\end{math} is unable to handle deviations that are not foreseeable from the score, due to its provision of jumps at only the specific block boundaries available beforehand. The plots also indicate that \begin{math}\textit{NWTW}\end{math} is unable to jump back, and therefore align repeated segments. This can be attributed to its waiting mechanism that prevents the handling of backward jumps.
\par The dilated CNN model \emph{DCNN}${}_{2+3}$ is able to capture such jumps that are missed by \begin{math}NWTW\end{math} and \begin{math}\textit{JumpDTW}\end{math}, whether these are induced deliberately by extra repetitions or inadvertently by mistakes during the performance. \emph{DCNN}${}_{2+3}$ is also able to capture intra-block jumps as well as deviations that are not foreseeable from the score (whether intra-block or inter-block), which are missed by \begin{math}\textit{JumpDTW}\end{math} (Figures \ref{fig:comparison1} - \ref{fig:comparison3}). \emph{DCNN}${}_{2+3}$ is also able to capture forward jumps more effectively than  \begin{math}\textit{JumpDTW}\end{math} and \begin{math}NWTW\end{math} (Figure \ref{fig:comparison1}). 
\par Figure \ref{fig:comparison2} demonstrates an example where \emph{DCNN}${}_{2+3}$ misses 1 out of 4 consecutively occuring short backward jumps. Further manual inspection of more alignment plots confirmed that \emph{DCNN}${}_{2+3}$ sometimes misses a jump if it is surrounded by multiple deviations within a short time span, suggesting that this is a challenging application area for the \emph{DCNN} models. We speculate that this is due to the larger receptive fields of the dilated convolutions, which, while capturing greater context, are sometimes unable to capture multiple inflection points within a small context. It must however be noted that such multiple deviations are generally quite rare in the performance scenario, and only occur frequently in the practice scenario. A specific architecture with reduced dilation could benefit such a scenario if the test setting is known beforehand. 
\par Overall, the alignment plots predicted by the dilated CNN model \emph{DCNN}${}_{2+3}$ capture structural differences better than both \begin{math}\textit{JumpDTW}\end{math} and \begin{math}\textit{NWTW}\end{math} and are thereby the closest to the ground truth alignment plots among these structure-aware methods.
\section{Conclusion and further developments}\label{sec:conc}
This chapter presented a novel method for structure-aware alignment applicable to performance-score synchronisation 

The quantitative and qualitative analysis of the performance obtained by the proposed method on various datasets and the comparisons with previous approaches suggest the following:
\begin{itemize}
    \item A convolutional architecture applied to the cross-similarity matrix between the input representations is a promising approach to structure-aware alignment\blue{.}
    \item Combining standard convolutions with dilated convolutions is an effective method to detect structural differences and outperforms convolutional models trained without any dilation\blue{.}
    \item Progressively increasing dilation with network depth yields better results than standard convolutions or consistently dilated convolutions\blue{.}
    \item A dilation rate greater than three hinders results and performs worse than a baseline CNN trained without any dilation\blue{.}
    \item The proposed method successfully captures various kinds of structural differences, regardless of their source and type\blue{.}
    \item The proposed method outperforms previously proposed structure-aware methods without requiring large hand-annotated data, and noticeably outperforms these methods given a limited amount of annotated data.
    \item The learnt frame similarity proposed in Chapter 3 helps improve the performance of the dilated CNN models even further\blue{.}
\end{itemize}
\par The method proposed in this chapter offers a number of advantages. While the primary experiments use chroma-based features for score-performance audio pairs, the proposed method could also be used with raw or scanned images of sheet music using learnt features, for instance, using multimodal embeddings trained on audio and sheet image snippets, or using the learnt frame similarity proposed in Chapter 3. Another major advantage of the proposed method is that it does not require manually labelled block boundaries, and can effectively deal with deviations from the structure given in the score, whether deliberate or inadvertent, in both the forward and backward directions.
\par The method presented in this chapter could be extended in multiple ways. Experiments could be carried out using parallel dilation and merging the features learnt using these kernels at each layer. 
A quantitative analysis of model performance for specific duration of repeats and jumps could be performed, and dynamically selecting dilation depending upon the scenario could also be explored. Additionally, the generation of synthetic data using \blue{deep generative models like generative adversarial networks \citep{goodfellow2014generative} or variational autoencoders \citep{kingma2014stochastic}} could also prove to be a promising exploration.

\chapter{Towards End-to-End Neural Synchronisation}\label{ch:crossModal}
The previous chapters developed data-driven methods to aid DTW-based alignment. While these methods offer the ability to learn features directly from data, capture structural differences and adapt to the application setting, they are still reliant on DTW for the actual alignment computation. This chapter presents an endeavour towards the learning of alignments for the offline synchronisation task in a data-driven manner and eschews the reliance on Dynamic Time Warping, thereby enabling end-to-end training. 
\par The chapter begins with an introduction to data-driven alignment and presents existing approaches related to this direction in Section \ref{sec:ch5_intro}. The incremental development of the methodology proposed in this chapter is then detailed via a description of relevant neural architectures pertinent to end-to-end alignment in Section \ref{sec:ch5_archDev}. Section \ref{sec:ch5_failed} presents initial attempts to learn alignment using only convolution, which despite not yielding satisfactory performance paved the way for the proposed method. The next section, Section \ref{sec:ch5_proposed} presents the main contribution of the chapter, a novel neural method that combines different layer types and successfully computes robust alignments across multiple performance synchronisation settings such as audio-to-MIDI alignment as well as audio-to-image alignment, whilst also being able to capture structural differences between the input sequences.
The chapter then concludes with the key takeaways and briefly discusses potential further extensions of the proposed method in Section \ref{sec:ch5_conc}. 
\section{Introduction and Related Work}\label{sec:ch5_intro}
The limitations of traditional alignment algorithms based on Dynamic Time Warping and Hidden Markov Models were discussed in the previous chapters, and various methods were proposed to overcome these limitations. The inability to learn the feature representation directly from data and adapt to different test settings was addressed in Chapter 3, and the inability to capture structural differences between the performance and score sequences was addressed in Chapter 4. While both these chapters presented data-driven methods that leveraged neural networks at some stage in the alignment pipeline, the alignment computation in these methods is still carried out using Dynamic Time Warping. 

\par A number of other approaches developed in parallel explore data-driven methods for the alignment task. The majority of these also employ neural networks as precursors to DTW-based alignment. \citet{kwon2017audio} propose audio-to-score alignment of piano music using RNN-based automatic music transcription in combination with DTW. A similar method that employs neural pre-processing for DTW-based alignment proposes CNNs for the identification of measures in sheet images to aid cross-document alignment using DTW
\citep{waloschek2019identification}. It should be noted that this method generates a coarse alignment at the measure level across different pieces, as opposed to finer levels of alignment (for instance, note-level) generated by the majority of alignment algorithms.  The recently proposed Hierarchical-DTW \citep{shan2020improved} is another DTW-based alignment method that employs pre-trained neural models for score following. This method is reliant on an automatic music transcription system \citep{hawthorne2017onsets} and a pre-trained model to extract bootleg score representations \citep{tanprasert2019midi}. It struggles when this representation is inaccurate and on short pieces containing jumps.

 \begin{table}[ht]
 \vspace{0.5cm}
   \caption{A comparison of contemporary approaches to alignment}
   \begin{tabular}{ c c c  c } \toprule
       \textbf{Method}  & \textbf{End to end?} & \textbf{Modalities}& \textbf{Structure-aware?}   \\ \midrule
        \cite{dorfer2017learning}  &   No   &   Audio-Image & No  \\ \midrule
        \cite{dorfer2018learning2} &  Yes  &  Audio-Image  & No\\ \midrule
        \cite{tanprasert2019midi}  & No  &  Score-Image  & No\\ \midrule
        \cite{waloschek2019identification}  & No  &  Image & No \\ \midrule
         \cite{shan2020improved}  & No  &  Audio-Image & Yes \\ \midrule
          \cite{henkel2020learning}  & Yes  &  Audio-Image & No \\ \midrule
      Ch 3 \citep{agrawal2021learning} & No  & Audio-Score & No \\ \midrule
      Ch 4 \citep{agrawal2021structure} & No  & Audio-Score & Yes \\ \midrule
       Proposed method & Yes  &   \pbox{20cm}{Audio-Score,\\ Audio-Image} & Yes\\ \midrule
      \bottomrule
\end{tabular}
\label{tab:works}
\end{table}   
\par Another very recent track for score following research consists of approaches that are based on the estimation of the current position of the performance audio in the sheet image, using techniques such as reinforcement learning \citep{dorfer2018learning, dorfer2018learning2}, or instance-based segmentation \citep{henkel2019audio, henkel2020learning}. All these approaches assume complete structural agreement between the performance and the corresponding score, and hence are not robust to structural deviations from the score. The contemporary approaches towards data-driven alignment are summarised in Table \ref{tab:works}. 

\par While neural architectures have recently been explored for score following approaches that assume structural agreement (\citep{dorfer2018learning, henkel2020learning}), their application to structure-aware performance-score synchronisation remains relatively unexplored. Additionally, the majority of data-driven synchronisation methods still rely on DTW for the actual alignment computation. This chapter furthers the development of data-driven synchronisation approaches and proposes a neural architecture for \textit{learnt} alignment computation, thereby eschewing the limitations of DTW-based alignment. 

\section{Developing the architecture for \textit{learnt} alignment}\label{sec:ch5_archDev}

\begin{figure}[H]
\begin{center}
\includegraphics[width=4in]{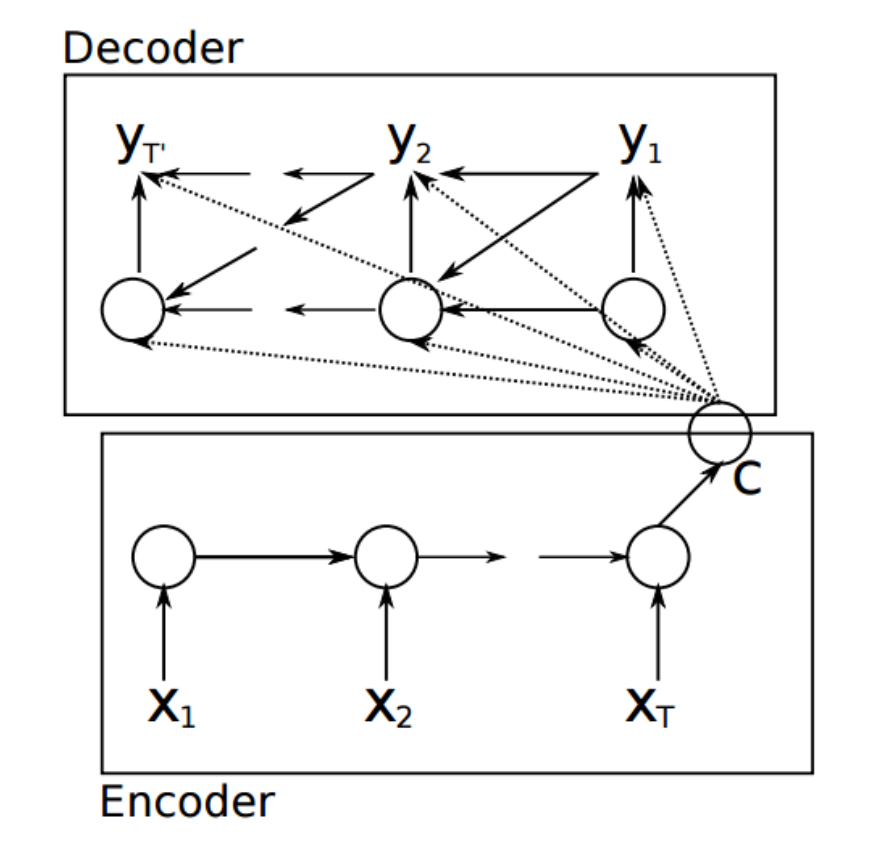}
\caption{A general encoder-decoder sequence to sequence architecture for the sequence transduction ($X = <x_1, x_2, .., x_T> \longrightarrow Y = <y_1, y_2, .., y_T>)$ task. The encoder encodes the input sequence into the context vector $C$, which is employed by the decoder to generate the output sequence. The figure demonstrates the step-by-step processing typically employed by recurrent networks}
\label{encdec}
\end{center}
\end{figure}
\par End-to-end performance synchronisation involves modeling the two input sequences (performance and score) and generating an output alignment sequence (corresponding mappings) with a single network. Sequence-to-sequence (seq2seq) models \citep{sutskever2014sequence} are a naturally suitable choice for this task. A general architecture for such a model is shown in Figure \ref{encdec}, where $x_1$, $x_2$, .., $x_T$ denotes the input sequence; $C$ denotes the hidden representation (also called context vector) and $y_1$, $y_2$, .., $y_T$ denotes the output (target) sequence. The output sequence is predicted one token at a time with information from the previously predicted outputs and the context vector $C$. Performance-score synchronisation using a seq2seq approach would entail a multi-encoder architecture to model the two separate input sequences (audio and score) and output the alignment sequence. 
\par Recurrent neural networks (such as GRUs or LSTMs) are a standard choice for sequence-to-sequence models, since they are inherently able to incorporate temporal information owing to the manner in which they are formulated. However, this capability is limited in the presence of very long input sequences, which is the case for performance synchronisation with inputs that could be thousands of frames long. Recurrent models struggle to capture long-term dependencies in these scenarios since their optimisation becomes difficult when computational graphs are too deep. The reason behind this is that RNNs apply the same operation repetitively at each time step for a long sequence. This results in gradients propagating over many stages, which leads to either a vanishing or exploding gradient that impedes training in the presence of long-term dependencies.
\par A different architecture for the end-to-end synchronisation task would thus be required to overcome this barrier in order to generate robust alignments for long input sequences. 
To this end, I present an endeavour towards developing a custom neural network architecture that combines different types of layers suitable for the alignment task. 
The major architecture choices along with their key features and computational complexities are briefly highlighted below to contextualise the development of the proposed architecture.
\vspace{0.2cm}
\\
\textbf{Convolutional layers}\\
These layers form the backbone of the CNN architecture, and are prominently employed for image analysis tasks such as recognition and classification.\\
\textit{Key features:}
\begin{itemize}
    \item Faster to train than RNNs
    \item Good at detecting geometric patterns
    \item Cannot capture temporality inherently, more suited to classification tasks than sequence prediction tasks
\end{itemize}
\textbf{Recurrent layers}\\
These layers form the backbone of the RNN/LSTM architecture, and are typically employed for natural language processing or time-series analysis tasks.\\
\textit{Key features:}
\begin{itemize}
    \item Can capture temporality quite well until a certain sequence length
    \item Vanishing gradient for long input sequences
    \item Difficult to parrallelise since current timestep depends on previous timestep computation
\end{itemize}
\textbf{Self-attention layers}\\
These layers form the backbone of the Transformer architecture \citep{vaswani2017attention} that has shown promising results for various sequence transduction tasks.\\
\textit{Key features:}
\begin{itemize}
    \item Do not have recurrence and can model long-term dependencies in theory
    \item Support for parallel processing since the representations are not computed in a sequential manner
    \item Require large GPU memory and can cause memory issues from a practical viewpoint if the input sequence is too long
\end{itemize}
\par A complexity comparison of the three layer types is presented in Table \ref{tab:layerTypes}, where $n$ is the length of the sequence, $d$ is the dimensionality of the representation, and $k$ is the kernel size.
\begin{table*}[ht]
\vspace{0.4cm}
   \caption{Complexity comparison for different architectures \citep{vaswani2017attention}}
   \centering
   \begin{tabular}{ c c c c} \toprule
      \textbf{Layer Type} & \textbf{Complexity per Layer} & \textbf{Seq. Operations} & \textbf{Max Path Length} \\ \midrule
      Convolutional  & $O(k \times n \times d^2)$  & $O(1)$ & $O(\log_k n)$ \\ \midrule
      Recurrent & $O(n \times d^2)$ & $O(n)$  & $O(n)$  \\ \midrule
      Self-Attention & $O(n^2 \times d)$  & $O(1)$ & $O(1)$  \\ \midrule
      \bottomrule
\end{tabular}
\label{tab:layerTypes}
\end{table*} 

\par Recurrent architectures would be unsuitable given the long input sequences as discussed earlier in the  section as well as their computational complexity (Table \ref{tab:layerTypes}). We explore the other two architectures in the next sections. We begin with a completely convolutional architecture, casting the alignment problem as a semantic segmentation task. This approach did not yield good results and is described briefly in Section \ref{sec:ch5_failed}. 
We then combine the convolutional and self-attention layers in a strategic manner. This method yielded promising results across multiple test settings and is described in detail in Section \ref{sec:ch5_proposed}. 
\section{Initial approaches using convolution}\label{sec:ch5_failed}
\vspace{0.2cm}
The initial exploration for learnt alignment computation was carried out using a completely convolutional framework. Chapter 4 demonstrated how the cross-similarity matrix served as a viable input choice for the dilated CNN models, simplifying the multiple sequence modelling task while maintaining the capability to capture temporality. The experimentation using dilated CNNs demonstrated that the convolutions are able to capture inflection points from the cross-similarity matrix for structure-aware alignment. It could be speculated based on this observation that convolutions could capture not just the inflection points, but the entire set of points that make up the alignment path. To this end, an exploration was carried out towards the alignment learning task using CNNs by casting it as a semantic segmentation problem. 

\vspace{0.1cm}
\par Semantic segmentation is the task of classifying each and every pixel in an image into a class. For the alignment task, this entails classifying each frame pair in the cross-similarity matrix as either \textit{belonging} or \textit{not belonging} to the alignment path. Three different convolutional models that have demonstrated promising results for the semantic segmentation task were trained and tested on the Mazurka dataset \citep{sapp2007comparative} to generate the pixel-level classification from the cross-similarity matrix. These include the U-net model \citep{ronneberger2015u}, the SegNet model \citep{badrinarayanan2017segnet} and the HRNet model \citep{wang2020deep}. \blue{The architecture of the three networks is kept the same as in the originally proposed methods. The models were trained for 40 epochs, using the Dice coefficient \citep{milletari2016v} loss function, which is commonly used for semantic segmentation tasks. }
\vspace{0.2cm}

\begin{figure*}[ht]
\vspace{0.1cm}
    \centering
       \subfloat[Input]
  {{\includegraphics[width=5cm, height=5cm]{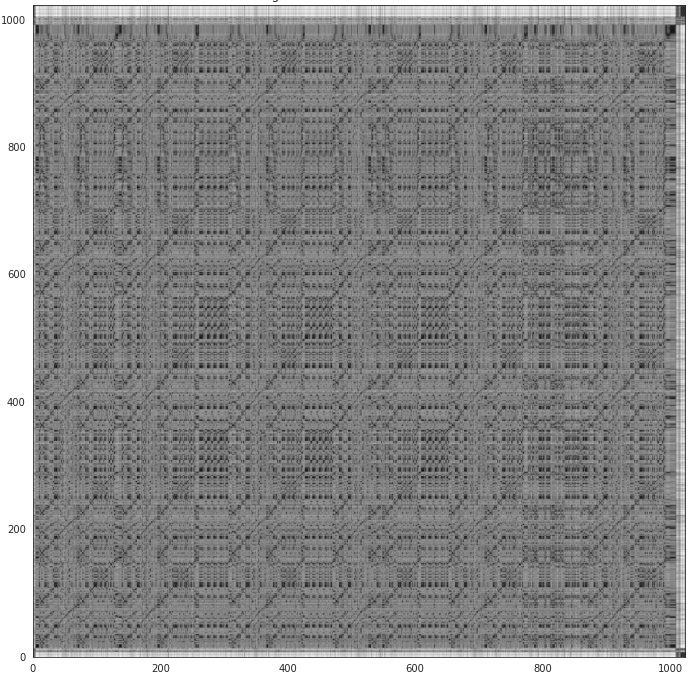} }}%
    \qquad
        \subfloat[Ground Truth]
    {{\includegraphics[width=5cm, height=5cm]{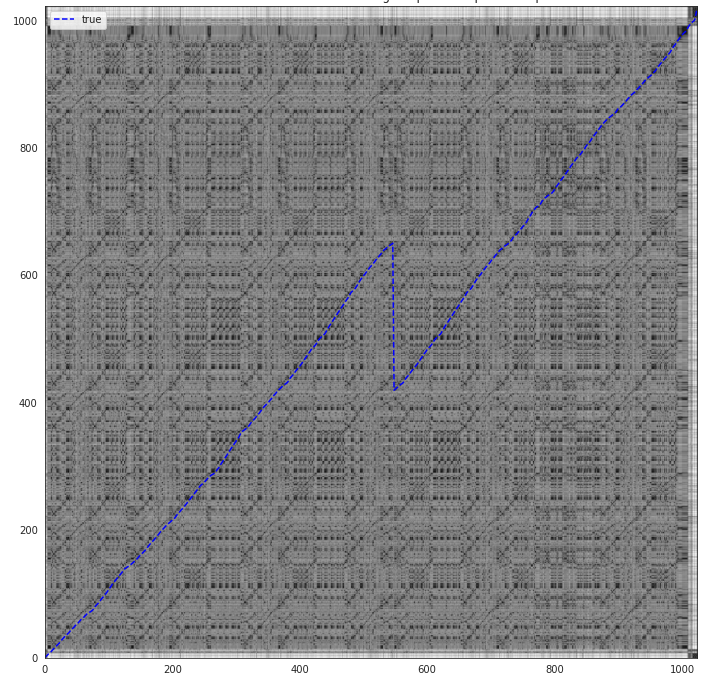} }}%
    \\
        \subfloat[Network output\\ \textit{Ground Truth in Blue}]
{{\includegraphics[width=5cm, height=5cm]{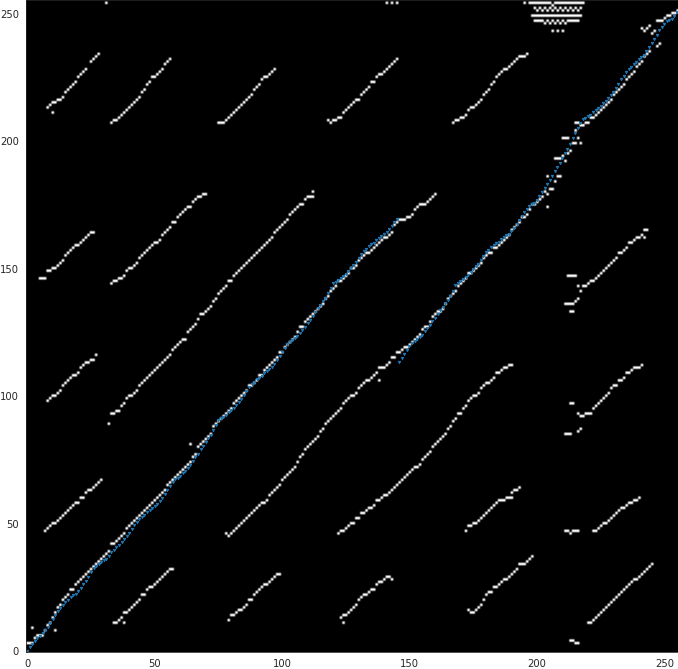} }}%
    \qquad
    \subfloat[Post-processed output\\\textit{Ground Truth in Blue}]
    {{\includegraphics[width=5cm, height=5cm]{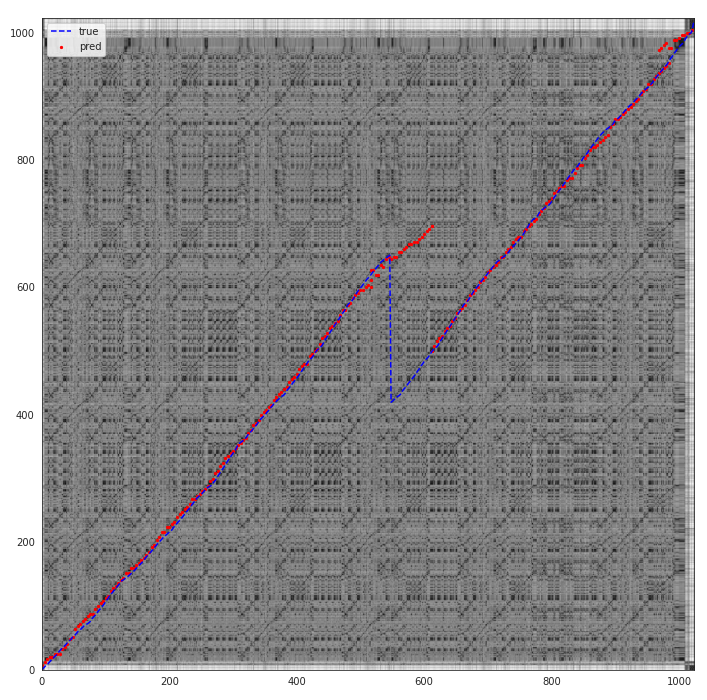} }}%
        \caption{An example of the alignment generated by the semantic segmentation method \blue{(U-net)}, and the need for post-processing. \\Input: Cross-similarity matrix between the performance and the score \\
    X-axis: Frame index (performance), Y-axis: Frame index (score)}
    \vspace{0.2cm}
    \label{fig:semSeg}%
\end{figure*}
\par The results obtained using this method were not satisfactory. Two problems were encountered with this approach, namely the imbalanced training data (i.e. a much higher number of negative samples than positive ones) and the need for post-processing. The models were not able to distinguish between matching frame pairs that belong to the alignment path and those that lie outside the alignment path, and post-processing was required to generate the alignment path from the predictions of the segmentation networks. \blue{This is depicted in Figure \ref{fig:semSeg}, which demonstrates the predictions of the U-net after 40 epochs of training (shown in white) and the post-processed output (shown in red), compared with the ground truth alignment path (shown in blue)}. Since these methods did not yield satisfactory performance without requiring post-processing, the details of the architectures and experiments are not provided, rather an alternative approach towards end-to-end alignment that yielded promising results is presented in greater detail in the next section.
\section{Proposed Methodology}\label{sec:ch5_proposed}
This section presents a novel neural method for performance-score synchronisation that is also robust to structural differences between the performance and the score.
I model the performance-score synchronisation task as a sequence prediction task, given the two input sequences corresponding to the performance and score respectively. However, rather than relying on recurrent neural networks or Transformers \citep{vaswani2017attention} and predicting the output sequence one token at a time, I propose a convolutional-attentional architecture that predicts the entire alignment path in a one-shot fashion. This allows the model to capture long-term dependencies and also handle structural differences between the performance and score sequences.
\par The proposed convolutional-attentional architecture has an encoder-decoder framework, with the encoder based on a convolutional stem and the decoder based on a stand-alone self-attention block \citep{ramachandran2019stand}. The intuition behind the combination is that the convolutions would detect the synchronous subpaths, and the self-attention layers would yield the alignments by capturing pairwise relations whilst incorporating the global context.  

\begin{sidewaysfigure}[htbp]
  \centering
\includegraphics[width=8in]{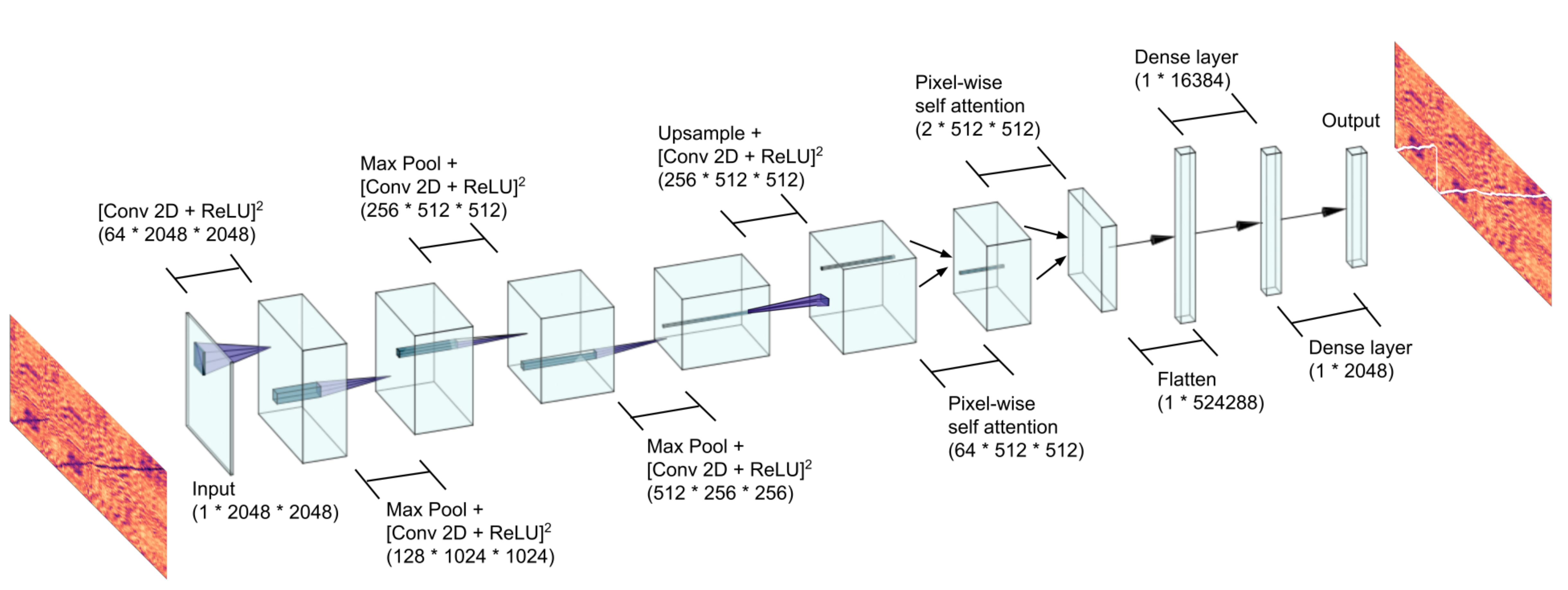}
  \caption{Schematic diagram illustrating the proposed method for learning alignment. \\ The output alignment path is plotted against the distance matrix for a simple example to aid visualisation.} 
  \label{fig:ch5_pipeline}
\end{sidewaysfigure}

The stand-alone self-attention block employs local self-attention layers and overcomes the limitations of global attention layers that are typically used on reduced versions of input images. 
The motivation behind employing the stand-alone self-attention block, as opposed to the more commonly used approach of an attention computation on top of the convolution operation, is that the stand-alone self-attention layers have proven to be effective at capturing global relations in vision tasks when employed in later stages of a convolutional network \citep{ramachandran2019stand}. The proposed method is illustrated in Figure \ref{fig:ch5_pipeline}.\\

\par The network operates on the cross-similarity matrix between the performance and score feature sequences and predicts the ($x$, $y$) co-ordinates corresponding to the frame indices that make up the optimal alignment path. Since the $X$-axis of the matrix corresponds to the performance, it progresses linearly in the alignment and the goal is essentially to predict the sequence of $y$ co-ordinates (i.e. frame indices on the score axis) that determine the alignment path. Formally, let $X$= $(x_1, x_2,..., x_p)$ and $Y$ = $(y_1, y_2,..., y_q)$ be the feature sequences corresponding to the performance and score respectively. The network is trained to predict the sequence of frame indices $\hat{Y}_p$= $(\hat{y}_1, \hat{y}_2,..., \hat{y}_p)$, denoting the path taken by the performance $X$ through the score $Y$. 

\par In addition to proposing a novel architecture, a customised time-series divergence loss function based on the differentiable soft-DTW computation \citep{cuturi2017soft} is employed to train our models. Experiments are conducted for two synchronisation tasks involving different score modalities, namely audio-to-MIDI alignment, i.e. aligning audio to symbolic music representations, and audio-to-image alignment, i.e. aligning audio to scanned images of sheet music. The results demonstrate that the proposed method generates robust alignments in both settings.  The next section describes the model architecture in detail.
\section{Model Architecture}
The general architecture of the proposed model is depicted in Figure \ref{fig:ch5_pipeline}. The network has an encoder-decoder architecture, with the encoder comprising four convolutional and downsampling blocks, and the decoder comprising an upsampling block, a stand-alone self-attention (hereafter abbreviated as \emph{SASA}) block  and a fully connected block. The decoder upsamples the encoded values and passes the output to the \emph{SASA} block. The \emph{SASA} block comprises two stand-alone self-attention layers that compute the pixel-level attention values and pass the output through a fully connected block with two dense layers. 

\subsection{The stem}
The convolutional stem comprises four convolutional and subsampling blocks.
Each convolution block consists of two convolutional and pooling layers, with 2D batch normalisation \citep{ioffe2015batch} and Rectified Linear Unit (ReLU) applied after each 2D convolution, before being passed on to the max-pooling layer. The locations of the maximum values obtained during the max-pooling operation are stored. These are employed by the \textit{max-unpooling} operation described next.
\subsection{Upsampling strategy}
As part of our upsampling strategy, we employ the max-unpooling operation \citep{zeiler2014visualizing} as opposed to the transposed convolution, which has been shown to result in artifacts \citep{odena2016deconvolution}. To perform the max-unpooling, the indices of the highest activations during the pooling stages within each filter window are stored by a mask. These recorded locations are passed to the upsampling block, where the unpooling places each element in the unpooled map according to the mask, instead of assigning it to the upper-left pixel. The upsampling block based on max-unpooling results in lower computational complexity and is faster to train than the transposed convolution filters. \\

\subsection{The SASA block}
\begin{figure*}
  \centering
\includegraphics[width=6.5in]{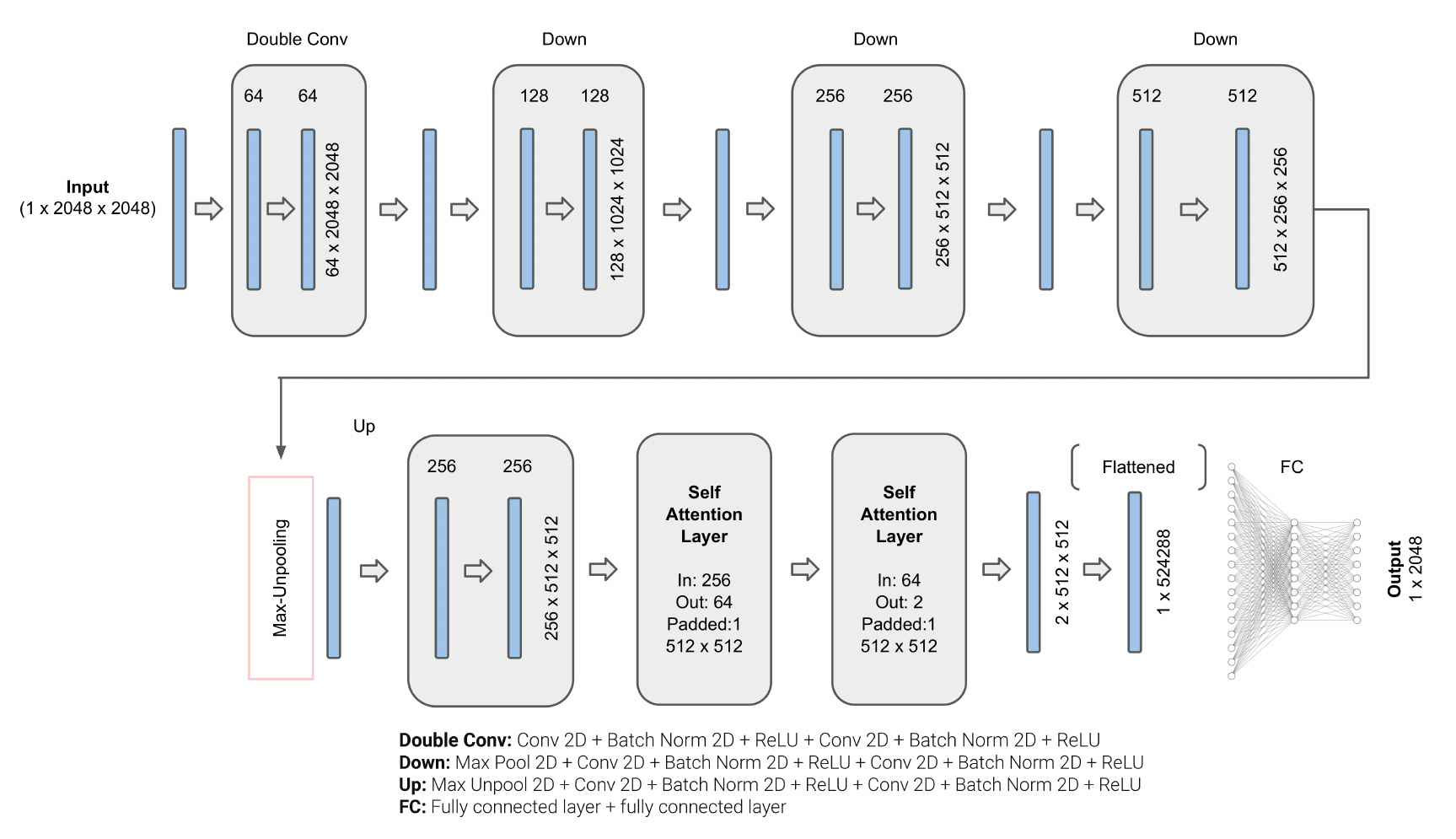}
  \caption{The detailed architecture of the proposed model}
  \label{fig:detailedPipeline}
  \vspace{1cm}
\end{figure*}

The upsampled output is passed on to the \emph{SASA} block, comprising two stand-alone self-attention layers. For each pixel $(i, j)$ in the upsampled output, the self-attention is computed relative to the memory block $M_{k}(i, j)$, which is a neighbourhood with spatial extent $k$ centred around $(i, j)$, as follows:
\begin{equation}
y_{ij} = \hspace{-0.2cm} \sum\limits_{a, b \in M_{k}(i, j)} \hspace{-0.4cm}\mathit{softmax_{ab}}  (q_{ij}^ \intercal k_{ab} + q_{ij}^ \intercal r_{a-i, b-j}) \hspace{0.1cm}  v_{ab}
\end{equation}
where $q_{ij} = W_q x_{ij}$ are the queries, $k_{ab} = W_k x_{ab}$ the keys and $v_{ab} = W_v x_{ab}$ the values computed as linear transformations from the activations at the $(i, j)^{th}$ pixel and its memory block. The displacements from the current position $(i, j)$ to the neighborhood pixel $(a, b)$ are encoded by row and column offset embeddings given by $r_{a-i}$ and $r_{b-j}$ respectively, which are concatenated to form $r_{a-i, b-j}$\blue{.}  The architecture employs four attention heads and splits the pixel features depthwise into four groups of the same size.  The attention is then computed on each group individually with different matrices $W$ and the results are concatenated to yield the pixel-wise attention values $y_{ij}$. This computation is repeated twice and the output is passed through a fully connected block with two dense layers to predict the alignment path. A graphic elaboration can be found in Figure \ref{fig:detailedPipeline}.
\par It must be noted that the \emph{SASA} block employed by the decoder is different from the commonly explored combination of an attention computation applied on top of a convolution operation \citep{oktay2018attention}, or the self-attention layer from the sequence to sequence Transformer architecture \citep{vaswani2017attention}.
The \emph{SASA} block borrows ideas from both convolution and self-attention, and is able to replace spatial convolutions completely and effectively integrate global information while reducing the computational complexity \citep{ramachandran2019stand}.
\subsection{Differentiable divergence loss}\label{sec:ch5_loss}
The proposed method employs a custom time-series divergence loss function to train the models, as opposed to a \blue{cross-entropy} loss. The primary motivation behind this loss is that it allows the model to minimise the overall cost of aligning the performance and score feature sequences by comparing the \textit{paths} rather than the \textit{feature sequences} using a positive definite divergence.

\par Let $X$= $(x_1, x_2,..., x_p)$ be the feature sequence corresponding to the performance and $Y$ = $(y_1, y_2,..., y_q)$ be the feature sequence corresponding to the score. 
The proposed loss function captures the divergence between the predicted and ground truth alignment sequences, based on the soft-DTW \citep{cuturi2017soft} distance. This distance is employed since it offers a differentiable measure of the discrepancy between the two sequences. 
\par Given the predicted alignment sequence $\hat{Y}$= $(\hat{y}_1, \hat{y}_2,..., \hat{y}_p)$ and the ground truth alignment sequence $Y$ = $(y_1, y_2,..., y_p)$, the soft-DTW distance $D_\lambda (a, b)$ at $(a, b)$ is computed as follows:
\begin{equation}\label{eq:softDTW}
    D_\lambda (a, b) = e(a, b) + min_{\lambda} \begin{cases}
    D_\lambda (a, b-1) \\ D_\lambda (a-1, b) \\  D_\lambda (a-1,  b-1) \\
    \end{cases}
\end{equation}
where $e(a, b)$ is the Euclidean distance between points $\hat{y}_a$ and $y_b$, and $min_{\lambda}$ is the soft-min operator parametrized by a smoothing factor \begin{math}\lambda \end{math}, replacing the hard minimum operation of a standard DTW computation, as follows: \\
\begin{equation}
    min_{\lambda}\{m_1, m_2, ..., m_n\} = \begin{cases}
min\{m_1, m_2, ..., m_n\} 
\hspace{0.5cm} \lambda=0 \\ -\lambda \log \sum_{i=1}^{i=n} e^{-m_i/\lambda} \\
\end{cases}
\end{equation}
\par The hard min operator is replaced by the \textit{soft} min operator in order to enable differentiability and thereby allow its usage as a loss function to train the networks. However, $D_\lambda$ is not strictly a distance (unlike DTW) and $D_\lambda (X, X) \neq 0$, since it considers all possible alignments weighted by their probability under the Gibbs distribution \citep{cuturi2017soft}. $D_\lambda$ can thus take on a negative value and it is not minimised when the time series are equal due to the bias introduced by entropic regularization. 
In order to address this,  $D_\lambda$ is normalised, thereby making it a 
positive definite divergence \citep{blondel2021differentiable}, as follows: 
\begin{equation} \label{eq:divergence}
SD_\lambda (\hat{Y}, Y) = D_\lambda (\hat{Y}, Y) - 1/2(D_\lambda (\hat{Y}, \hat{Y}) + D_\lambda (Y, Y))
\end{equation}
This ensures that $SD_\lambda(\hat{Y}, Y)>0$ for $\hat{Y}\neq Y$  and $SD_\lambda(\hat{Y}, Y) = 0$ for $\hat{Y} = Y$, yielding a completely learnable framework since $SD_\lambda(\hat{Y}, Y)$ is non-negative and differentiable at all points. 
\par It must be noted that rather than doing the DTW computation on the feature sequences to \textit{generate} the alignment path, the soft-DTW computation described in this section is used on the predicted alignment sequences to \textit{compare} the alignment paths. Additionally, only the the distance metric $D_\lambda (a, b)$ from the soft-DTW computation in Equation \ref{eq:softDTW} is employed by the models, and not the alignment path (between the predicted and ground truth alignment paths) that minimises its value. The alignment computation between the input sequences is carried out by our neural framework, by minimizing the custom divergence loss $SD_\lambda(\hat{Y}, Y)$. 

\subsection{A note on the relation to the Transformer architecture}
It would be noteworthy to present a juxtaposition of the proposed model architecture with the classic Transformer architecture proposed by \citet{vaswani2017attention}.  
The Transformer architecture, initially proposed for language translation, combines self-attention and cross-attention and outputs the target sequence one token at a time. 
While the Transformer captures long-term dependencies better than recurrent architectures, it struggles with very long input sequences \citep{vaswani2017attention}. Additionally any index-based information such as the relative position of the tokens must be incorporated into the embeddings for the Transformer by means of sinusoidal positional encodings. The combination of stand-alone attention layers and convolutional layers has been demonstrated to yield better performance than fully attentional or fully convolutional models \citep{ramachandran2019stand} for vision tasks.
\par The performance synchronisation task entails modelling two heterogeneous sequences, each of which could be thousands of tokens long, with possible structural differences between the inputs. This motivated the development of a novel architecture that approaches the synchronisation problem in a different fashion than the seq2seq Transformer method, and predicts the alignment vector in a one-shot manner using a convolutional-attentional network. The proposed architecture also equips the model to capture possible structural differences between the performance and score sequences. 
It must also be noted that the proposed method differs from the commonly explored combination of convolution with attention, since the former employs the stand-alone self-attention layer, as opposed to the standard self-attention employed by the latter. The reader is referred to  \citet{cordonnier2019relationship} for a detailed analysis of the relationship between convolution and attention.

\section{Experiments and Results}\label{experiments}
\subsection{Experimental Setup}\label{setup}
Experiments are conducted for two alignment tasks pertinent to different score modalities, namely audio-to-MIDI alignment and audio-to-image alignment. Two publicly available datasets are employed for our experiments on the two alignment tasks.
For each performance-score pair, the cross-similarity matrix is computed using the Euclidean distance between the chromagrams for audio-to-MIDI alignment, and the Euclidean distance between learnt cross-modal embeddings \citep{dorfer2017learning} for audio-to-image alignment. The computation of the chromagrams as well as cross-similarity matrices is carried out using librosa \citep{mcfee2015librosa}. It must be noted that these feature representations are employed to facilitate comparison with previous approaches, however, the proposed method is compatible with other feature representations too. A sampling rate of 22050 Hz, a frame length of 2048 samples and a hop length of 512 samples is employed for the chromagram computation.

\par The encoder-decoder architecture of the proposed method was illustrated in the previous section. On the encoder side, the output of each 2D convolution is batch normalized \citep{ioffe2015batch} and passed through a Rectified Linear Unit (ReLU) non-linearity, before being passed on to max-pooling.

A dropout of 0.4 is employed for the fully connected layers to avoid overfitting. The output of the final layer is a vector of length 2048, encoding the \blue{$\hat{y}$-indices} making up the predicted alignment path \blue{$\hat{Y}$}. \blue{During training and testing, each performance and score feature sequence is scaled to length 2048. These are then rescaled back to the original dimensions for comparing the predicted alignment with the ground truth.}
It must be noted that the output vector is sufficient to capture the length of all pieces in the data, since the audio-to-MIDI task has beat-level annotations (less than 2048 per piece), and the audio-to-image task has notehead-level annotations (also less than 2048 per piece).
The output vector is compared with the ground truth using the custom loss $SD_\lambda(\hat{Y}, Y)$, as explained in Section \ref{sec:ch5_loss}.
The proposed model is abbreviated as \emph{CA$_{\textit{custom}}$} in the upcoming studies.

\subsection{Study 1: Results for audio-to-MIDI alignment}

\begin{table*}[ht]
\vspace{0.5cm}
   \centering
\begin{tabular}{ccccc} \toprule
\hline 
\multirow{2}{*}{\textbf{Model}} & 
\multicolumn{3}{c}{\textit{Error margin}} 

\tabularnewline
  & \textbf{$<$50 ms}& \textbf{$<$100 ms} & \textbf{$<$200 ms}
  \\
\midrule 
 \begin{math}\textit{MATCH}\end{math} \citep{dixon2005line} &  74.6* & 79.5* & 85.2*  \\
\midrule

 \begin{math}\textit{JumpDTW}\end{math} \citep{Fremerey2010handling} &  75.2* & 80.4* & 86.7*  \\
\midrule


 \emph{SiameseDTW} \citep{agrawal2021learning}  & \underline{77.9} & \underline{83.3}* & 89.5*  \\
\midrule 
  \begin{math}\textit{DeepCTW}\end{math} \citep{trigeorgis2017deep}  & 76.1* & 81.6* & 88.9*  \\
\midrule 
 \begin{math}\textit{DilatedCNN}\end{math} \citep{agrawal2021structure} & 77.5* & 82.4* & \underline{90.4}*   \\
\midrule 

  \emph{CA$_{\textit{custom}}$} & \textbf{78.7} & \textbf{85.2} & \textbf{92.6}   \\
\midrule 
\bottomrule
\end{tabular}
\vspace{0.2cm}
\caption{Audio-to-MIDI alignment accuracy in \% on the \emph{Mazurka-BL} dataset. Best in bold, second best underlined.\\$*$: significant differences from \emph{CA$_{\textit{custom}}$}, $p < 0.05$}
\label{tab:ch5_results_score}
\end{table*}
The experimentation for the audio-to-MIDI alignment task is carried out on the publicly available Mazurka-BL dataset \citep{kosta2018mazurkabl}. This dataset comprises 2000 recordings with annotated alignments at the beat level. The recordings correspond to performances of Chopin’s Mazurkas dating from 1902 to the early 2000s, and
 span various acoustic settings. This set is randomly divided into sets of 1500, 250 and 250 recordings respectively, forming the training, validation and testing sets. 

The results obtained by the proposed model are compared with \begin{math} \textit{MATCH}\end{math} \citep{dixon2005line}, \begin{math}\textit{JumpDTW}\end{math} \citep{Fremerey2010handling}, \begin{math}\textit{SiameseDTW}\end{math} \citep{agrawal2021learning},  \begin{math}\textit{DilatedCNN}\end{math} \citep{agrawal2021structure}, and the \begin{math}\textit{Deep Canonical Time Warping}\end{math} \emph{(DeepCTW)} method \citep{trigeorgis2017deep}.
The percentage of beats aligned correctly within error margins of 50, 100 and 200 ms respectively is computed for each piece, and the alignment accuracy \citep{cont2007evaluation} obtained by each model averaged over the entire test set is reported. The two best performing models for each evaluation setup are highlighted, with the best marked by bold and second best marked by underline.
Significance testing is also conducted using the Diebold-Mariano test \citep{harvey1997testing} and pairwise comparisons of all model predictions with the \emph{CA}$_{\textit{custom}}$ predictions for each error margin are performed to examine the statistical significance of the results. \blue{All the convolutional-attentional models were trained on a GeForce RTX 2080 Ti GPU card containing 11 GB GPU memory, with the NVIDIA driver version 418.87.00 and CUDA version 10.1. The training for all models was completed in less than 12 hours when trained on a single GPU core.}

The results obtained by the proposed method for audio-to-MIDI alignment are given in Table \ref{tab:ch5_results_score}.
Overall accuracy on the Mazurka-BL dataset suggests that the proposed model \emph{CA$_{\textit{custom}}$} outperforms the DTW-based frameworks \emph{MATCH}, \emph{JumpDTW} and \emph{SiameseDTW} by up to 7\% (rows 1-3) as well as the neural frameworks \begin{math}\mathit{DeepCTW}\end{math} and \begin{math}\mathit{Dilated} \mathit{CNN}\end{math} by up to 4\% (rows 4-5) for all error margins. 
 The comparison with contemporary approaches reveals that our method yields higher improvement over the state-of-the art for coarse alignment (Error margins $>$ 50ms) than for fine-grained alignment 
 (Error margin $<$ 50ms), however overall performance improves over previous approaches in both the cases.

\subsection{Study 2: Results on audio-to-image alignment}

\begin{table*}[ht]
   \centering
\begin{tabular}{cccc} \toprule
\hline 
\multirow{2}{*}{\textbf{Model}} & 
\multicolumn{3}{c}{\textit{Error margin}} 
\tabularnewline
  & \textbf{$<$0.5 s}& \textbf{$<$1 s} & \textbf{$<$2 s}
  \\
\midrule 
 \citet{dorfer2017learning} & 73.5* & 81.2* & 84.7*   \\
\midrule
  \citet{dorfer2018learning2} & 76.4* & 84.5* & 89.3*   \\
\midrule 
   Audio-conditioned U-net \citep{henkel2020learning}  & \underline{84.6} & \underline{88.4}* & \underline{90.1}* \\
\midrule 
  \emph{CA$_{\textit{custom}}$}  & \textbf{85.2} & \textbf{91.5} & \textbf{92.9}   \\
\midrule 
\bottomrule
\end{tabular}
\vspace{0.2cm}
\caption{Audio-to-Image alignment accuracy in \% on the \emph{MSMD} dataset. Best in bold, second best underlined.\\$*$: significant differences from \emph{CA$_{\textit{custom}}$}, $p < 0.05$}  \label{tab:ch5_results_image}
\end{table*}

The experimentation for the audio-to-image alignment task employs the Multimodal Sheet Music Dataset (MSMD) \citep{dorfer2018learning}, a standard dataset  for sheet image alignment analysis. MSMD comprises polyphonic piano music for 495 classical pieces, with notehead-level annotations linking the audio files to the sheet images.  
This set is randomly divided into sets of 400, 50 and 45 recordings respectively, forming the training, validation and testing sets. The results obtained by the proposed model are compared with contemporary audio-to-image alignment methods \citet{dorfer2017learning},  \citet{dorfer2018learning2}, and the  
audio-conditioned U-net model \citep{henkel2020learning}.
\par For comparison with \citet{henkel2020learning}, the predicted sheet image co-ordinates using their method are extrapolated to the time domain from the ground truth alignment between the note onsets and the corresponding notehead co-ordinates in the sheet images. The alignment accuracy obtained by each model is then computed by calculating the percentage of onsets aligned correctly within the error margins of 500 ms, 1 s and 2 s respectively. The results averaged over the test set along with the significance tests are reported in Table \ref{tab:ch5_results_image}. Note that the same feature representation \citep{dorfer2017learning} is employed for all audio-to-image methods. 
\par The experimentation for audio-to-image alignment reveals trends similar to the ones presented for audio-to-MIDI alignment. The results demonstrate that \emph{CA}$_{\textit{custom}}$ outperforms \citet{dorfer2017learning} and \citet{dorfer2018learning2}  
in overall alignment accuracy by 2-10\% and  \citet{henkel2020learning} by 1-4\% for all error margins. A further advantage of the proposed method over \citet{henkel2020learning} is the ability to work with pieces containing several pages of sheet music, as opposed to only one. 

\subsection{Study 3: Results on structure-aware alignment}
\begin{table*}[ht]
\vspace{0.5cm}
   \centering
\begin{tabular}{ccccc} \toprule
\hline 
\multirow{2}{*}{\textbf{Model}} & 
\multicolumn{3}{c}{\textit{Error margin}} 

\tabularnewline
  & \textbf{$<$50 ms}& \textbf{$<$100 ms} & \textbf{$<$200 ms} 
  \\
\midrule 
 \begin{math}\textit{MATCH}\end{math} \citep{dixon2005line} &  61.8* & 67.4* & 74.6* \\
\midrule
 \begin{math}\textit{JumpDTW}\end{math} \citep{Fremerey2010handling} &  72.5* & 76.2* & 82.0*  \\
\midrule
 \emph{SiameseDTW} \citep{agrawal2021learning}  & 70.7* & 72.8* & 80.3*   \\
\midrule 
  \begin{math}\textit{DeepCTW}\end{math} \citep{trigeorgis2017deep}  & 71.2* &  75.6* & 80.8*  \\
\midrule 
 \begin{math}\textit{DilatedCNN}\end{math} \citep{agrawal2021structure} & \textbf{76.4}* & \textbf{80.3}* & \underline{84.2}*  \\
\midrule 
    \emph{CA$_{\textit{custom}}$} & \underline{75.8} & \underline{79.5} & \textbf{84.9}  \\
\midrule 
\bottomrule
\end{tabular}
\vspace{0.2cm}
\caption{Structure-aware Audio-to-MIDI alignment accuracy in \% on the \emph{Mazurka-BL} dataset. Best in bold, second best underlined.\\$*$: significant differences from \emph{CA$_{\textit{custom}}$}, $p < 0.05$}
\vspace{0.5cm}
\label{results_score_structure}
\end{table*}

\begin{table*}[ht]
   \centering
\begin{tabular}{cccc} \toprule
\hline 
\multirow{2}{*}{\textbf{Model}} & 
\multicolumn{3}{c}{\textit{Error margin}} 
\tabularnewline
  & \textbf{$<$0.5 s}& \textbf{$<$1 s} & \textbf{$<$2 s} 
  \\
\midrule 
 \citet{dorfer2017learning} & 68.4* & 67.8* & 77.6*   \\
\midrule
  \citet{dorfer2018learning2} & 69.2* & 70.3* & 80.4*    \\
\midrule 
   Audio-conditioned U-net \citep{henkel2020learning}  & \underline{70.6} & \underline{72.1}* & \underline{81.1}* \\
\midrule 
  \emph{CA$_{\textit{custom}}$}  & \textbf{75.4} & \textbf{77.4} & \textbf{89.5}   \\
\midrule 
\bottomrule
\end{tabular}
\vspace{0.2cm}
\caption{Structure-aware Audio-to-Image alignment accuracy in \% on the \emph{MSMD} dataset. Best in bold, second best underlined.\\$*$: significant differences from \emph{CA$_{\textit{custom}}$}, $p < 0.05$}  \label{results_image_structure}
\end{table*}
\vspace{0.2cm}
\par  The alignment of performances that deviate structurally from the score is a known limitation of the majority of alignment methods \citep{dixon2005match, arzt2016flexible, dorfer2018learning2, henkel2020learning, agrawal2021learning}.  In addition to the overall accuracy on the test sets, it would be desirable to report alignment results for structurally different performance-score pairs for both datasets. In order to specifically test the model performance on structure-aware alignment for both the tasks, 20\% additional samples that contain structural differences between the score and the performance are generated via a randomized split-join operation using the audio from the respective datasets. 50\% of these samples are appended to the training sets and the other 50\% are employed as the test sets. The ground truth alignments are extrapolated from the original alignments using the split-join locations. 
\par The experimentation on structure-aware audio-to-MIDI alignment demonstrates that the proposed method outperforms all approaches except \begin{math}\textit{DilatedCNN}\end{math} \citep{agrawal2021structure} by 3-12\%.
Additionally, the proposed model \emph{CA}$_{\textit{custom}}$ demonstrates comparable results to \begin{math}\textit{DilatedCNN}\end{math} \citep{agrawal2021structure}, without explicitly modeling structure, and while being trained on limited structure-aware data. The author recommends that \emph{CA}$_{\textit{custom}}$ is employed for a test setting that entails a heterogeneous dataset, whereas \begin{math}\textit{DilatedCNN}\end{math} could be employed in settings where structural differences between the performances and scores are inevitable and abundant, for instance in recordings of rehearsals.
\vspace{0.1cm}
\par The results obtained by the models on structure-aware audio-to-image alignment similarly suggests that \emph{CA}$_{\textit{custom}}$ outperforms all contemporary approaches by 5-12\%, a higher margin than that for audio-to-MIDI alignment. This suggests that \emph{CA}$_{\textit{custom}}$ is able to handle structural deviations from the score even when it is presented in the image domain, which is a limitation of the contemporary audio-to-image alignment approaches, including \citet{henkel2020learning}. 

\subsection{Study 4: Ablative analyses}

\begin{table*}[ht]
\vspace{0.5cm}
   \centering
\begin{tabular}{cccccc} \toprule
\hline 
\multirow{2}{*}{\textbf{Model}} & 
\multicolumn{3}{c}{\textit{Overall}} 
& \textit{Structure}
\tabularnewline
  & \textbf{$<$50 ms}& \textbf{$<$100 ms} & \textbf{$<$200 ms} & \textbf{$<$100 ms} 
  \\
\midrule 
 \begin{math}\textit{DilatedCNN}\end{math} \citep{agrawal2021structure} & 77.5* & 82.4* & 90.4* & \underline{80.3}  \\
\midrule 
  \emph{CD$_{\textit{CE}}$} & 72.8* & 80.1* & 85.3* & 71.9*  \\
  \midrule 
  \emph{CD$_{\textit{custom}}$} & 74.1* & 81.7* & 87.5* & 74.2*  \\
  \midrule 
  \emph{CA$_{\textit{CE}}$} & 76.4* & 84.1* & \underline{90.9}* & 76.8*  \\
  \midrule
  \emph{CA$_{\textit{custom}}$} & \underline{78.7} & \underline{85.2} & \underline{92.6} & \underline{79.5}  \\
\midrule 
\emph{CA$_{\textit{custom-L}}$} & \textbf{80.4}* & \textbf{87.5}* & \textbf{93.8}* & \textbf{81.2}  \\
\midrule 
\bottomrule
\end{tabular}
\vspace{0.2cm}
\caption{Ablation studies for Audio-to-MIDI alignment. Accuracy reported in \% on the \emph{Mazurka-BL} dataset. Best in bold, second best underlined.\\$*$: significant differences from \emph{CA$_{\textit{custom}}$}, $p < 0.05$}
\label{ablation_score}
\end{table*}

\begin{table*}[ht]
\vspace{0.2cm}
   \centering
\begin{tabular}{ccccc} \toprule
\hline 
\multirow{2}{*}{\textbf{Model}} & 
\multicolumn{3}{c}{\textit{Overall}} &
\textit{Structure}
\tabularnewline
  & \textbf{$<$0.5 s}& \textbf{$<$1 s} & \textbf{$<$2 s} & \textbf{$<$1 s} 
  \\
\midrule 

   Audio-conditioned U-net \citep{henkel2020learning}  & \underline{84.6} & \underline{88.4}* & 90.1* & 72.1* \\
\midrule 
 \emph{CD$_{\textit{CE}}$} & 80.8* & 84.1* & 87.3* & 71.9*  \\
  \midrule 
  \emph{CD$_{\textit{custom}}$} & 81.1* & 85.7* & 88.5* & 73.2*  \\
  \midrule 
  \emph{CA$_{\textit{CE}}$} & 83.4* & 88.1* & \underline{91.3}* & \underline{76.9}*  \\
  \midrule
  \emph{CA$_{\textit{custom}}$}  & \textbf{85.2} & \textbf{91.5} & \textbf{92.9} & \textbf{77.4}  \\
\midrule 
\bottomrule
\end{tabular}
\vspace{0.2cm}
\caption{Ablation studies for Audio-to-Image alignment. Accuracy reported in \% on the \emph{MSMD} dataset. Best in bold, second best underlined.\\$*$: significant differences from \emph{CA$_{\textit{custom}}$}, $p < 0.05$} 
\vspace{0.2cm}
\label{ablation_image}
\end{table*}

\begin{figure}[ht]
    \centering
       \subfloat[Input]
  {{\includegraphics[width=5cm, height=5cm]{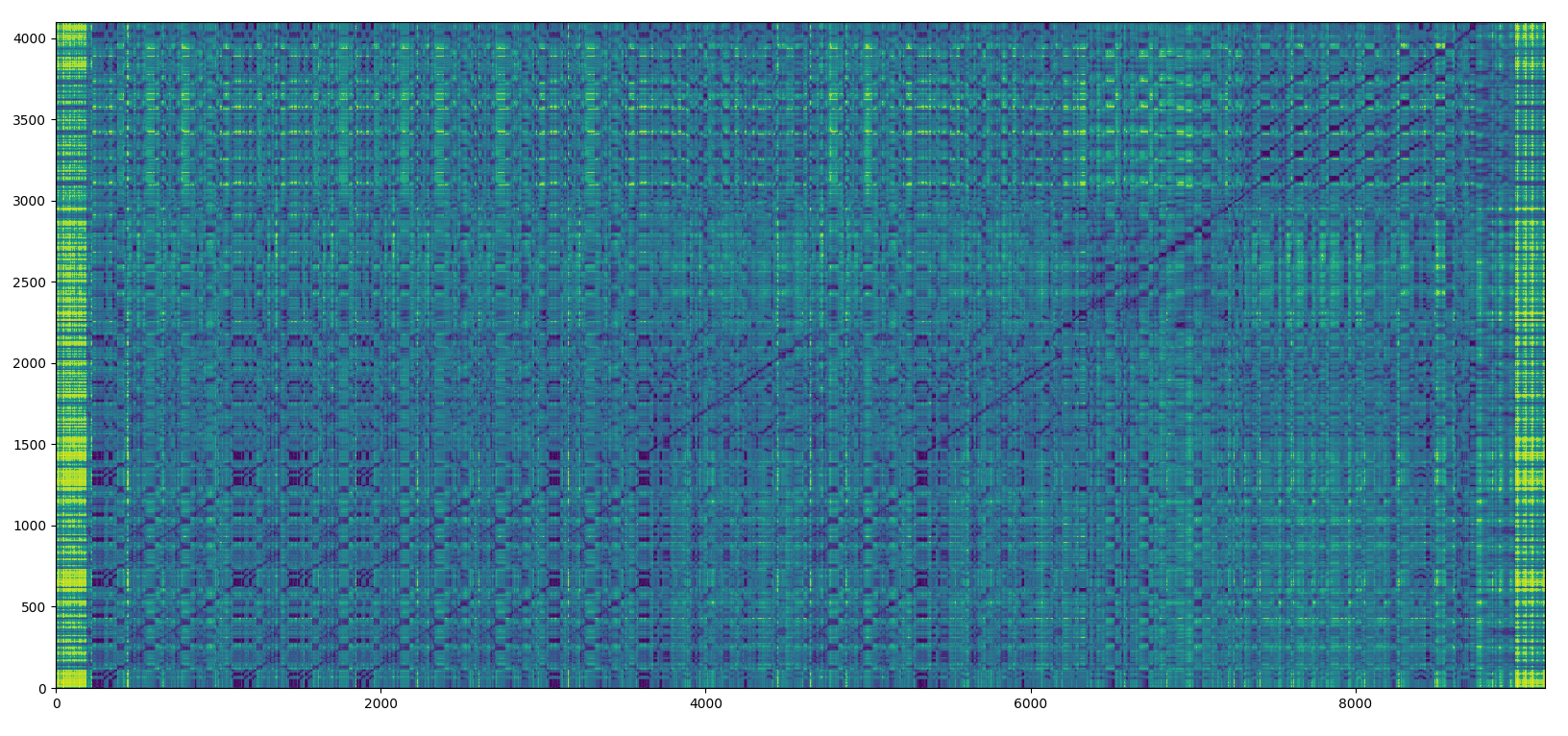} }}%
    \qquad
        \subfloat[\begin{math}\textit{SiameseDTW}\end{math}]
    {{\includegraphics[width=5cm, height=5cm]{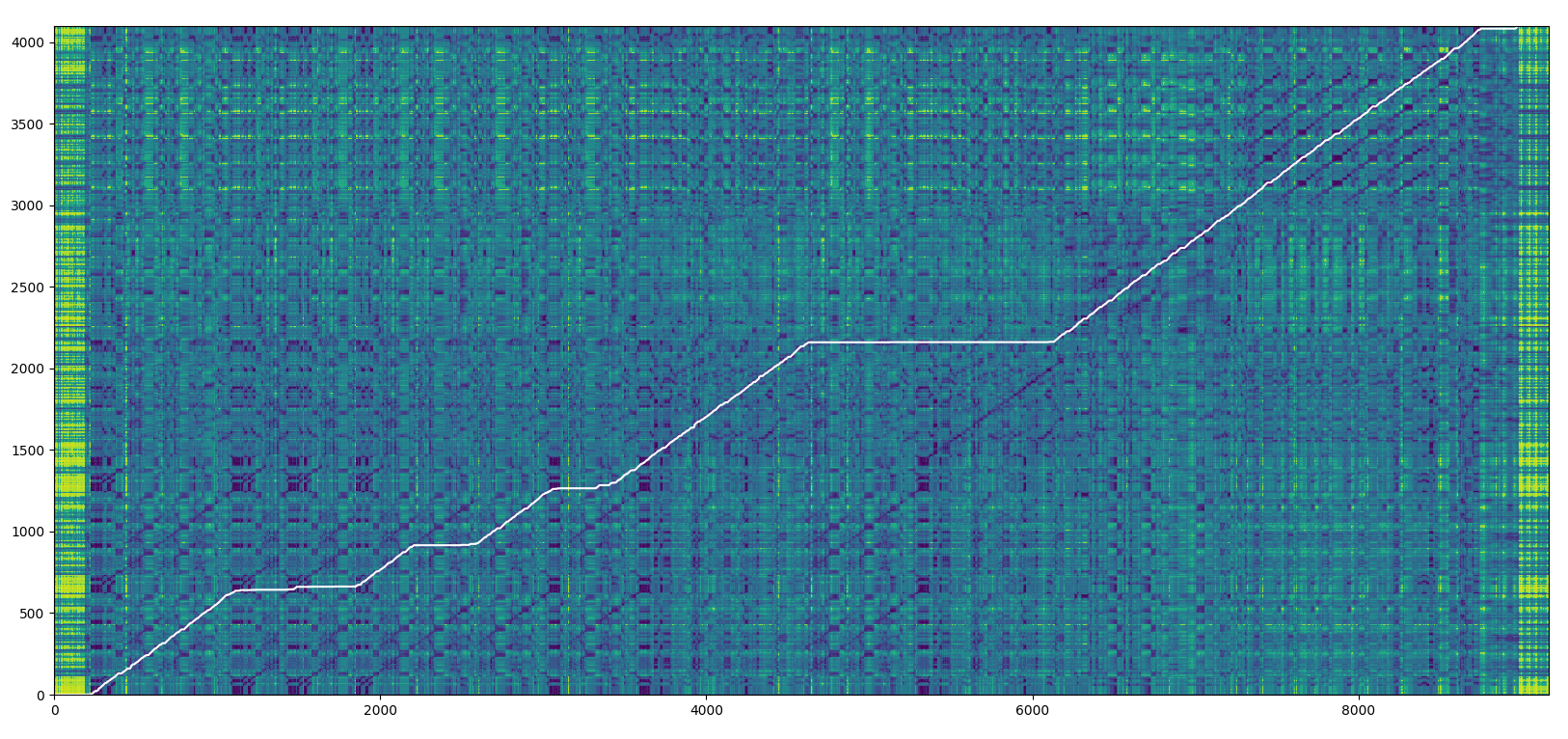} }}%
    \\
    \subfloat[\emph{CA}${}_{custom}$]
    {{\includegraphics[width=5cm, height=5cm]{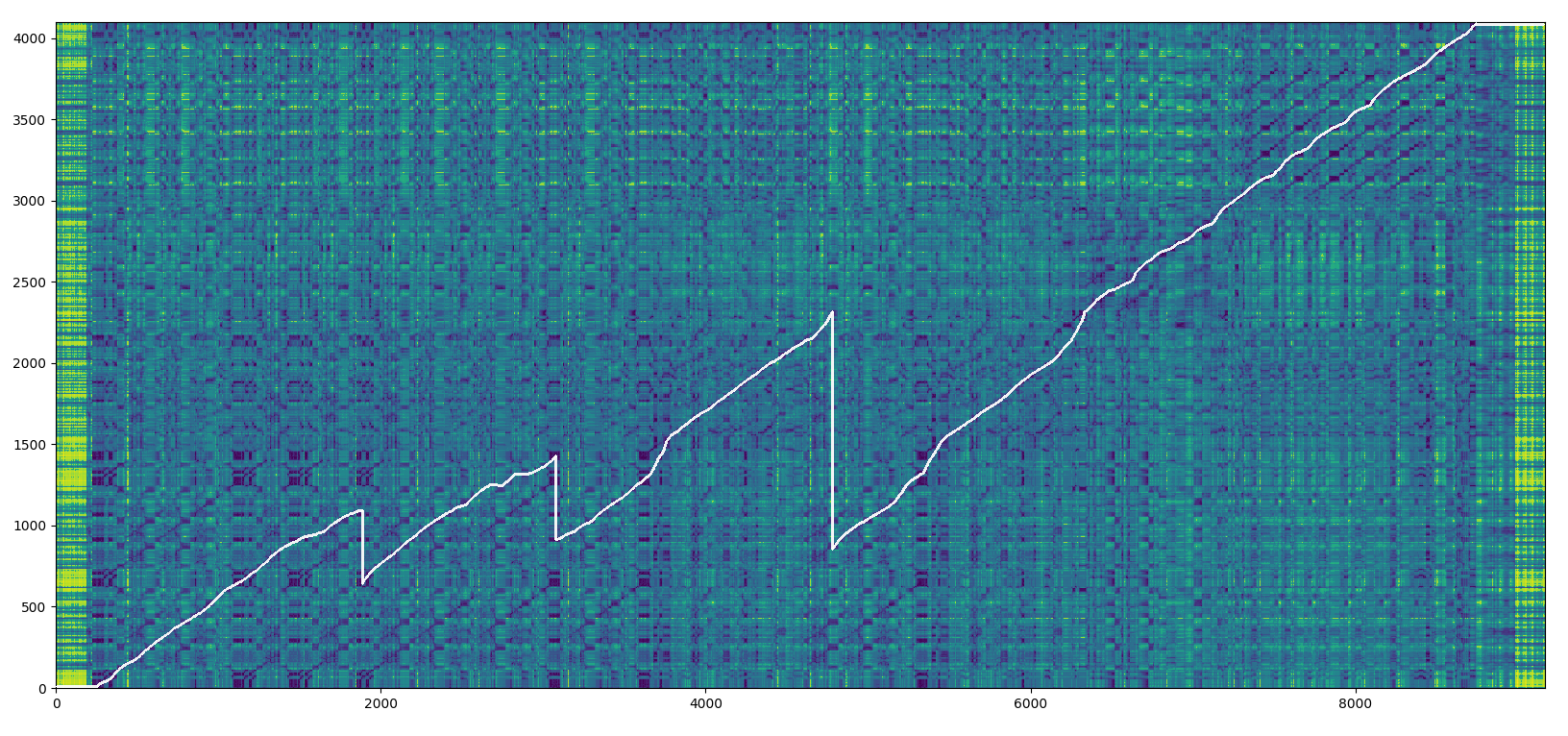} }}%
    \qquad
        \subfloat[Ground Truth]
    {{\includegraphics[width=5cm, height=5cm]{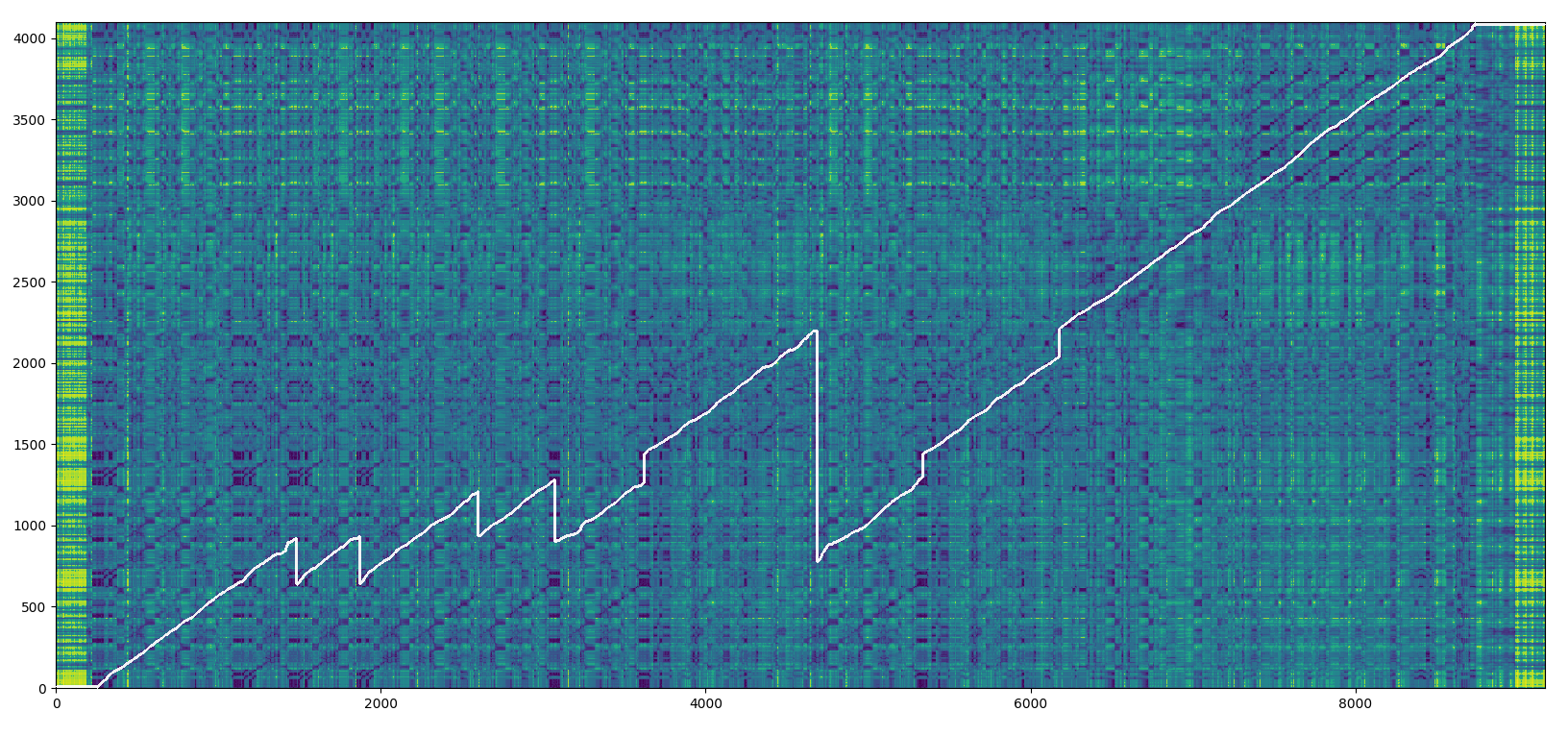} }}%
    \\
        \caption{Examples of alignment plots:\\(a) Input: Cross-similarity matrix between the performance and the score \\(b) Predictions of the SiameseDTW model \\ (c) Predictions of the \emph{CA}${}_{custom}$ model and \\ (d) The ground truth alignment path\\
    X-axis: Frame index (performance), Y-axis: Frame index (score)}
    \label{fig:comparison_1}%
\end{figure}

\begin{figure}[ht]
    \centering
       \subfloat[Input]
  {{\includegraphics[width=5cm, height=5cm]{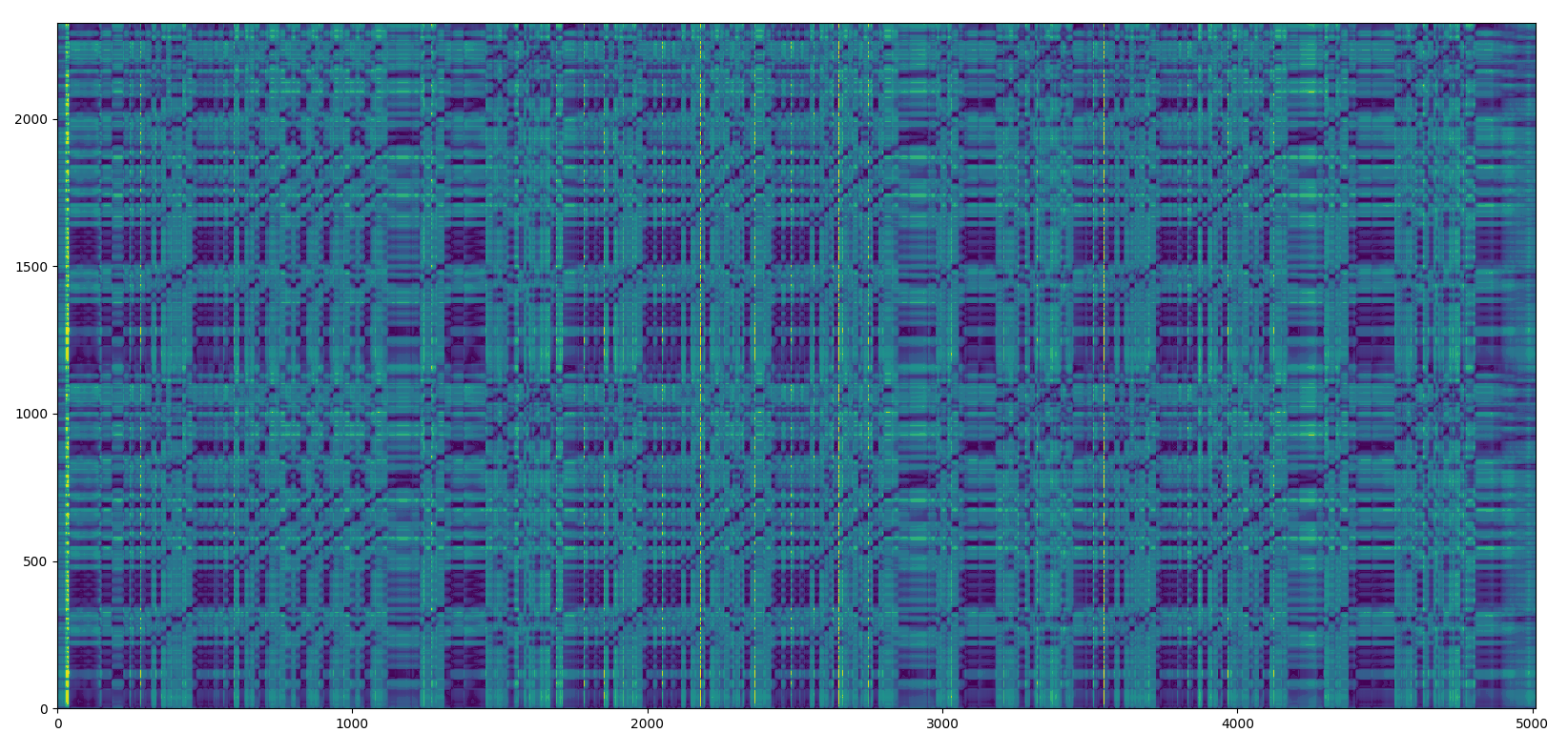} }}%
    \qquad
        \subfloat[\begin{math}\textit{SiameseDTW}\end{math}]
    {{\includegraphics[width=5cm, height=5cm]{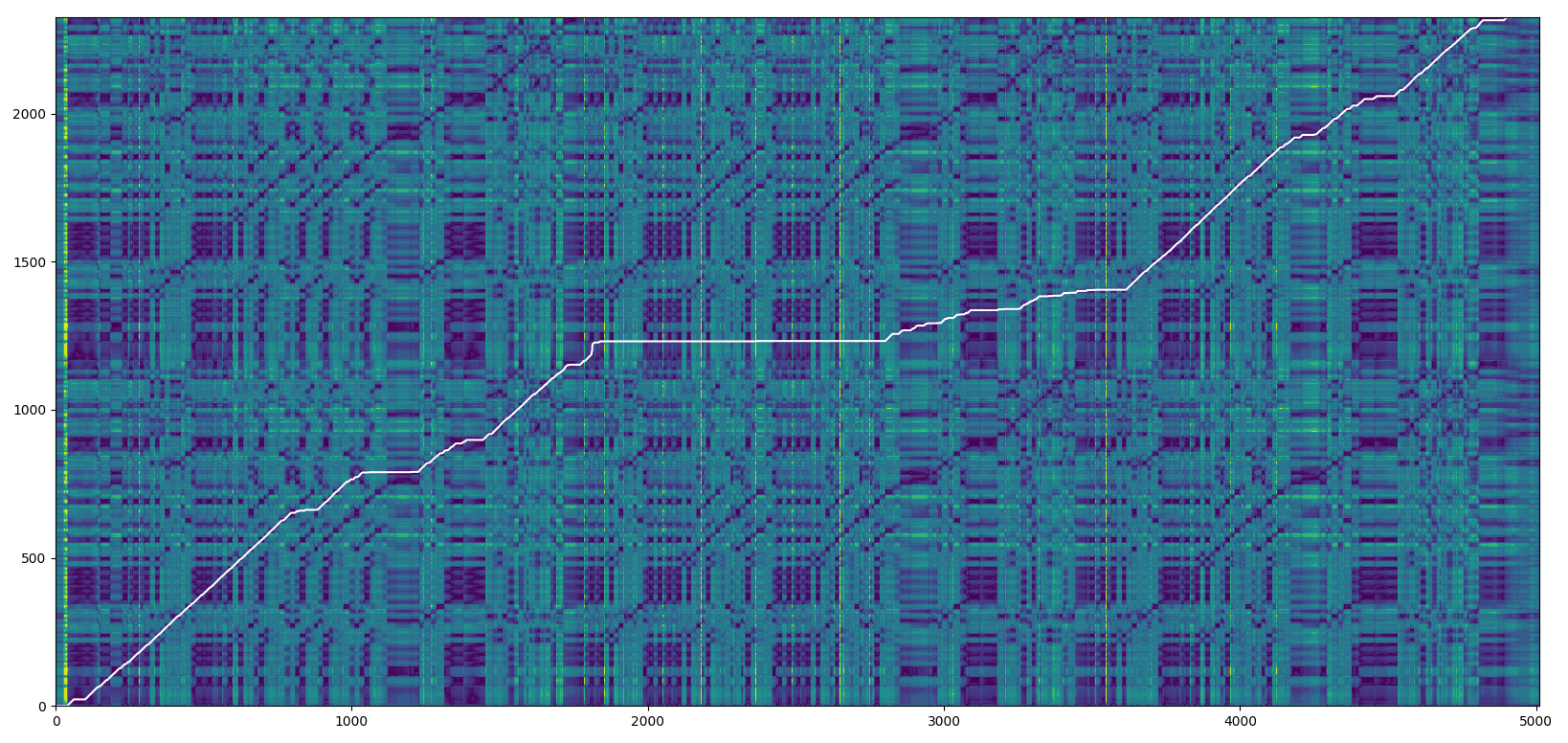} }}%
    \\
    \subfloat[\emph{CA}${}_{custom}$]
    {{\includegraphics[width=5cm, height=5cm]{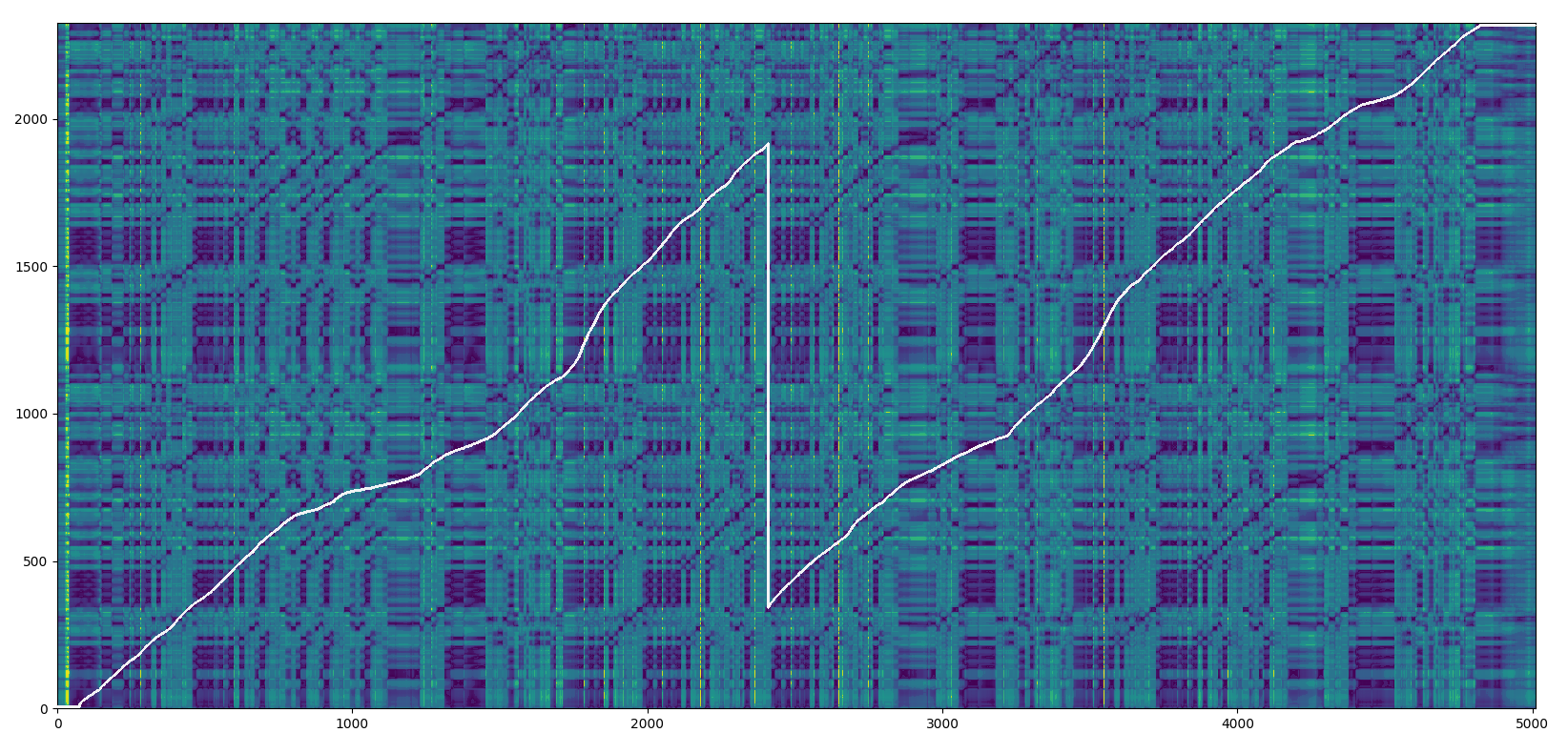} }}%
    \qquad
        \subfloat[Ground Truth]
    {{\includegraphics[width=5cm, height=5cm]{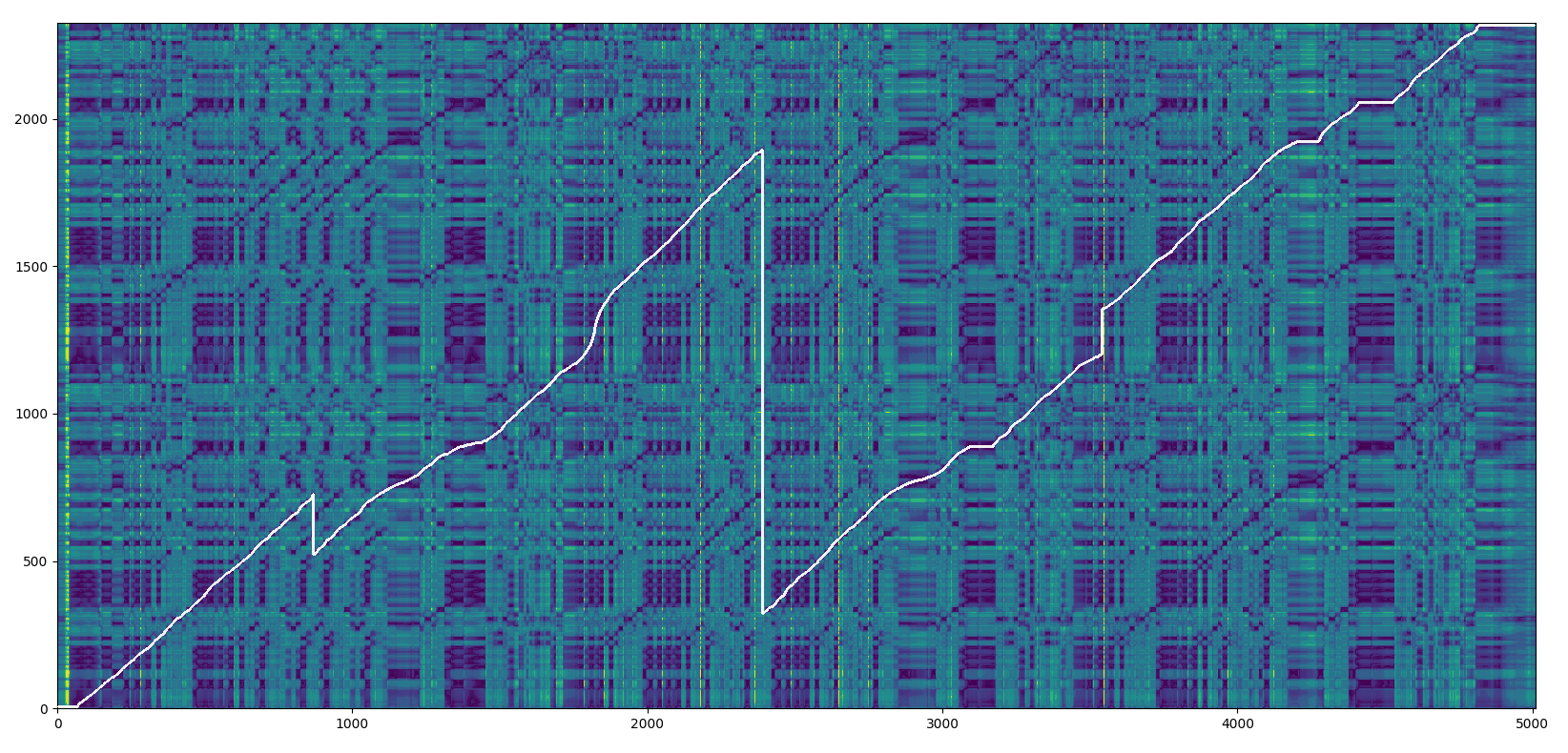} }}%
    \\
        \caption{Examples of alignment plots:\\(a) Input: Cross-similarity matrix between the performance and the score \\(b) Predictions of the SiameseDTW model \\ (c) Predictions of the \emph{CA}${}_{custom}$ model and \\ (d) The ground truth alignment path\\
    X-axis: Frame index (performance), Y-axis: Frame index (score)}
    \vspace{0.5cm}
    \label{fig:comparison_2}%
\end{figure}
\vspace{0.2cm}
This subsection presents the ablation studies that are conducted in order to assess the specific improvements obtained by employing stand-alone self-attention and the custom loss function in our architecture. To this end, the \emph{SASA} layers in the convolutional-attentional model \emph{CA$_{\textit{custom}}$} are replaced with convolutional layers, keeping the input and output dimensionalities constant. The resulting model has a conv-deconv architecture. Experiments are conducted for both the convolutional-attentional and conv-deconv architectures using a cross-entropy loss and the custom loss presented in Section \ref{sec:ch5_loss}. 
\par The Conv-Deconv models are abbreviated as \emph{CD}$_{x}$, with $x$ denoting the loss function employed, i.e. \emph{CE} for the cross-entropy loss and \emph{custom} for the custom loss. The convolutional-attentional models are similarly abbreviated as \emph{CA}$_{x}$. Additionally, experiments are carried out for audio-to-MIDI alignment on the same train and test sets, but using the cross similarity matrix generated using the method proposed in Chapter 3, in order to assess the effect of learnt similarity on the proposed method. This model is abbreviated as \emph{CA$_{\textit{custom-L}}$}.

\par The results of the ablation studies are reported in Table \ref{ablation_score} for audio-to-MIDI alignment and in Table \ref{ablation_image} for audio-to-image alignment. The ablative studies suggest that the convolutional-attentional architecture (\emph{CA}) outperforms the conv-deconv architecture (\emph{CD}) for all error margins for both the tasks. The results demonstrate that the stand-alone self-attention layer yields a 3-5\% improvement in alignment accuracy over a Conv-Deconv architecture without attention.  
 
\par Additionally, the custom loss yields an improvement of 1-3\% over the cross-entropy loss, for both the \emph{CD} and \emph{CA} architectures (rows 6-9), with the \emph{CA}$_{\textit{custom}}$ model yielding the best overall performance among the four configurations. Employing learnt frame similarity as proposed in Chapter 3 \blue{boosts} performance even further (\emph{CA}$_{\textit{custom-L}}$), and is recommended especially in non-standard acoustic conditions \blue{and instrumentation settings, wherein additional information about the target domain such as the presence of artefacts and tuning abnormalities could be leveraged by the models during training and therefore learnt to be handled at test time.}

\par The ablative analysis demonstrates that the \emph{CA}$_{x}$ models also outperform the \emph{CD}$_{x}$ models for structure-aware alignment by 4-6\% (a higher improvement observed on structurally different pieces than overall accuracy), confirming that the stand-alone self-attention layers in the decoder facilitate long-term contextual incorporation. Manual inspection of the alignment plots corroborated that \emph{CA}$_{\textit{custom}}$ was able to capture structural deviations such as jumps and repeats. The reader can find such examples in Figures \ref{fig:comparison_1} and \ref{fig:comparison_2}.

\subsection{A note on multi-modality and end-to-end learning}
 The proposed method is compatible with both uni-modal and multi-modal data, since the similarity and alignment computations are carried our separately. 
While this separation hinders complete end-to-end training, it allows the method to be integrated with learnt feature representations such as cross-modal embeddings (as demonstrated in the audio-to-image alignment experiments), in addition to being applicable in tasks with scarce data but readily available robust feature representations, such as chromagrams.
Similarly, while chroma-based features were used for the primary experimentation for the audio-to-MIDI alignment task, the proposed method can also be used with learnt frame similarity (\emph{CA}$_{\textit{custom-L}}$).
\par It must therefore be noted that while the proposed method is not strictly \textit{end-to-end} in that it does not work directly on raw audio or raw sheet images, however it is \textit{end-to-end} in the more general sense since a single network is trained to generate the alignment paths with a single objective (as opposed to other data-driven methods that compute alignment using DTW after RNN/CNN based pre-processing).


\section{Conclusion and further developments}\label{sec:ch5_conc}
This chapter presented a novel data-driven method for learning alignments for structure-aware performance-score synchronisation. A convolutional-attentional architecture trained with a custom loss based on time-series divergence is proposed. Experiments are conducted for the audio-to-MIDI and audio-to-image alignment tasks pertained to different score modalities. The effectiveness of the proposed architecture is validated via ablation studies and comparisons with state-of-the-art alignment approaches. The results of this experimentation suggest the following:
The results obtained by the proposed method across various settings and the comparisons with previous approaches suggest the following:
\begin{itemize}
    \item The proposed method generates accurate alignment of heterogeneous sequences without reliance on Dynamic Time Warping.
     \item The proposed method is able to capture temporality without requiring recurrence.
     \item The proposed method yields robust performance for alignment tasks involving different score modalities, i.e. audio-to-MIDI and audio-to-image alignment.
    \item Combining stand-alone self-attention layers with a convolutional stem outperforms contemporary alignment approaches as well as a conv-deconv framework across different test settings.
    \item The proposed method effectively handles performances containing structural deviations from the score and is able to deal with long-term dependencies. 
    \item The proposed method is compatible with different feature representations and can therefore be customised to the test setting.
    \item The custom soft-DTW based divergence is an effective loss function for training performance-score synchronisation models.
\end{itemize}

\par The proposed method is thus a promising framework for performance-score synchronisation. In the future, an exploration of multi-modal methods that work directly with raw data could be conducted. The proposed method employed a neural architecture with a fixed output size of 2048. \blue{This limits the size of the performances that the network can model (depending upon the sampling rate).  A detailed quantitative analysis of alignment granularity and sequence length could prove to be a promising exploration to further optimise the models for very long input sequences. While the proposed architecture is able to effectively handle the pieces contained in the test set, a combination of the inflection point detection method along with the convolutional-attentional architecture could prove to be useful for very long performances. Dynamic neural methods that can adjust to the alignment granularity needed for the task at hand could also be explored.} 

\chapter{Conclusion and future work}\label{ch:conclustion}
As the culmination of my thesis, I deliver a synopsis of the main findings and contributions presented in all the chapters. I also posit directions for future work and open problems that serve as promising research directions for further advancements in data-driven alignment. 
\section{Conclusions}
This thesis proposes \textit{context-aware, data-driven} methods for performance-score synchronisation. The chapters incrementally present three machine learning approaches that assist or replace the standard alignment pipeline and generate robust performance in real world settings for various data modalities. The thesis begins with an exploration of representation learning to learn task-specific representations for audio-to-score alignment. Simple experiments with autoencoder models demonstrated the viability of this approach, despite yielding only slight improvements over handcrafted features, and paved the way for the metric learning approach to learn spectral similarity at the frame level, presented in Chapter 3. This chapter proposed Siamese Convolutional Neural Networks to learn the frame similarity matrix, which was then used in the DTW computation to generate the fine alignments. This method proved to be able to learn the similarity values for DTW directly from data without requiring large hand-annotated datasets. This chapter also demonstrated the applicability of the learnt frame similarity across multiple acoustic settings and highlighted the greater domain coverage and adaptability offered by the method over handcrafted features. It furthermore presented a study to analyse the data needs of the model and demonstrated that deep salience representations and data augmentation are effective techniques to improve alignment accuracy in data-scarce conditions. Since the alignment computation in this approach was still being carried out using standard DTW, which only allows for a monotonic alignment path, the performance of the models for structurally different performance-score pairs showed ample scope for improvement. 
\par Drawing from this motivation, Chapter 4 presented a progressively dilated Convolutional Neural Network architecture for structure-aware performance synchronisation. The proposed method incorporated varying dilation rates at different layers of the network to capture both short-term and long-term context and detect inflection points marking the structural mismatches between the performance-performance or performance-score pairs, thereby enabling structure-aware alignment. This chapter also presented experimentation across multiple test settings and conducted ablative analyses to delineate the improvements on structure-aware and overall alignment and to determine the optimal levels of dilation. The results presented in this chapter suggested that progressively increasing dilation with network depth yielded better results than standard convolutions or consistently dilated convolutions, and the proposed method successfully captured various kinds of structural differences regardless of their source and type, outperforming previously proposed structure-aware methods without requiring a large hand-annotated dataset. While the proposed method improved results noticeably for structure-aware alignment, the performance for monotonic alignment also improved slightly, which could be attributed to the handling of small jumps that are not encoded as structural differences in the test annotations. This emphasised the applicability of the method for performance synchronisation regardless of the structural agreement of the test data. This chapter additionally demonstrated that employing the learnt frame similarity proposed in Chapter 3 improved the performance of the dilated CNN models even further.
\par Chapters 3 and 4 demonstrated the effectiveness of neural frameworks to assist DTW for \textit{context-aware} alignment, however the reliance on DTW still hindered end-to-end learning in a completely data-driven manner. To this end, Chapter 5 developed a novel neural architecture that enabled learning the alignments (and not just the representations) from the data itself too, thereby eschewing the reliance on DTW and furthering the development of data-driven alignment. This chapter proposed a convolutional-attentional neural framework trained with a custom loss based on time-series divergence for performance-score synchronisation. Experiments were conducted for the audio-to-MIDI and audio-to-image alignment tasks pertaining to different score modalities and the effectiveness of the proposed architecture was validated via ablation studies and comparisons with state-of-the-art alignment approaches. The results presented in this chapter demonstrated that the proposed approach outperforms previous synchronisation methods for a variety of test settings across score modalities and acoustic conditions. The ablative analyses confirmed that combining stand-alone self-attention layers with a convolutional stem outperforms contemporary alignment approaches as well as a conv-deconv framework for both audio-to-MIDI and audio-to-image alignment and that the custom loss based on time-series divergence is an effective loss function for training performance synchronisation models. It also showed that the method is also robust to structural differences between the performance-score pairs, which is a common limitation of standard alignment approaches. 
\section{Future work}

Having summarised the main findings of the thesis, 
I would now like to present an outlook for the future by highlighting two research directions that offer promise for further exploration.
\subsection{Balancing alignment granularity and input length}
This thesis focused on offline alignment and the methods proposed herein predict the entire alignment path at once, as opposed to predicting the path token by token, as is done by sequence to sequence models such as LSTMs and Transformers. 
While this formulation was motivated by the need for structure-aware alignment and the proposed architectures were successful at capturing multi-scale context and structural differences, it limits the length of the inputs that can be captured. 
\par The problem of building models that can yield precise fine alignments while being able to model very long performances offers multiple avenues for further development. To this end, an exhaustive analysis of fine-grained alignment performance for long inputs and the development of neural methods aimed specifically at generating fine alignments could be explored. This requires an abundance of data containing precise fine-grained ground truth alignment annotations. The creation of such datasets is therefore an important bottleneck for future developments and is the next logical step in my opinion to foster further research in this direction. 
Hierarchical neural frameworks drawing inspiration from \citet{muller2006efficient} and \citet{shan2020improved} could also prove to be an effective endeavour.
\par While the methods presented in Chapter 3 and 4 worked with cross-similarity matrices as the input representations, future endeavours could be carried out using multi-encoder convolutional-attentional models, and instead of one-shot prediction, predicting one token at a time could be explored further. The inputs for such a model could be sliding spectral windows (say spanning 10 seconds) with the model trained to predict the location in the second sliding window corresponding to the central frame of the first window at each timestamp. Possible challenges to overcome with this approach would be to ensure synchronous beginning and end of the inputs and optimising the window length to prevent error propagation and ensure that the aligned frame-pairs actually belong to the given input windows despite tempo changes in the performances. The incorporation of structure would also be a challenging problem to tackle with this formulation. Additionally, as with sequential models, another important challenge to address would be the handling of long input sequences. Combining such a convolutional-attentional framework in a hierarchical manner drawing inspiration from \citet{muller2006efficient} and \citet{shan2020improved} could prove to be effective at overcoming some of these challenges. 
\subsection{Intelligent adaptive systems}\label{sec:adaptiveAlignment}
Until large-scale, cross-modal datasets with high quality fine-grained annotations are made available, the development of intelligent adaptive systems that could learn from limited data is a promising direction. 
As an example, such a system would be able to continuously update the network parameters by learning from user corrections, and create a positive feedback loop which jointly optimises both parameter tuning as well as the user experience.
\begin{figure}[H]
\vspace{1cm}
\begin{center}
\includegraphics[width=6in]{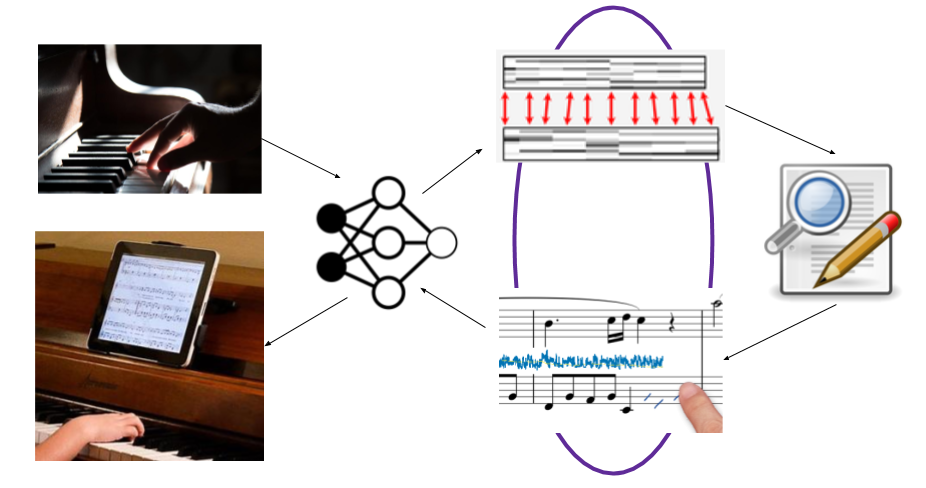}
\vspace{0.3cm}
\caption{A vision for the future - Creating a positive feedback loop for adaptive alignment systems}
\label{asma}
\end{center}
\end{figure}
\par Such corrections could come from the users either in the form of annotations on top of an automatically generated alignment (for instance one provided by commercial software) by means of providing anchor points constraining the alignment path to pass through them. Another possible source of user inputs could be global annotations marking the salient blocks, either through manual inputs or using user data such as eye tracking. 
While automatic post-editing has shown promising results in Neural Machine Translation \citep{tebbifakhr2018multi}, automatically improving alignment performance by learning from human corrections (or user context) and adapting to the user's acoustic conditions and instrumentation settings is something that is relatively unexplored for music alignment. An example of such a positive feedback loop is presented in Figure \ref{asma}, providing a vision for future work. To this end, research on optimal ways to effectively leverage user context as well as developing effective techniques for adapting an existing model to a stream of manual corrections could be explored.
Instance-based adaptation is another promising technique to be investigated, which could offer multiple advantages over vanilla domain adaptation. This would entail querying for a (small) set of training observations similar to the current one and fine tuning the parameters of the network on these observations. 
\par The impact of the research carried out as part of this PhD would be twofold. 
We foresee a significant impact on both the scientific community and the music market in the long run. The scientific community interested in music synchronisation and alignment would benefit from this research, and will hopefully develop it further after the PhD. There is an unprecedented amount of interest generated by automatic music processing tools, coupled with the advancement in artificial intelligence, making it possible to build systems which are capable to cut the costs of human intervention on difficult tasks like transcription and alignment. The preparation of training data for automatic transcription is hugely valuable, and robust alignment methods offer avenues for large-scale dataset creation by reducing human effort. The methods proposed in this thesis could be employed for building robust systems to aid digital music education, wherein 
alignments could be used to better demonstrate musical concepts, for automatic assessment and also to indicate to the students where their performance deviates from indicated score markings. 
Additionally, robust alignment can also aid audio editing and analysis where selecting a measure in the score could automatically select the corresponding audio, enabling convenient navigation. 
From a cultural perspective, the envisaged reduction of the costs of automatic alignment aims to promote the diffusion of content (hence knowledge and culture) across musical genres, even for niche music material for which, at current costs, the industry would be scarcely inclined to invest. 

\par I hope that the research presented in this thesis bolsters the endeavour towards intelligent, adaptive, data-driven alignment models that are capable of learning meaningful relationships from raw data even in data-scarce conditions. The methods proposed in this thesis could also impact related branches of music information processing such as cross-modal retrieval of audio from images of sheet music and vice versa, and structure-aware music generation. Lastly, while the methods presented in this thesis focussed on the \textit{performance-score} alignment scenario, some of the proposed techniques could be tweaked or extended to build generic sequence alignment methods 
for other domains of research such as automatic video captioning, subtitle synchronisation and protein sequencing. The adaptations needed would be dependent on factors such as the need for structure-aware alignment, desired alignment granularity and the data modalities to be dealt with. The convolutional-attentional architecture proposed in Chapter 5 could then be modified to cater to the application setting, for instance by insertion or removal of convolutional and/or stand-alone self-attention layers, or modifying the input and output  dimensionalities.

\newpage
\nocite{agrawal2017integrating, agrawal2017towards, agrawal2017building, agrawal2017experiments, agrawal2017vis, amirhossein2018multi, tebbifakhr2018multi, amirhossein2018multi2, tebbifakhr2018multi1}

\bibliography{refs}
\bibliographystyle{apalike}
\fancyhead[L]{Bibliography}
  
\cleardoublepage
\end{document}